\pdfoutput=1
\documentclass[11pt]{article}
\usepackage{jcapmod}
\usepackage{shorthand}

\usepackage{mathtools}
\usepackage{booktabs}
\usepackage[english]{babel}
\usepackage{amsmath,amssymb,amsbsy,amstext, amsthm, simplewick, amsfonts}
\usepackage{hyperref}
\usepackage{graphicx}
\usepackage[small]{caption}
\usepackage{siunitx}
\usepackage{upgreek}
\usepackage{framed}
\usepackage{wrapfig}
\usepackage{multirow}
\usepackage{bbm}
\usepackage[utf8x]{inputenc}
\usepackage{selinput}
\usepackage{bm}
\usepackage{float}
\usepackage{geometry}
\usepackage{hyperref}
\usepackage{yfonts}
\usepackage{caption}
\usepackage{subcaption}
\usepackage{sidecap}
\usepackage{longtable}
\usepackage{anyfontsize}
\usepackage{dsfont}

\usepackage[framemethod=default]{mdframed}
\newmdenv[skipabove=7pt,
skipbelow=7pt,
rightline=false,
leftline=false,
topline=false,
bottomline=false,
backgroundcolor=gray!10,
linecolor=gray,
innerleftmargin=5pt,
innerrightmargin=5pt,
innertopmargin=5pt,
innerbottommargin=5pt,
leftmargin=0cm,
rightmargin=0cm,
linewidth=4pt]{eBox}
\newmdenv[skipabove=7pt,
skipbelow=7pt,
rightline=false,
leftline=false,
topline=false,
bottomline=false,
backgroundcolor=gray!10,
linecolor=gray,
innerleftmargin=5pt,
innerrightmargin=5pt,
innertopmargin=-5pt,
innerbottommargin=5pt,
leftmargin=0cm,
rightmargin=0cm,
linewidth=4pt]{eBox2}

\definecolor{blue3}{RGB}{31, 119, 180}
\definecolor{red3}{RGB}{	214, 39, 40}
\definecolor{orange3}{RGB}{255, 127, 14}
\definecolor{green3}{RGB}{44, 160, 44}
\definecolor{repBlue}{RGB}{31, 119, 180}
\definecolor{repRed}{RGB}{	214, 39, 40}
\definecolor{repGreen}{RGB}{44, 160, 44}

\def\be{\begin{equation}}
\def\ee{\end{equation}}
\newcommand{\bea}{\begin{eqnarray}}
\newcommand{\eea}{\end{eqnarray}}

\def\vp{\varphi}

\def\bge{\begin{equation}}
\def\ede{\end{equation}}
\def\bga{\begin{aligned}}
\def\eda{\end{aligned}}
\def\bgb{\begin{bmatrix}}
\def\edb{\end{bmatrix}}
\def\bgp{\begin{pmatrix}}
\def\edp{\end{pmatrix}}
\def\bgm{\begin{matrix}}
\def\edm{\end{matrix}}
\def\bgs{\begin{subequations}}
\def\eds{\end{subequations}}

\def\di{{\mathrm{d}}}

\def\pd{\partial}

\def\la{\langle}\def\ra{\rangle}

\def\to{\rightarrow}

\def\ii{\mathrm{i}}

\def\be{\beta}

\def\lam{\lambda}

\def\si{\sigma}

\def\Re{\mathrm{Re}\,}
\def\Im{\mathrm{Im}\,}

\def\2F1{{}_2\mathrm{F}_1}
\def\3F2{{}_3\mathrm{F}_2}

\newcommand{\wh}[1]{\mkern 2mu \widehat{\mkern-2mu#1\mkern-2mu}\mkern 2mu}
\def\tf{{\eta_0}}
\newcommand{\n}{\nonumber}

\usepackage{colortbl}
\definecolor{lightgreen}{cmyk}{0.2, 0, 0.2, 0.2}
\definecolor{lightgray}{cmyk}{0.1,0.2,0,0.1}
\definecolor{lightgray2}{cmyk}{0.1,0.1,0,0.1}

\makeatletter
\newlength{\apb@width}
\newcommand{\autoparbox}[2][c]{\settowidth{\apb@width}{#2}\parbox[#1]{\apb@width}{#2}}

\makeatother


\def\beq{\begin{equation}}
\def\eeq{\end{equation}}

\allowdisplaybreaks[1]
\begin{document}

\newgeometry{top=2cm, bottom=2cm, left=2.9cm, right=2.9cm}

\begin{titlepage}
\setcounter{page}{1} \baselineskip=15.5pt 
\thispagestyle{empty}

\begin{center}
{\fontsize{20}{18} \bf  $\lambda\phi^4$ as an Effective Theory in de Sitter }\\ 
\end{center}

\vskip 30pt

\begin{center}
\noindent
{\fontsize{12}{18}\selectfont Sebasti\'an C\'espedes$^1$,
Zhehan Qin$^{2,4}$
and Dong-Gang Wang$^{3,4}$}
\end{center}

\vskip 20pt

\begin{center}
  \vskip8pt
   {$^1$ \fontsize{12}{18}\it Abdus Salam Centre for Theoretical Physics, \\Imperial College, London, SW7 2AZ, UK
}

  \vskip8pt
  {$^2$ \fontsize{12}{18}\it Department of Physics,  Tsinghua University, Beijing, 100084, China
}

  \vskip8pt
{$^3$ \fontsize{12}{18}\it Institute for Advanced Study and Department of Physics,\\ Hong Kong University of Science and Technology, Clear Water Bay, HK, China
}

  \vskip8pt
{$^4$ \fontsize{12}{18}\it Department of Applied Mathematics and Theoretical Physics,\\ University of Cambridge,
Wilberforce Road, Cambridge, CB3 0WA, UK}

\end{center}

%
%

%=========================================
\vspace{0.4cm}
 \begin{center}{\bf Abstract} 
 \end{center}
 \noindent
Effective field theories (EFTs) provide a powerful framework to parametrise unknown aspects of possible ultraviolet (UV) physics. For scalar fields in de Sitter space, however, new emergent phenomena can arise when the cut-off scale of the theory lies below the horizon scale $H$, as seen in the stochastic formalism of inflation. In this work, we study EFTs that, at leading order, reproduce the standard quartic theory in de Sitter, but with a variable cut-off identified with the mass of an integrated-out hidden sector. We perform the complete analytic computation for the tree- and loop-level matching between the effective $\lambda\phi^4$ theory and two possible UV realisations. We find that when the cut-off is much larger than the horizon, the theory admits a unitary description, up to exponentially suppressed corrections. In contrast, when the cut-off is lowered below $H$, the system evolves into a mixed state and diffusive effects emerge. Nevertheless, at leading order, the EFT remains local and reproduces the same effective quartic coefficient as in the unitary regime. Furthermore, for the EFT matching at the loop-level, the effective quartic coupling changes sign and becomes negative as the cut-off decreases, in agreement with the result obtained from the stochastic formalism. 
In general, for cosmological EFTs, our findings highlight the role of non-unitary effects and illustrate their regimes of validity, within and beyond perturbation theory.

\noindent

\end{titlepage}

\newpage

\restoregeometry
\setcounter{tocdepth}{3}
\setcounter{page}{1}
\tableofcontents

\newpage
%=======================================
\section{Introduction}
%=======================================
\addcontentsline{}{}{}

Effective field theories (EFTs) play a fundamental role in contemporary physics. They offer a systematic framework for parameterising the influence of short-wavelength/ultraviolet (UV) degrees of freedom on long-wavelength/infrared (IR) observables. From the Wilsonian perspective, integrating out high-energy modes yields a low-energy effective action organised as an expansion in higher-dimensional local operators, each suppressed by powers of a physical cut-off scale~\cite{Weinberg:1978kz,Polchinski:1992ed,Georgi:1993mps,Burgess:2007pt}. Such a framework is particularly powerful in cosmology, where access to early-universe physics is limited to indirect observations. As a result, UV information must be inferred by measuring the coefficients of EFT operators at late times~\cite{Cheung:2007st,Weinberg:2008hq,Green:2022ovz}.

However, our ability to construct such EFTs is constrained by the limited understanding of quantum field theory (QFT) in curved spacetimes, especially in de Sitter space. Even the simplest models involving light scalar fields, whether with self-interactions or interactions with spectator fields, pose significant challenges, and basic observables — such as the four-point functions — have only recently been fully understood. This progress has required the development of conceptual and advanced analytical computational tools, including cosmological bootstrap~\cite{Arkani-Hamed:2015bza,Arkani-Hamed:2018kmz,Baumann:2019oyu,Pajer:2020wxk,Pimentel:2022fsc,Jazayeri:2022kjy,Qin:2022fbv,Wang:2022eop,Qin:2023ejc,Aoki:2024uyi,Liu:2024str,Qin:2025xct}, Mellin-Barnes techniques~\cite{Sleight:2019hfp,Sleight:2019mgd,Premkumar:2021mlz,Qin:2022lva,Qin:2022fbv,Xianyu:2023ytd,Qin:2024gtr}, spectral decomposition~\cite{Marolf:2010zp,Xianyu:2022jwk,Loparco:2023rug,Werth:2024mjg,Qin:2025xct,Zhang:2025nzd}, dispersion integral~\cite{Liu:2024xyi}, diagrammatic  methods \cite{Baumgart:2019clc, Benincasa:2019vqr,Benincasa:2024ptf,Arkani-Hamed:2024jbp} and new techniques to deal with loop integrals and their renormalisation~\cite{Xianyu:2022jwk,Lee:2023jby,Qin:2023bjk,Qin:2023nhv,Huenupi:2024ksc,Baumann:2024mvm,Benincasa:2024ptf,Ballesteros:2024cef,Qin:2024gtr,Bhowmick:2025mxh,Braglia:2025cee,Zhang:2025nzd,Wang:2025qfh,Jain:2025maa}.

Most cosmological EFTs have been constructed by analogy with flat-space field theory. A paradigmatic example is the EFT of inflation~\cite{Cheung:2007st}, which describes the Goldstone boson arising from spontaneous breaking of the time translation symmetry. This theory includes all operators invariant under non-linearly realised time diffeomorphism and is valid below a cut-off scale,
which is typically much above the Hubble scale.
However, this framework is tailored to the inflationary background and does not capture the full structure of QFT in de Sitter ~\cite{Salcedo:2022aal,Green:2024cmx,Burgess:2024heo}. For instance, even a simple theory with quartic interactions on a fixed de Sitter background exhibits qualitatively different behaviour on subhorizon and superhorizon scales.

Let us take a closer look at a scalar field theory in de Sitter space with potential interactions. On subhorizon scales, the dynamics closely resemble those of flat space, and perturbation theory remains well-controlled. However, once modes cross the horizon, perturbative expansions become increasingly unreliable, and qualitatively new phenomena emerge. In the absence of shift symmetries, light scalar fields can lead to the breakdown of perturbation theory at late times due to the accumulation of IR effects. In this regime, there is an emergent stochastic approach which offers a powerful non-perturbative framework, effectively capturing the late-time dynamics by coarse-graining over superhorizon modes and treating the long-wavelength sector semiclassically~\cite{Starobinsky:1986fx,Starobinsky:1994bd,Baumgart:2019clc,Gorbenko:2019rza,Mirbabayi:2019qtx,Cespedes:2023aal}.

This emergent regime is characterised by modified power-counting rules, akin to those in large-occupancy many-body systems, and allows for a controlled resumation of secular IR effects~\cite{Burgess:2009bs,Burgess:2014eoa,Baumgart:2019clc,Cespedes:2023aal}. Crucially, the resulting effective description does not necessarily preserve unitarity. This is a consequence of the fact that, on superhorizon scales, the system effectively behaves classically, and quantum coherence is lost through the accumulation of squeezed states. The appropriate language in this limit is that of an open quantum system, where long-wavelength modes interact with short-wavelength degrees of freedom that have been traced out \cite{Burgess:2015ajz,Li:2025azq}. Another analogous point of view is that the stochastic equation emerges as a time-dependent UV cut-off scale is taken to smear out the short-wavelength modes. This effect is controlled by a renormalisation group (RG) equation that can be shown to be equivalent to {the diffusion term} in the Fokker-Planck equation~\cite{Cespedes:2023aal}.

Although our present focus is on IR-induced effects, we note that there are two distinct sources of subtlety in constructing EFTs in de Sitter space: (1) the breakdown of unitarity due to decoherence and classicalisation of long modes as well as the expansion of spacetime background itself, and (2) the enhanced role of IR divergences and the need to resum secular growth. Both aspects are intertwined but conceptually distinct, and recent work has begun to clarify how they manifest themselves in local EFTs for cosmology and general open systems~\cite{Gao:2018bxz,Salcedo:2025ezu,Colas:2025app,CCoEFT}.

These considerations motivate a broader notion of EFT in cosmology—one that extends beyond standard Lagrangian approaches. In time-dependent, out-of-equilibrium backgrounds, the appropriate object to describe the system is the {\it reduced density matrix} rather than the wavefunction. Integrating out short modes then leads not only to corrections in long-wavelength dynamics, but also to open-system effects such as dissipation, decoherence, and diffusion~\cite{CCoEFT,Frangi:2025xss,Balasubramanian:2011wt,Salcedo:2024smn,Li:2025azq}.

In this paper, we revisit the construction of an EFT for a light scalar field in de Sitter, using $\lambda\phi^4$ theory as a benchmark. We consider a light scalar field $\phi$ directly coupled to another \emph{massive} scalar sector $\sigma$, and derive an effective description in which the quartic interaction of $\phi$ dominates at leading order. By analysing the four-point function of $\phi$, which displays an IR-divergent contribution that becomes dominant on superhorizon scales, we explore how different EFT regimes emerge depending on the mass of the $\sigma$ fields. This analysis allows us to organise the operator matching to the UV theory, track the breakdown of perturbative control, and identify the onset of semiclassical behaviour. 

In particular, we trace how the effective action for the density matrix\footnote{The interaction part is known as the influence functional~\cite{Feynman:1963fq}.} evolves as the cut-off scale is lowered toward superhorizon momenta. In this regime diffusive effects dominate in a similar way to the case of stochastic inflation. This offers an insight into the emergence of this description and clarifies the regime of validity of the resulting effective theory. Our approach provides a unified framework for understanding how local, unitary EFTs evolve into semiclassical, stochastic descriptions as physical scales redshift across the Hubble radius.

\paragraph{Summary of results}
We consider interactions of the form,
\begin{align} \label{action}
    S=\int d^4x\sqrt{-g}
    \left[-\frac12(\partial_\mu\phi)^2-\frac12(\partial_\mu\sigma)^2-\frac12m^2\sigma^2-\frac12\phi^2f(\sigma)-V(\phi,\sigma)\right] \ ,
\end{align}
where $g = \det g_{\mu\nu}$ is the determinant of the de Sitter metric, $f(\sigma)$ is a polynomial function of a second scalar field $\sigma$ with mass $m$, and $V(\phi, \sigma)$ is a potential containing interaction terms that are at least cubic in $\phi$ and $\sigma$. Upon integrating out $\sigma$, the leading contribution is a quartic interaction for $\phi$, resulting in an effective theory. The corresponding four-point function exhibits an IR divergence at tree level, since the interaction remains active even after the modes cross the horizon,
\begin{align}
\langle\phi_{\bm{k}_1}\phi_{\bm{k}_2}\phi_{\bm{k}_3}\phi_{\bm{k}_4}\rangle'
\supset \frac{\lambda_{\mathrm{eff}}}{4} \frac{H^4}{\prod k_i^3}
\sum k_i^3 \log\left(-\sum k_i \eta_0\right) \ ,
\label{eq:4pt}
\end{align}
where $\lambda_{\rm eff}$ is the effective quartic coupling that depends on the parameters of the UV model,  $\bm k_i$ are the external momenta, $\eta_0$ is the conformal time at which the correlation function is evaluated, and by a prime we denote that the momentum conservation $\delta$ function has been dropped off.

At sufficiently late times, perturbation theory eventually breaks down, and a non-perturbative resummation via stochastic methods becomes necessary. However, before this breakdown, the IR-divergent term in \eqref{eq:4pt} dominates the tree-level four-point function, and can be used to organise the effective field theory description.

We construct an effective action for the density matrix that reproduces \eqref{eq:4pt} at leading order, where the coefficient is obtained via a standard matching process. In the case where $\sigma$ is heavy, the resulting theory is described by a unitary action on a two-fold Schwinger-Keldysh contour, with higher-derivative and non-local corrections suppressed by at least inverse powers of the mass. In particular, interactions between the forward and backward contours are exponentially suppressed, and thus the EFT remains unitary~\cite{Gwyn:2012mw,Jazayeri:2022kjy,Jazayeri:2023xcj,Cespedes:2025dnq}. As a result, the effective action contains two vertices at leading order—one on each branch of the contour—and exhibits the correct pole structure~\cite{Pajer:2020wnj}. This result also aligns with our intuition: in the heavy mass regime, the curvature of spacetime background is negligible, so we should recover the traditional unitary EFT as in flat space.

As the mass is lowered, the spacetime expansion gains importance, and the non-unitary contributions from cross-contour interactions become significant, causing the unitary EFT description for the whole system to break down. However, in the regime where $m^2\ll H^2$, we demonstrate that a non-unitary EFT can still be constructed, accurately reproducing \eqref{eq:4pt}. We propose the EFT on superhorizon scales to have the following form,
\begin{align}
      S_{\mathrm{EFT}}&=-\int d^3x\int dt\sqrt{-g} \left[\partial_\mu\phi_a\partial^\mu\phi_r+\lambda_{\mathrm{eff}}\phi_a\phi_r^3+\bar\lambda_{\mathrm{eff}}\phi_a^3\phi_r\right]\nonumber\\&\qquad+i\int d^3x\int d^3y \int dt a^3(t)  \ \phi_a^2(\bm{x},t)\lambda_\mathrm{NL}(\vert\bm{x}-\bm{y}\vert)\phi_r^2(\bm{y},t) \ ,
      \label{eq:EFTsuperhorizon}
\end{align}
where the fields in the effective action are expressed in the $a/r$ basis of the Schwinger–Keldysh contour, and the real parameters $\lambda$ represent the distinct quartic couplings determined by matching to the corresponding four-point functions.
This EFT describes a mixed state, valid away from the squeezed limit, and its interaction part is dominated by a single vertex, $\phi_a\phi_r^3$, since the scalings of fields $\phi_r$ and $\phi_a$ are different on superhorizon scales. The non-local kernel encodes the degree of decoherence of the system on superhorizon scales, and is also subdominant when computing IR part of the four-point function.

We perform explicit matching of the EFT~\eqref{eq:EFTsuperhorizon} in two representative cases: when $f(\sigma)$ is linear and when it is quadratic in $\sigma$. In the linear case, the resulting Wilson coefficient  $\lambda_\text{eff}$  for the dominant vertex $\phi_a\phi_r^3$ is always negative, as expected from the structure of the cubic interaction. In the quadratic case, we find that this effective quartic coupling transitions from positive (for large mass $m^2\gg H^2$ ) to negative (for small mass $m^2\ll H^2$). We show that this negative coupling in the small mass regime matches the result obtained from a two-field stochastic analysis at equilibrium.

\begin{table}[ht]
    \centering
    \begin{tabular}{| c | c | c | c | c |}
    %\hline
    \hline
    %\vspace{5pt}
                \multirow{2}{*}{\textbf{UV models}} & \multicolumn{2}{c|}{$\alpha\phi^2\sigma$} & \multicolumn{2}{c|}{$g\phi^2\sigma^2$}  \\
                \cline{2-5}
   ~ & pert. & non-pert. & pert. & non-pert. \\
    \hline
   % \hline
    %\vspace{5pt}
    $m\gg H$ & $\lambda_{\rm eff}<0$ & $\times$ & $\lambda_{\rm eff}>0$ & $\lambda_{\rm eff}>0$ with $\phi< \frac{H}{\lambda_{\rm eff}^{1/4}}$ \\ 
     \hline
     % \vspace{5pt}
    $m\ll H$ & $\lambda_{\rm eff}<0$ & $\times$ & $\lambda_{\rm eff}<0$ & Not $\lambda\phi^4$, heavy tail \eqref{PDF:quartic} \\
    \hline
    \end{tabular}
    \caption{Summary of two possible UV realisations of $\lambda \phi^4$ in different regimes.}
    \label{tab:summary}
\end{table}

The paper is structured as follows. In Sec.~\ref{sec:DensityMatrix}, we describe how to define and compute density matrices in cosmology, and in Sec.~\ref{sec:phi4} we focus on the case of a single massless scalar field in de Sitter with quartic interactions. In Sec.~\ref{sec:EFTdS}, we derive an effective description for the density matrix after integrating out a gapped spectator field in a cubic interaction. We compute the tree-level four-point function and identify distinct EFT regimes depending on the mass of the spectator field. In Sec.~\ref{sec:loop}, we turn to analyse a quartic coupling and extract the effective density matrix via a loop-level matching, tracking the sign change of the effective quartic coupling with varying of the mass. In Sec.~\ref{sec:bpt}, we study a two-field stochastic system at equilibrium and compare its predictions with the perturbative results. We conclude in Sec.~\ref{sec:conclussions}. We provide several appendices with all the technical details.

\paragraph{Notation}
We work in the inflationary patch of de Sitter spacetime and use the mostly-plus metric $ds^2=-dt^2 + a^2dx^2$, with the scale factor $a=e^{Ht}$ and $H$ is the Hubble constant. For computational convenience, we also often switch to the conformal coordinates and use the conformal time $\eta \equiv -e^{Ht}/H$, and the metric becomes $ds^2=a^2(-d\eta^2+dx^2)$ with $a=-1/(H\eta)$.
Throughout this paper, we use $\phi(t,{\bf x})$ and $\sigma(t,{\bf x})$ as the bulk fields in dS, and $\varphi({\bf x})=\phi(t_0,{\bf x})$ and $\varsigma({\bf x})=\sigma(t_0,{\bf x})$ are the corresponding profiles on the future boundary at time $t_0$ (or $\eta_0$ in conformal time).
The field variables with $+$ and $-$ indices are on the forward and backward branches of the Schwinger-Keldysh contour respectively.
We also use the advanced/retarded basis and denote the field variables as $\phi_a$ and $\phi_r$ respectively.
For a scalar field in de Sitter with mass $m$, we define dimensionless mass parameters $\mu\equiv \sqrt{m^2/H^2-9/4}$ and $\nu\equiv \sqrt{9/4-m^2/H^2}$ to denote the mass, which are positive real for the scalar in the principal series ($m>3H/2$) and complementary series ($m<3H/2$), respectively.

%=======================================
\section{Density Matrices in Cosmology}
\label{sec:DensityMatrix}
%=======================================

An appropriate formalism to deal with time-dependent phenomena is the Schwinger-Keldysh formalism~\cite{Schwinger:1960qe,Feynman:1963fq,Keldysh:1964ud,Weinberg:2005vy}. In the cases we will be interested in, the quantum state of the system evolves in time from an initial vacuum state at $t=-\infty$  towards a boundary state defined at $t=t_0$, to then evolves back into the past to the same initial state. This allows us to take into account the dynamics at the time $t_0$ without having to make assumptions about the asymptotic future, as is normally done for scattering amplitudes. The price to pay is that one needs to double the fields and introduce additional propagators. If the initial state is the vacuum $\vert\Omega\rangle$ with the density matrix $\rho_0(t=-\infty)=\vert\Omega\rangle\langle\Omega\vert$, the evolution of an operator $\mathcal{O}$ is given by,
\begin{align}
    \langle\mathcal{O}(t)\rangle=\mathrm{Tr}\left(\rho(t)\mathcal{O}\right)=\mathrm{Tr}(U(t)\rho_0U^\dagger(t)\mathcal{O}) \ ,
\end{align}
where $U(t)$ is the evolution operator starting from $t=-\infty$. These can be described in terms of path integrals with boundaries.  We specify the portion of the contour of the field by labeling;
$\phi_+$ is the bulk field defined in the time-ordered part of the contour, while $\phi_-$ is in the anti-time-ordered part. The prescription over the initial state can be implemented as a boundary condition over the bulk path integral. Assuming the evolution of operators are derived from an action $S[\phi]$, the density matrix is given by,
\begin{align}
\rho[\varphi_+,\varphi_-]=\int_{\Omega}^{\phi_+(t_0)=\varphi_+}\mathcal{D}\phi_+\int_{\Omega}^{\phi_-(t_0)=\varphi_-}\mathcal{D}\phi_- e^{iS[\phi_+]-iS[\phi_-]} \ ,
\end{align}
where $\phi$ is the field $\phi$ defined at $t_0$. Given that the system is unitary, the density matrix can be equivalently written in terms of the wavefunction of the universe as (c.f. Eq.~\eqref{eq_defWF})
\begin{align}
\rho[\varphi_+,\varphi_-]=\Psi[\varphi_+]\Psi^*[\varphi_-] \ .
\label{eq_rho=wv2}
\end{align}
The factorisation of the density matrix assumes unitary evolution from the vacuum up to time $t_0$. While this is not generally valid—interactions can induce off-diagonal components—we will work under the approximation that the system begins in a factorised vacuum state and evolves unitarily up to $t_0$, so that factorisation holds throughout.

To evaluate the path integrals, we follow the same procedure used for the wavefunctional: given the saddle-point approximation, we express the classical solution in terms of propagators satisfying the boundary conditions imposed by the path integral.\footnote{These are usually called \textit{bulk-to-boundary} and \textit{bulk-to-bulk} propagators, see more details in App.~\ref{app:propagators}. In this semi-classical way all the tree diagrams are generated. Using the functional quantisation, we can go beyond the saddle-point approximation to derive these propagators, which are also able to form loop diagrams~\cite{Cespedes:2023aal}.} This leads to the following schematic form for the density matrix:
\begin{align}
    \rho[\varphi_+,\varphi_-]=\exp\left(\frac{1}{2}\int_{\bm{k}}\psi_2\varphi_+(\bm{k})\varphi_+(-\bm{k})+\frac{1}{2}\int_{\bm{k}}\psi^*_2\varphi_-(\bm{k})\varphi_-(-\bm{k})+\cdots\right) \ ,
\end{align}
where the dots indicate terms of higher order in the field profiles. In the case that the density matrix is non-unitary there will be mixing terms like powers of $\phi_+\phi_-$. Notice that when computing correlation functions of  $\phi$ we would need to take the trace of the density matrix and thus any phase of the wavefunction coefficients will not contribute. 
\paragraph{Effective field theories}
A central idea in quantum field theory is that when a system contains fields with dynamics governed by vastly different energy scales, it becomes possible to decouple the high energy physics and construct an EFT for the light degrees of freedom. If we consider two fields $\phi$ and $\sigma$, where the former is massless and the latter has a mass $M$, we can in general integrate out $\sigma$ and obtain an action for $\phi$  given by
\begin{align}
    e^{iS_{\mathrm{eff}}[\phi]}=\int\mathcal{D}\sigma e^{iS[\phi,\sigma]} \ .
    \label{sef:example}
\end{align}
In general, this procedure generates non-local terms in $S_{\mathrm{eff}}[\phi]$, as the effects of the heavy field $\sigma$ manifest as non-local interactions among light fields, mediated by virtual $\sigma$ exchanges. However, when the mass of $\sigma$,  $M$, is large compared to the characteristic energy scales of $\phi$, the non-local terms can be expanded in a series of higher-dimensional local operators, organised in a derivative expansion and suppressed by inverse powers of the heavy mass scale $M$.\footnote{This argument is mostly based on scattering amplitudes in flat spacetime. For equal-time correlators (even in flat spacetime), there are non-negligible contributions from non-unitary processes, which can also be organised by inverse powers of $M$~\cite{Burgess:2024heo,Green:2024cmx,CCoEFT}.}

The EFT includes, in principle, all operators consistent with the symmetries of the system. Its predictive power lies in determining the coefficients of these operators, known as Wilson coefficients. These are fixed through a matching procedure, in which physical observables—such as scattering amplitudes or correlation functions—are computed in both the full theory and the EFT and then equated at a common renormalisation scale (typically near the heavy mass scale $M$). This procedure ensures that the EFT accurately reproduces the low-energy behaviour of the full theory, capturing all relevant physical effects below $M$ while significantly simplifying the description by integrating out the heavy degrees of freedom.

Importantly, the effective action in Eq.~\eqref{sef:example} describes a unitary theory in standard quantum field theory in flat space, where time evolution is Hamiltonian and the path integral is evaluated along a single time contour. However, in time-dependent or non-equilibrium settings, such as those encountered in cosmology the appropriate formalism is the Schwinger–Keldysh or in-in formalism, which, discussed previously, doubles the degrees of freedom and involves a contour running both forward and backward in time. 

In this framework, integrating out a heavy field such as $\sigma$ generically induces interactions between the forward and backward branches of the contour, leading to an effective action that is non-local in time and not necessarily derivable from a unitary Hamiltonian. Therefore, the dynamics of the light fields after coarse-graining over short-distance modes can become non-unitary, reflecting the emergence of dissipation, decoherence, and memory effects. Nevertheless, the structure of the effective action remains constrained by the unitarity and locality of the underlying UV theory, imposing the following non-perturbative conditions:
\begin{align}
    &S[\phi,\phi]=0\ ,\qquad \mathrm{Im} S[\phi_+,\phi_-]\geq 0\ ,\nonumber\\
    &S[\phi_+,\phi_-]=-S^*[\phi_+,\phi_-] \ .
\end{align}
These constraints will fix some of the parameters and field contents of the theory restricting the allowed EFT.
\paragraph{Advanced/retarded basis}
It is customary to recast the theory in the \textit{advanced/retarded} (sometimes also called the Keldysh) basis:
\begin{align}
    \phi_r=\frac{1}{2}(\phi_++\phi_-)\ ,\qquad \phi_a=\phi_+-\phi_-
    \label{eq_+-2ar} \ ,
\end{align}
where $\phi_r$ captures the average (classical) configuration and $\phi_a$ acts as an auxiliary source encoding quantum and statistical fluctuations. In this basis the effective action $S[\phi_a,\phi_r]$ is for the density matrix, not the unitary action. The constraints on the effective action become:
\begin{align}
    &S[\phi_r,\phi_a=0]=0\ ,\qquad \mathrm{Im} S[\phi_r,\phi_a]\geq 0\ ,\nonumber\\
    &S[\phi_r,\phi_a]=-S^*[\phi_r,-\phi_a] \ .
    \label{constr_ar}
\end{align}
This formulation makes it manifest that $\phi_a$ is not an independent degree of freedom, and that all physical observables can be derived from correlation functions of $\phi_r$. The advantage of this basis is especially clear in the presence of dissipation and noise, which can be directly read off from the structure of the effective action. It also provides a natural framework for studying the classical limit of quantum field theories.

\section{Single Field $\lambda\phi^4$ in dS}
\label{sec:phi4}
\begin{figure}[t]
\centering
\includegraphics[trim={4cm 14cm 2cm 0cm},scale=.6]{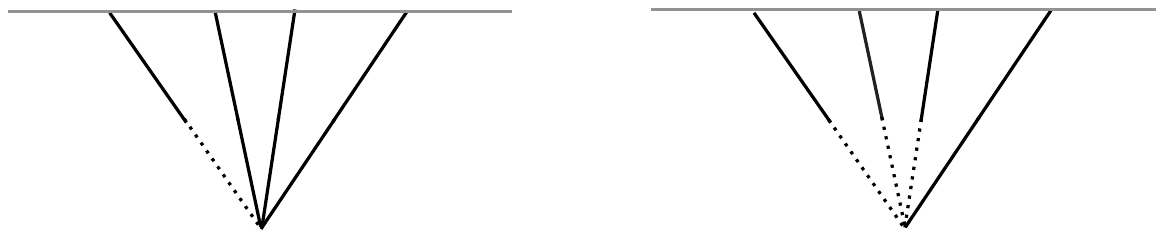}
\caption{\small Quartic vertices appearing at the tree level: Dashed lines represent $\phi_a$ fields and solid black lines represent $\phi_r$ fields. A contraction between two $\phi_r$ fields corresponds to a Keldysh propagator $G_K$, while a contraction between $\phi_r$ and $\phi_a$ corresponds to a retarded or advanced propagator, $G_{R}$ or $G_{A}$.
} \label{Fig:correlatorar}
\end{figure}

Let us first illustrate the utility of this basis with the main case we are interested in understanding: $\lambda\phi^4$ in a de Sitter background with the action,
\begin{align}
    S=\int d^4x \sqrt{-g}\left[-\frac{1}{2}(\partial_\mu\phi)^2-\frac{\lambda}{4}\phi^4\right] \ .
\end{align}
In the advanced/retarded bases, the action for the density matrix becomes,
\begin{align}
    S[\phi_a,\phi_r]=\int d^4 x\sqrt{-g}\left[-\partial_\mu\phi_a\partial^\mu\phi_r-\lambda\phi_a\phi_r^3-\frac{\lambda}{4}\phi_a^3\phi_r
    \right] \ ,
\end{align}
where the two quartic vertices reflect different time-orderings in the in-in formalism. Notice that the action satisfies the unitarity constraint as both vertices are proportional to odd powers of $\phi_a$, whereas the third constraint in \eqref{constr_ar} implies that $\lambda$ is real. Written in this way the classical equations of motion are obtained by varying the action with respect to $\phi_a$ and the setting $\phi_a=0$,
\begin{align}
    \left.\frac{1}{\sqrt{-g}}\frac{\delta}{\delta\phi_a} S[\phi_r,\phi_a]\right\vert_{\phi_a=0,\phi_r=\varphi}= (\square-\lambda\phi^2)\phi=0 \ .
\end{align}
This will become useful later when dealing with including quantum effects systematically.
There are three propagators in this basis, which can be written in terms of the quadratic action as,
\begin{align}
    S_0= -\frac12\int_{\bm{k}}\int dt \ (\phi_r\  \phi_a) \begin{pmatrix}
0 & D_A \\
D_R & -2i D_K 
\end{pmatrix}\begin{pmatrix}
\phi_r \\
\phi_a
\end{pmatrix}  \ ,
\end{align}
where the differential operators are defined in terms of the propagators as,
\begin{align}
    \begin{pmatrix}
0 & D_A \\
D_R & -2iD_K 
\end{pmatrix}\cdot\begin{pmatrix}
G_K& G_R \\
G_A & 0
\end{pmatrix}=\begin{pmatrix}
1& 0\\
0 & 1
\end{pmatrix} \ .  
\end{align}
Notice that the propagators satisfy the constraints on the action. Actually the $G_{A/R}$ are just the response to a source whereas the Keldysh propagator describes the statistical properties of the system. For a free theory we have that the retarded and Keldysh propagators are given by the commutator and anti-commutator of field operators, c.f. Eqs.~\eqref{eq_GR}-\eqref{eq_GK},\footnote{One subtlety here is that, to restore the correct form of $G_K$, one must include the correct $i\epsilon$-prescriptions, see App.~\ref{app:propagators}. In particular, $D_K$ is proportional to $\epsilon$ while $G_K$ remains finite when taking $\epsilon\to 0$. We encourage unfamiliar readers to refer to~\cite{Salcedo:2024smn} for more details.}
\begin{align}
G_R(\bm{k},t,t')&\equiv i\langle\phi_r(t,\bm{k})\phi_a(t',-\bm{k})\rangle'= i \langle[\phi_{\bm{k}},(t),\phi_{-\bm{k}}(t')]\rangle'\theta(t-t') \ ,\\
   G_K(\bm{k},t,t')&\equiv i\langle\phi_r(t,\bm{k})\phi_r(t',-\bm{k})\rangle'= \frac i2 \langle\{\phi_{\bm{k}}(t),\phi_{-\bm{k}}(t')\}\rangle' \ .
\end{align}
We will usually write explicit expressions in conformal time $\eta$. In particular, the retarded and Keldysh propagators for a massless field in dS are:
\begin{align}
&G_R(\bm k,\eta,\eta') = \frac{H^2\theta(\eta-\eta')}{k^3}\left[(1+k^2\eta\eta')\sin k(\eta-\eta')-k(\eta-\eta')\cos k(\eta-\eta')\right], \label{eq_GRmassless}\\
&G_K(\bm k,\eta,\eta') = \frac{iH^2}{2k^3}\left[(1+k^2\eta\eta')\cos k(\eta-\eta') + k(\eta-\eta')\sin k(\eta-\eta')\right].\label{eq_GKmassless}
\end{align}
As shown in Fig.~\ref{Fig:correlatorar}, we represent $\phi_a$ fields with dotted lines and $\phi_r$ fields with solid lines. In this notation, propagators connecting two solid lines correspond to contractions between $\phi_r$ fields and represent the Keldysh propagator $G_K$. Propagators with one solid line and one dotted line correspond to contractions between $\phi_a$ and $\phi_r$, and thus represent the retarded or advanced propagators, $G_R$ or $G_A$.
When computing equal time correlators at a late time $\eta_0$ the following prescription allows too expand around the classical solution,
\begin{align}
\langle\varphi_{\bm{k}_1}\cdots\varphi_{\bm{k}_n}\rangle&=\int d\varphi_k\varphi_{\bm{k}_1}\cdots\varphi_{\bm{k}_n}\rho[\varphi]\qquad\nonumber\\
\mathrm{where}\qquad\rho[\varphi]&=\int^{\phi_r(\eta_0)=\varphi}\mathcal{D}\phi_r\int^{\phi_a(\eta_0)=0}\mathcal{D}\phi_a e^{iS[\phi_r,\phi_a]} \ ,
\end{align} 
is the diagonal part of the density matrix, which is equivalent to the square of the wavefunction for unitary theories (c.f. Eq.~\eqref{eq_rho=wv2}). In practice this prescription implies that only diagrams with external $\phi_r$ legs contribute to the correlation function. For the $\lambda \phi^4$ theory, the two quartic vertices shown in Fig.~\ref{Fig:correlatorar} contribute separately, but their effects are qualitatively different. Importantly, the correct singularity structure of the four-point function only emerges when both contributions are combined. The first vertex, $\phi_a \phi_r^3$, generates a four-point function built from three Keldysh propagators and one retarded propagator:
\begin{small}\begin{align}
\langle\varphi_{\bm{k}_1}\varphi_{\bm{k}_2}\varphi_{\bm{k}_3}\varphi_{\bm{k}_4}\rangle'&\supset -6i\lambda\int ^{\eta_0}_{-\infty} d\eta a^4(\eta)G_R(k_1,\eta_0,\eta)G_K(k_2,\eta_0,\eta)G_K(k_3,\eta_0,\eta)G_K(k_4,\eta_0,\eta)+3~\mathrm{perms}\nonumber\\
&=\frac{\lambda}4\frac{H^4}{\prod k_i^3} \sum k_i^3 \log(-\eta_0) + \mathcal A(k_i) \ .
\end{align}
\end{small}Notice that the factor $6$ comes from permutations between $G_K(k_{2,3,4},\eta_0,\eta)$, and we have taken the late-time limit $\eta_0\to0$ and kept only nonvanishing terms. The function $\mathcal A(k_i)$ depends on all the momenta, but not on $\eta_0$.

The second vertex, by contrast, involves three retarded propagators and one Keldysh propagator
\begin{small}\begin{align}
\langle\varphi_{\bm{k}_1}\varphi_{\bm{k}_2}\varphi_{\bm{k}_3}\varphi_{\bm{k}_4}\rangle'&\supset \frac{-6i}{4}\lambda\int ^{\eta_0}_{-\infty} d\eta a^4(\eta)G_R(k_1,\eta_0,\eta)G_R(k_2,\eta_0,\eta)G_R(k_3,\eta_0,\eta)G_K(k_4,\eta_0,\eta)+3~\mathrm{perms}\nonumber\\
&=\mathcal B(k_i)~,
\end{align}
\end{small}with the function $\mathcal B(k_i)$ also independent of $\eta_0$.
Summing both contributions, together with all permutations of the external legs, the total four-point function becomes,\footnote{This result is more easily derived from the in-in calculation, see App.\ \ref{app:propagators}:
\begin{align}
    \langle \phi_{\bm k_1}\phi_{\bm k_2}\phi_{\bm k_3}\phi_{\bm k_4}\rangle'=&~
    \sum_{\pm}(\pm i)\times 4!\times \frac{-\lambda}{4}\times \int_{-\infty}^{\eta_0} d\eta a^4(\eta)\prod_{i=1}^4K_{\pm}(k_i,\eta)\nonumber\\
    =&~\frac{\lambda H^4}{4\prod k_i^3}
    \sum k_i^3\log(-k_T\eta_0)\nonumber\\
    & - \frac{\lambda H^4}{4\prod k_i^3k_T}\left\{ (1-\gamma)\sum k_i^4+\sum_{i\neq j}\left[(2-\gamma) k_i^3k_j +k_i^2k_j^2\right] + \frac12\sum_{i\neq j\neq \ell}k_i^2k_jk_\ell- \prod k_i\right\}\ .
    \label{full4pt}
\end{align}
}
\begin{align}
\langle\varphi_{\bm{k}_1}\varphi_{\bm{k}_2}\varphi_{\bm{k}_3}\varphi_{\bm{k}_4}\rangle’
= \frac{\lambda}{4} \frac{H^4}{\prod k_i^3}
\left( \sum k_i^3 \log(-k_T \eta_0) + \text{Poly}(k_i) \right) \ ,
\label{LO4pt}
\end{align}
where $k_T \equiv k_1 + k_2 + k_3 + k_4$ is the total energy.

In general, each vertex alone contains both the total-energy pole and the folded singularities in the functions $\mathcal A$ and $\mathcal B$. This can be seen by sending the integrand to the folded configuration, where there is a piece that does not oscillate in the early time, and thus the integral diverges. However, as the contributions are summed over, the folded singularities cancel, leaving only the physical total-energy pole in the logarithm, which finally combines with the late-time divergence in the first vertex contribution to give $\log(-k_T\eta_0)$.
This cancellation of unphysical singularities is a direct consequence of the unitarity of the underlying Schwinger–Keldysh action, which fixes the relative coefficients of the two vertices. Similar cancellations in different interaction structures were recently highlighted in~\cite{Salcedo:2024smn,CCoEFT}. 

Another lesson we learn from this simple case is that different vertices can have different scalings when approaching the future boundary, with the $\phi_a\phi_r^3$ vertex giving the dominant contribution.
The different late-time growth of the two quartic vertices is also related to the different scalings of the propagators and can be understood from a simple argument: For a massless field, the propagators \eqref{eq_GRmassless} and \eqref{eq_GKmassless} in the long-wavelength limit scale as
\begin{align}
G_K(k,\eta,\eta’) &\sim \frac{H^2}{2k^3}\left(1 + \mathcal{O}(k^2 \eta^2)\right)\ , \qquad
G_R(k,\eta,\eta’) \sim H^2\left(\eta^3 - \eta’^3\right)\left(1 + \mathcal{O}(k^2 \eta^2)\right)\ .
\label{action_nonunitterm}
\end{align}
From this we find that,
\begin{align}
\frac{\phi_r}{\phi_a} \sim \frac{\langle \phi_r^2 \rangle}{\langle \phi_a \phi_r \rangle} = \frac{G_K(k,\eta,\eta’)}{G_R(k,\eta,\eta’)} \sim \frac{1}{(k \eta)^3} \ .
\label{scaling:fields}
\end{align}
Thus, as modes evolve on superhorizon scales, interactions involving a larger number of $\phi_r$ fields become dominant. This behaviour is consistent with the emergence of a semiclassical regime at late times, since the equation of motion involves only a single power of $\phi_a$, indicating that this term governs the classical saddle-point behaviour. Rescaling the fields as in~\cite{Glorioso:2017fpd},
\begin{align}
\phi_r \to \phi_r, \qquad \phi_a \to \hbar \phi_a \ ,
\end{align}
will make this structure explicit: in the $\hbar \to 0$ limit, only vertices with a single $\phi_a$ survive, precisely corresponding to the dominant contribution identified in~\cite{Cespedes:2023aal}.

For convenience, let us consider a massive scalar field with potential given by $V(\phi)=m^2\phi^2/2+\lambda\phi^4/4$ in de Sitter. 
Notice that, as the mass of the field has to be compared with the horizon, there are three different interesting regimes for this theory.
The EoM is given by
\begin{align}
\ddot\phi +3H \dot\phi +\frac{k^2}{a^2} \phi - m_{\rm eff}^2\phi =0~~~~~~{\rm with}~~~~~~m_{\rm eff}^2 =  m^2+\lambda \phi^2 \ .
\end{align}
If this scalar field is light but not exactly massless, with $m_{\rm eff}\ll H$, we find three regimes for the time evolution of one Fourier mode $\phi_{\bm k}$. The corresponding approximated EoMs are as follows.
\begin{itemize}
    \item Subhorizon regime: $k>a(t) H$
    \begin{equation}
    \ddot\phi  +\frac{k^2}{a^2} \phi   \simeq 0 \ .
    \label{point1}
    \end{equation}
    \item Intermediate regime: $m_{\rm eff}<k/a(t)< H$
    \begin{equation}
    \ddot\phi + 3H\dot\phi +\frac{k^2}{a^2} \phi  \simeq 0 \ .
    \end{equation}
    This is the regime after Hubble horizon exit, but before mass horizon crossing. The mass term is still negligible, but the Hubble friction starts to become dominant. 
     \item Stochastic regime: $k/a(t)< m_{\rm eff}$
    \begin{equation}
     3H\dot\phi \simeq m^2_{\rm eff} \phi \ .
    \end{equation}
    After mass horizon crossing, this long wavelength perturbation is approximately described by the Hubble friction and the mass term, like a slow-roll evolution. Upon this moment, the quantum diffusion becomes important, which contributes to a noise term to the above classical equation. This first-orderness is the key for a light scalar to become stochastic.\footnote{Notice that in this regime perturbation theory will eventually break down. For instance, in a $\lambda\phi^4$ theory the perturbative expansion remains valid only while
$\lambda\log(-k\eta) \lesssim 1$ .
The usefulness of the stochastic description is that it provides a framework to go beyond this limitation: by rewriting the problem in terms of a Fokker–Planck equation, valid at all times, one can resum the secular growth and obtain a controlled description of the long-time dynamics.}
\end{itemize}

Now we consider coarse-graining by introducing a Window function with a physical UV cut-off $\epsilon H$, and focus on the long-wavelength perturbation $k_l<\Lambda= \epsilon a(t)H $. If $\epsilon H<m_{\rm eff}$, this means that all the $k_l$ modes are in the stochastic regime and we can apply the Fokker-Planck equation to describe its probability distribution.
If $m_{\rm eff}<\epsilon H$, some of the long wavelength perturbations are still within the intermediate regime, and we do not have a stochastic description.

The above analysis assumed a unitary theory to start with, but it is expected that when writing an effective description for the long modes, the action will become non-unitary as the scale is lowered (see~\cite{Li:2025azq}). As we will now see, when integrating out a field, there appear off-diagonal components in the density matrix, which encode interference between the forward and backward branches of the contour and can play a crucial role. If the initial state is entangled or if heavy fields remain partially coupled, these components can induce observable non-unitary effects. In particular, infrared divergences at late times in de Sitter space can drive large-scale decoherence, suppressing interference and leading to an effective stochastic description. Accurately incorporating these effects is essential for building a reliable EFT for the long-wavelength sector.

%%%%%%%%%%%%%%%%%%%%%%%%%%%%%%%%%%%%%%%%%%%%%%%%
\section{Tree-Level Matching: $\alpha\phi^2\sigma\rightarrow \lambda\phi^4$}
%%%%%%%%%%%%%%%%%%%%%%%%%%%%%%%%%%%%%%%%%%%%%%%%
\label{sec:EFTdS}
To illustrate these issues concretely, we consider a simple two-field model on a fixed de Sitter background. In this setup, one field, $\phi$, is massless and evolves on superhorizon scales, while the second field, $\sigma$, is heavy and will be integrated out. As a starting point, we focus on a cubic interaction between $\phi$ and $\sigma$, which already captures the essential features we wish to highlight. More general interactions will be explored later in the paper. The model is defined by the action:
\begin{align}
    S=\int d^4x \sqrt{-g}\left(-\frac{1}{2}(\partial_\mu\phi)^2-\frac{1}{2}(\partial_\mu\sigma)^2-\frac{m^2}{2}\sigma^2-\frac{1}{2}\phi^2 f(\sigma)-V(\phi,\sigma)\right) \ ,
\end{align}
where $f(\sigma)$ is a general coupling function between the fields, and $V(\phi, \sigma)$ is a potential containing interaction terms that are at least cubic in $\phi$ and $\sigma$. Our goal is to integrate out the heavy field $\sigma$ and derive an effective action for the light field $\phi$.

\subsection{Effective action of the density matrix}
To begin the analysis, we take $f(\sigma) = \alpha \sigma$ and set $V(\phi, \sigma) = 0$, where $\alpha$ is a coupling constant with the dimension of the mass, that is,
\begin{align}
S=\int d^4x \sqrt{-g}\left(-\frac{1}{2}(\partial_\mu\phi)^2-\frac{1}{2}(\partial_\mu\sigma)^2-\frac{m^2}{2}\sigma^2-\frac{\alpha}{2}\phi^2 \sigma\right)\ .
\label{action:2fiedllinear}
\end{align}
This simplified interaction will serve as our baseline example; more general cases will be considered in later sections.
As previously discussed, the proper framework to integrate out degrees of freedom in a time-dependent background is the Schwinger–Keldysh formalism, which tracks the full evolution of the density matrix. The full density matrix for the system is given by:
\begin{align}
    \rho[\varphi_+,\varphi_-,\varsigma_+,\varsigma_-]=\int^{\varphi_+} \mathcal{D}\phi_+\int^{\varphi_-}\mathcal{D}\phi_-\int^{\varsigma_+}\mathcal{D}\sigma_+\int^{\varsigma_-}\mathcal{D}\sigma_-e^{iS[\phi_+,\sigma_+]-iS[\phi_-,\sigma_-]}  \ .\label{density_matrix}
\end{align}

To obtain the reduced density matrix for $\varphi$, we trace over the field $\varsigma$, that is,
\begin{align}
    \rho[\varphi_+,\varphi_-]&=\mathrm{Tr}_\varsigma\   \rho[\varphi_+,\varphi_-,\varsigma_+,\varsigma_-]\nonumber\\
&=\int d\varsigma \int^{\varphi_+} \mathcal{D}\phi_+\int^{\varphi_-}\mathcal{D}\phi_-\int^{\varsigma}\mathcal{D}\sigma_+\int^{\varsigma}\mathcal{D}\sigma_-e^{iS[\phi_+,\sigma_+]-iS[\phi_-,\sigma_-]} \ ,
    \label{red_densitym}
\end{align}
which corresponds to computing the path integrals for $\sigma_+$ and $\sigma_-$ with the common final boundary condition $\varsigma$, and then integrating over all possible $\varsigma$. Operationally, this means solving the EoM for $\sigma_\pm $ given sources from the $\varphi_\pm$ configurations, evaluating the resulting classical on-shell actions, and then performing the final Gaussian integral over the boundary field $\varsigma$.

Since the action is quadratic in $\varsigma$, the integration over $\sigma$ is Gaussian and can be performed exactly. The result is:
\begin{align}
    \rho[\varphi_+,\varphi_-]&=\int^{\varphi_+} \mathcal{D}\phi_+\int^{\varphi_-}\mathcal{D}\phi_- \exp\left(iS_0[\phi_+]-iS_0[\phi_-]\right.\nonumber\\
   & +\frac{\alpha^2}{4}\int_{\bm{k}_1,\cdots,\bm{k}_4}\int dt\sqrt{-g}\int dt'\sqrt{-g}\left.\sum_{\pm,\pm}\phi_{\bm{k}_1}^\pm(t)\phi_{\bm{k}_2}^\pm(t)G^\sigma_{\pm,\pm}(s,t,t')\phi_{\bm{k}_3}^\pm(t')\phi_{\bm{k}_4}^\pm(t')\right.\nonumber\\
   &\left.+~t-\mathrm{and}\  u- \mathrm{channels}\right) \ ,
   \label{effective_linear}
\end{align}
where $S_0[\phi_\pm]$ is the free action for the massless field $\phi$, and $G^\sigma_{\pm,\pm}(k, t, t’)$ are the in-in propagators of the heavy field $\sigma$ (see App. \ref{app:propagators} for their definition), with momentum dependence suppressed for brevity. The additional contributions from $t$ and $u$ channels arise from contracting the interaction vertex in different pairings.

The effective action in Eq.~\eqref{effective_linear} already exhibits many of the structural features discussed earlier. Without making any approximations, it is manifestly non-local and non-Markovian, and contains non-vanishing off-diagonal terms that mix the forward and backward segments of the Schwinger–Keldysh contour. This is expected: the original theory includes the exchange of a massive particle, and the non-local behaviour aligns with known results in the context of cosmological collider physics. Several of these features directly originate from integrating out the heavy field, which induces both temporal non-locality and correlations between forward and backward time branches~\cite{CCoEFT}.

An important question is whether the effective action \eqref{effective_linear} can be approximated by a local expansion in terms of effective operators. Based on flat-space intuition, one expects that in the limit $m^2 \gg H^2$, the theory simplifies and the dominant term in the effective action corresponds to a local, unitary interaction of the form $\lambda \phi^4$. However, the presence of off-diagonal terms in the Schwinger–Keldysh contour introduces subtleties: these components must be shown to be suppressed compared to the local diagonal terms in the appropriate limits. Furthermore, one may ask whether there exist other regimes, beyond the large-mass limit, where a local description remains valid.

To address these questions, one needs to perform a matching between correlation functions computed in the UV theory (Eq.~\eqref{effective_linear}) and those obtained from a local EFT. This matching is subtle, as the full correlators receive contributions from all three structures in Eq.~\eqref{effective_linear}, and these are generally entangled in a way that prevents direct identification with specific wavefunction components. For this reason, it is useful to rewrite the action in the advanced/retarded ($a/r$) basis, where causal structure and unitarity constraints become more transparent.
In this basis, the effective action takes the form,
\begin{align}
S[\phi_a,\phi_r]&=\int d^4x \sqrt{-g}(-\partial_\mu\phi_a\partial^\mu\phi_r)\nonumber\\
&+\alpha^2\int d^4x \sqrt{-g}\int d^4 y\sqrt{-g}\left[\frac{1}{8}\phi^2_a(x)G^\sigma_R(x,y)\phi_a(y)\phi_r(y)+\frac{1}{2}\phi_a(x)\phi_r(x)G^\sigma_R(x,y)\phi^2_r(y)\right.\nonumber\\
&\qquad\qquad\qquad\qquad\qquad\qquad\qquad\left.+\frac12\phi_a(x)\phi_r(x)G^\sigma_K(x,y)\phi_a(y)\phi_r(y)\right] \ ,
\label{action:areff}
\end{align}
where $G_R^\sigma$ and $G_K^\sigma$ denote the retarded and Keldysh propagators of the heavy field $\sigma$, respectively.

This form makes several key properties manifest. First, it satisfies the constraints listed in Eq.~\eqref{constr_ar}: the action vanishes when $\phi_a = 0$, all terms are consistent with causality, and no unphysical (pure-$\phi_r$) contributions appear. In addition, we find that although a $\phi_a^4$ term is allowed by symmetry, it vanishes due to the specific structure of the propagators—mirroring the cancellation that also eliminates the $\phi_r^4$ term. As we will show later, such cancellations persist even in the presence of nonlinearities and play a crucial role in preserving causality.

\paragraph{Four-point correlation function}

To better understand the structure of the possible effective field theories, let us begin by computing the four-point correlator generated by the exchange of the heavy field $\sigma$. It will be useful to write the relevant interaction in the following form,
\begin{align}
    \frac12\left(\frac{-i\alpha}{2}\phi^2\right) \frac{-i}{-\Box+m^2} \left(\frac{-i\alpha}{2}\phi^2\right) =&~ \frac{i\alpha^2}{8m^2}\phi^4 + \frac{i\alpha^2}{8m^2}\phi^2 \frac{1}{-\Box+m^2}\Box\phi^2\nonumber\\
    =&~\frac{i\alpha^2}{8m^2}\phi^4 + \frac{i\alpha^2}{8m^4}\phi^2\Box\phi^2
    +\frac12\times\frac{i\alpha^2}{4m^4}\Box\phi^2\frac{1}{-\Box+m^2}\Box\phi^2.
    \label{propagator_expansion2}
\end{align}
This expression makes explicit that the exchange generates a local directly-coupled quartic vertex, with the remaining terms involving derivative interactions that are IR-finite. Importantly, no expansion in large mass has been performed; the result remains exact for all non-vanishing $m$.

Using this structure, we can rewrite the action in Eq.~\eqref{action:2fiedllinear} as:\footnote{See Appendix \ref{app:UVcomputation} for further details.}
\begin{equation}
    S[\phi,\sigma] = S_0[\phi,\sigma] + \int d^4x\sqrt{-g}\left[\frac{\alpha^2}{8m^2}\phi^4+\frac{\alpha^2}{4m^4}\phi^2(\partial_\mu\phi)^2\right]
    -\frac{\alpha}{m^2}\int d^4x\sqrt{-g}(\partial_\mu\phi)^2\sigma \ ,
    \label{action:2fiedlrew}
\end{equation}
with $S_0$ denoting the free part of the action, and we have applied the equation of motion of $\phi$. This action preserves the structure described above. In particular, when integrating out $\sigma$, the only contribution to the off-diagonal terms in the Schwinger–Keldysh contour comes from the last interaction term, while the contact terms remain diagonal. Using this reformulated action, the tree-level four-point function can be naturally decomposed into three contributions:
\begin{align}
    \langle \varphi_{\bm k_1}\varphi_{\bm k_2}\varphi_{\bm k_3}\varphi_{\bm k_4}\rangle' = \mathcal I_1 + \mathcal I_2 + \mathcal I_3.
\end{align}
The first term, $\mathcal I_1$, is the  $\phi^4$ contact interaction. Similar to Eq.~\eqref{full4pt}, it leads to the standard IR-divergent behaviour:
\begin{align}
    \mathcal I_1 = &-\frac{\alpha^2H^4}{8m^2\prod k_i^3}
    \sum k_i^3\log(-k_T\eta_0)\nonumber\\
    & + \frac{\alpha^2H^4}{8m^2\prod k_i^3k_T}\left\{ (1-\gamma)\sum k_i^4+\sum_{i\neq j}\left[(2-\gamma) k_i^3k_j +k_i^2k_j^2\right] + \frac12\sum_{i\neq j\neq \ell}k_i^2k_jk_\ell- \prod k_i\right\}\ .
    \label{int:I1}
\end{align}
The second term $\mathcal I_2$ comes from the contact $\phi^2(\partial_\mu\phi)^2$ interaction.  Since this term involves derivatives, it is IR-finite and additionally suppressed by an extra power of $H^2/m^2$:
\begin{align}
    \mathcal I_2 
    =&~\frac{\alpha^2H^6\sum k_i^3}{16m^4\prod k_i^3} \ .
    \label{int:I2}
\end{align}
Finally, $\mathcal I_3$ comes from the exchange diagram with the $(\partial_\mu\phi)^2\sigma$ interaction:
\begin{align}
    \mathcal I_3  =&~\frac{\alpha^2H^6}{4m^4\prod k_i^3 s}\mathcal O_{12}\mathcal O_{34} \mathcal I\left(\frac{s}{k_{12}},\frac{s}{k_{34}}\right)+~t-\mathrm{and}\  u- \mathrm{channels}  \ .
      \label{int:I3}
\end{align}
Here we have defined the weight-shifting operators:
\begin{equation}
    \mathcal O_{ij} \equiv k_i^2k_j^2\partial_{k_{ij}}^2 - \bm k_i\cdot\bm k_j\left(1-k_{ij}\partial_{k_{ij}} + k_ik_j\partial_{k_{ij}}^2\right),
\end{equation}
and the seed integral $\mathcal I$ is given by~\cite{Arkani-Hamed:2018kmz,Qin:2022fbv}:
\begin{align}
    \mathcal I(u,v) \equiv~&\frac{1+i\sinh\pi\mu}{2\pi}\left[
    \mathbf F_+(u)\mathbf F_+(v) + \mathbf F_+(u)\mathbf F_-(v)\right]+ (\mu\to-\mu)\nonumber\\
    &+\sum_{p,q=0}^\infty\frac{(-1)^q(q+1)_{2p}}{2^{2p+1}(\frac q2+\frac14+\frac{i\mu}2)_{p+1}(\frac q2+\frac14-\frac{i\mu}2)_{p+1}}u^{2p+1}\left(\frac{u}{v}\right)^{q}\ ,
    \label{int:I3seed}
\end{align}
where we have defined the mass parameter $\mu\equiv \sqrt{m^2/H^2-9/4}$ and assumed $k_{12}>k_{34}$ for the convergence of the series representation. We have also introduced the homogeneous solution to the bootstrap equation:
\begin{equation}
    \mathbf F_\pm(r) \equiv r^{1/2}\left(\frac{r}{2}\right)^{\pm i\mu}\Gamma\left(\frac12\pm i\mu\right)\Gamma\left(\mp i\mu\right) {}_2\mathrm F_1\left[
    \begin{matrix}
        \frac14\pm\frac{i\mu}2,\frac34\pm\frac{i\mu}2\\
        1\pm i\mu
    \end{matrix}\middle|r^2
    \right]\ .
    \label{homoF}
\end{equation}
This final term receives contributions from both the unitary process and the non-unitary effect such as particle productions, as captured in the cosmological collider signal~\cite{Baumann:2019oyu}.\footnote{As noted in \cite{DuasoPueyo:2025lmq}, the unitary contribution admits an asymptotic expansion in terms of $H^2/m^2$, while the non-unitary contribution will emerge when resuming this asymptotic series.} A key point is that this term is the only one contributing to the off-diagonal components of the density matrix in Eq.~\eqref{effective_linear}. This becomes manifest in the reformulation of the action, where the expansion in the mass of the heavy field generates an asymptotic series expansion of local operators of increasing dimension, involving only powers and derivatives of $\phi$. These terms can be consistently incorporated into a unitary effective action. In contrast, the interaction term $(\partial_\mu\phi)^2\sigma$ explicitly couples to $\sigma$ and, upon integrating out $\sigma$, yields a genuinely off-diagonal contribution—reflecting the non-unitary, non-local nature of the underlying dynamics, as discussed above.
 
Let us now analyze the behaviour of the four-point function in different mass regimes.
When the spectator field $\sigma$ is heavy, $m^2 \gg H^2$, the leading contribution comes from \eqref{int:I1}. The second term, \eqref{int:I2}, is suppressed by an additional factor of $H^2/m^2$. For the third term \eqref{int:I3}, we can analyse the large mass behaviour of the seed integral \eqref{int:I3seed}. One finds the first line of \eqref{int:I3seed} is exponentially suppressed by the mass, whereas the second line is suppressed by at least a factor of $H^2/m^2$ (for the term with $p=q=0$). Altogether, in the regime $m^2/H^2 \gg 1$, the correlator reduces to that of a unitary $\lambda\phi^4$ theory, with higher-order effects under control, as expected from the large-mass expansion.

As the mass decreases, another interesting limit emerges. When $m^2/H^2 \ll 1$, the dominant contribution still arises from \eqref{int:I1}. This is the only term that exhibits an IR divergence, and at late times its contribution grows logarithmically. Consequently, even in this regime the effective description is once again dominated by a quartic interaction—though now this interaction originates from long-time evolution rather than from a large-mass expansion. This regime is valid only away from the squeezed limit: there the $\mathcal{I}_3$ term dominates, since it contains factors scaling as $(s/k_{12})^{-\nu}$ and $(s/k_{34})^{-\nu}$ with $\nu\equiv\sqrt{9/4-m^2/H^2}$, which could grow faster than the other contributions so long as $\nu>0$.\footnote{Such kinematic factors come from $(s/k_{12})^{\pm i\mu}$ and $(s/k_{34})^{\pm i\mu}$ with $\mu\equiv\sqrt{m^2/H^2-9/4}$ in the homogeneous solution \eqref{homoF}. When $m>3H/2$, these factors are always of order 1 and oscillate on the logarithmic scale, thus generating the so-called cosmological collider signal.} To avoid this complication, we restrict our analysis to momentum configurations closer to the equilateral limit.

From an EFT perspective, this behaviour can be understood in terms of coarse-graining. The heavy mass $m$ sets the effective cut-off of the theory.
\begin{itemize}
	\item	When $m^2 \gg H^2$, we integrate out modes well inside the horizon, leading to a local, unitary, and weakly coupled EFT. The usual EFT intuition from large-mass expansions applies.
	\item	As the cut-off approaches $m^2 \sim H^2$, particle production becomes significant, and the EFT develops non-local and non-unitary features—the exchange of the heavy field can no longer be approximated as instantaneous.
	\item	For $H^2/m^2 \gg 1$, a large-mass expansion fails, but so long as $\log(-k_T\eta_0) \gg H^2/m^2$, the first term still dominates and the IR admits a local EFT description. In this limit the separation between long and short modes follows from superhorizon coarse-graining and is directly analogous to stochastic inflation (see \eqref{point1}), where the long-wavelength dynamics is governed by a Fokker–Planck equation and the coarse-graining scale is set by $\Lambda \sim \epsilon H$ with $\epsilon \ll 1$. Perturbation theory in this regime requires
$\alpha \,\frac{H^2}{m^2}\,\log(-k_T\eta_0)\ll 1$,
otherwise it breaks down.
\end{itemize}

A central subtlety in cosmological EFTs is the choice of matching scale. In flat-space QFT, one matches the EFT to the UV theory at scales close to the heavy mass $M$ in order to minimize large logarithms. In inflation, however, observables are defined at horizon crossing, where the physical scale is $H$. This mismatch introduces complications.

For models such as $\lambda\phi^4$, which lack non-linearly realised symmetries, correlation functions continue to evolve after horizon exit and exhibit secular growth. This raises the question of whether matching should be performed at horizon crossing or at later times, after additional subhorizon dynamics have been integrated out. A similar ambiguity appears in stochastic inflation, where the coarse-graining scale is set by $\Lambda \sim \epsilon H$, and modes with $k/a > \Lambda$ are integrated out. To address these issues, we will match at different energy scales and examine how this choice maps onto different EFT regimes. Our objective is to rewrite the non-local action in Eq.~\eqref{effective_linear} as a sum of local operators whose coefficients reproduce the correct correlators in the appropriate limits.

In summary, two controlled regimes emerge:
\begin{itemize}
\item $m^2/H^2 \gg 1$: the EFT is local, unitary, and weakly coupled, with non-local corrections exponentially suppressed. Higher order corrections are organised in term of local operators whose couplings are suppressed in powers of $H^2/m^2$.
\item $m^2/H^2 \ll 1$ with $\log(-k_T\eta_0)\gg H^2/m^2$: IR effects dominate, and a local stochastic description becomes valid. In this case higher order corrections are organised in powers of gradients $\nabla^2/(aH)^2$  which are suppressed on superhorizon scales. 
\end{itemize}

In the intermediate regime, where $m^2/H^2 \sim 1$, all terms in \eqref{effective_linear} contribute with comparable magnitude. In this situation, the effective action can no longer be reorganized as a local derivative expansion, and a simple EFT description ceases to exist. Nevertheless, at late times $(-k\eta \ll 1)$, the logarithmic enhancement of superhorizon modes remains dominant. Consequently, the quartic vertex continues to provide the leading contribution to the four-point function, capturing the main late-time behaviour even in the absence of a fully local EFT expansion for the remainder of the terms.

 \subsection{Heavy mass limit $m^2/H^2 \gg 1$.}
In the regime where the spectator field $\sigma$ decays well before horizon exit, its effects can be effectively captured by a local and unitary EFT valid up to energies around $m$. In this case, coarse-graining over scales much smaller than the Hubble radius results in a theory that is approximately Minkowskian, with no spontaneous particle production. This is reflected in the exponential suppression of the off-diagonal components of the density matrix, as seen in Eq.~\eqref{int:I3}.

At leading order in the $1/m^2$ expansion, the dominant contribution to the four-point function comes from the time-ordered (diagonal) terms. In this limit, the nonlocal time structure simplifies:
\begin{align}
\phi_{\bm{k}_1}^\pm(t)\phi_{\bm{k}_2}^\pm(t)G^\sigma_{\pm,\pm}(s,t,t')\phi_{\bm{k}_3}^\pm(t')\phi_{\bm{k}_4}^\pm(t')\to \phi_{\bm{k}_1}^\pm(t)\phi_{\bm{k}_2}^\pm(t)\phi_{\bm{k}_3}^\pm(t)\phi_{\bm{k}_4}^\pm(t) \ ,
\end{align}
with a coefficient that depends on the coupling $\alpha$ and the mass $m$. This indicates that the interaction becomes local in time and effectively Markovian. At leading order, the resulting dynamics reproduce those of a local $\lambda \phi^4$ theory, where the effective quartic coupling is generated by integrating out the heavy field $\sigma$. 

Corrections to this leading behaviour appear as an expansion in time derivatives and spatial gradients. Importantly, the off-diagonal components of the effective action are exponentially suppressed in $m/H$, signaling that the reduced theory approaches unitarity in the heavy-mass limit.

Indeed, using the following expansion we find the leading-order expansion of the nonlocal interaction:
\begin{align}
    \frac12\left(\frac{i\alpha}{2}\phi^2\right) \frac{-i}{-\Box+m^2} \left(\frac{i\alpha}{2}\phi^2\right) =&~\frac{i\alpha^2}{8m^2}\phi^4 + \frac{i\alpha^2}{8m^4}\phi^2\Box\phi^2
    +\mathcal{O}(1/m^6) \ ,
    \label{propagator_expansion}
\end{align}
which corresponds to the usual gradient expansion of the propagator. Already at this level, the effective theory becomes local, with higher-derivative corrections suppressed by powers of $1/m^2$.

Returning to the advanced/retarded ($a/r$) basis, the leading-order effective action can be expressed in the form of a unitary $\lambda \phi^4$ theory. Since the interaction arises from a single vertex, the matching enforces a fixed relation between the coefficients of the $\phi_a \phi_r^3$ and $\phi_a^3 \phi_r$ operators, which in the unitary theory implies that they appear with relative weight $1/4$. When the spectator mass is lowered, this relation may receive corrections, and the two coefficients need not remain tied to their unitary values. For this reason, we will keep them independent in what follows:
\begin{align}
    S_{\mathrm{EFT}}[\phi_a,\phi_r]\supset\int d^4 x\sqrt{-g}\left[-\partial_\mu\phi_a\partial^\mu\phi_r-\lambda_{\mathrm{eff}}\phi_a\phi_r^3-\bar\lambda_{\mathrm{eff}}\phi_a^3\phi_r
    \right] \ ,
\end{align}
At the next order in the derivative expansion, we obtain the following terms suppressed by gradients
\begin{align}
    S_{\mathrm{EFT}}[\phi_a,\phi_r]\supset-\int d^4 x \sqrt{-g}\left[+\frac{\lambda_{\mathrm{eff}}^{(2)}}{H^2}\phi_r^2\partial_\mu\phi_a\partial^\mu\phi_r
    +\frac{\bar\lambda_{\mathrm{eff}}^{(2)}}{H^{2}}\phi_a^2\partial_\mu\phi_a\partial^\mu\phi_r
    \right] \ ,
    \label{action:subleading}
\end{align}
where we have used integration by parts to simplify the structure of the operators. 
The coefficients are determined by matching to the full theory. We of course expect that $\lambda_{\mathrm{eff}},\bar\lambda_{\mathrm{eff}}\sim \alpha^2/m^2$ and $\lambda_{\mathrm{eff}}^{(2)},\bar\lambda_{\mathrm{eff}}^{(2)}\sim \alpha^2 H^2/m^4$ and the coefficients at each order as we will see by the matching.

\paragraph{EFT matching}
To perform the matching, we compute the four-point function derived from the effective action and compare it with the result obtained from the full two-field theory in Eq.~\eqref{int:I1}. At leading order in the $1/m^2$ expansion, the four-point correlator is given by
\begin{align}
\langle\varphi_r(\bm{k}_1)\varphi_r(\bm{k}_2)\varphi_r(\bm{k}_3)\varphi_r(\bm{k}_4)\rangle' =- \frac{\alpha^2H^4}{8m^2}\frac{k_1^3+k_2^3+k_3^3+k_4^3}{k_1^3k_2^3k_3^3k_4^3}\log(-k_T \eta_0) \ ,
\end{align}
This expression contains only a total-energy pole and no folded singularities, indicating that the effective theory remains unitary. In particular, this enforces the relation $\lambda_{\mathrm{eff}} = 4\bar\lambda_{\mathrm{eff}}$ among the effective couplings.

From this, we identify the leading-order quartic coupling matched to the two-field model as:
\begin{align}
    \lambda_{\mathrm{eff}}=-\frac{1}{2}\frac{\alpha^2}{m^2} \ .
\end{align}
The negative sign of the quartic coupling reflects the fact that the original cubic interaction is unbounded from below for large negative field values. However, this does not indicate an immediate instability, as the effective theory remains well-defined within the perturbative regime.

Importantly, this sign is not corrected by additional interactions appearing in the EFT: there are no higher powers of the field arising after integrating out $\sigma$, and the derivative corrections—such as those involving four gradients—yield coefficients with the same overall sign. This confirms the robustness of the result within the low-energy expansion.

At subleading order, the matching proceeds analogously. The effective action in Eq.~\eqref{action:subleading} gives rise to four-point functions that depend solely on the total energy, reinforcing the same structural relation among EFT coefficients 
\begin{align}
   \lambda_{\mathrm{eff}}^{(2)} = -8\bar\lambda_{\mathrm{eff}}^{(2)} = \frac{\alpha^2H^2}{2m^4} \ ,
\end{align}
which is suppressed by an additional power of $H^2/m^2$, as expected from the derivative expansion.\footnote{
The action includes $S \supset \int d^4x\sqrt{-g}\frac{\alpha^2}{4m^4}\phi^2(\partial_\mu\phi)^2$ at subleading order, and
\begin{align}
\phi_+^2(\partial_\mu\phi_+)^2 - \phi_-^2(\partial_\mu\phi_-)^2 =&~ 2\phi_a\phi_r(\partial_\mu\phi_r)^2 + 2(\partial_\mu\phi_a)(\partial^\mu\phi_r)\phi_r^2 + \frac12 \phi_a^2(\partial_\mu\phi_a)(\partial^\mu\phi_r) + \frac12(\partial_\mu\phi_a)^2\phi_a\phi_r\nonumber\\
=& -2(\partial_\mu\phi_a)(\partial^\mu\phi_r)\phi_r^2 + \frac14(\partial_\mu\phi_a)^2\phi_a\phi_r\ ,
\end{align}
where we have used integration by parts and EoM $\Box\phi_{a,r}=0$, so that $(\partial_\mu\phi_a)(\partial^\mu\phi_r)\phi_{a,r}^2 = -2(\partial_\mu\phi_a)(\partial^\mu\phi_r)\phi_{a,r}^2$ and $(\partial_\mu\phi_a)^2\phi_a\phi_r = (-1/2)(\partial_\mu\phi_a)(\partial^\mu\phi_r)\phi_a^2$. Therefore, we know:
\begin{align}
S[\phi_a,\phi_r] \supset -\frac{\alpha^2}{2m^4} \int d^4x\sqrt{-g} \Big(\phi_r^2\partial_\mu\phi_a\partial^\mu\phi_r - \frac18\phi_a^2\partial_\mu\phi_a\partial^\mu\phi_r\Big)\ .
\end{align}
}
%%%%%%%%%%%%%%%%%%%%%%%%%%%%%%%%%%%%%%%%%%%
\subsection{Stochastic regime}
\label{subsec:stochasticregime}
%%%%%%%%%%%%%%%%%%%%%%%%%%%%%%%%%%%%%%%%%%%
As previously discussed, there exists a regime where the infrared-divergent part of the four-point function, given in Eq.~\eqref{int:I1}, provides the dominant contribution—even when the mass of the spectator field is smaller than the Hubble scale, $m < H$. This occurs because the contribution from Eq.~\eqref{int:I1} grows logarithmically with time, eventually overtaking the other terms when $\log(-k\eta) \gg m^2/H^2$.

We now demonstrate that, in this regime, an effective field theory (EFT) emerges whose leading-order contributions to correlation functions are captured by a local interaction, while subleading effects appear as spatially non-local corrections. This can be shown both via explicit matching to the four-point function and by analyzing the structure of the full non-local action.

The EFT becomes valid on superhorizon scales, but for momenta such that $-k\eta\ll m/H\ll 1$, so that there remains a hierarchy between the energies of interest and the mass of the $\sigma$ field. Let us first isolate the leading IR contribution, which in momentum space takes the form
\begin{align}
S[\phi_a,\phi_r] \supset &~ \frac{\alpha^2}{2}\int_{-\infty}^{\eta_0}\frac{d\eta_1}{(-H\eta_1)^4}\int_{-\infty}^{\eta_0}\frac{d\eta_2}{(-H\eta_2)^4} \int_{\bm{k}_1,\cdots,\bm{k}_4,\bm{s}} \phi_a(\eta_1,\bm k_1)\phi_r(\eta_1,\bm k_2)\times G_R^\sigma(\bm s,\eta_1,\eta_2) \nonumber\\
&\times \phi_r(\eta_2,\bm k_3)\phi_r(\eta_2,\bm k_4) +~t-\mathrm{and}\  u- \mathrm{channels} \ .
\end{align}
In the superhorizon limit, we can use the late-time approximation for the retarded Green’s function (see Appendix for details):
\begin{align}
G_R^\sigma( s,\eta_1,\eta_2) =  \frac{H^2}{2\nu} \Big[ \Big(\frac{\eta_1}{\eta_2}\Big)^{-\nu} - \Big(\frac{\eta_2}{\eta_1}\Big)^{-\nu}\Big] (\eta_1\eta_2)^{3/2}\theta(\eta_1-\eta_2)(1+\mathcal{O}(\eta_2^2,\eta_1^2)) \ .
\end{align}
At leading order, the propagator is analytic in momentum $\bm s$ (and similarly for $\bm t$ and $\bm u$), implying that the interaction is effectively local in space. Subleading corrections introduce terms proportional to $k^2\eta^2$, leading to spatial gradient operators such as $\nabla^2\phi_{a/r}$ in the EFT, which are suppressed in the long-wavelength limit.

To examine whether the interaction is also local in time (i.e., Markovian), we evolve $\phi_{r/a}$ using the free equation of motion. To leading order in small $\eta$, one finds~\cite{CCoEFT,Colas:2022hlq}:
\begin{align}
\phi(\eta_2,\bm k) =&~ \Big[\frac{\eta_2}{\eta_1}\cos k(\eta_1-\eta_2) + \frac{1}{k\eta_1}\sin k(\eta_1-\eta_2)\Big]\phi(\eta_1,\bm k)\nonumber\\
&+\Big[\frac{\eta_1-\eta_2}{k^2\eta_1^2}\cos k(\eta_1-\eta_2) - \frac{1+k^2\eta_1\eta_2}{k^3\eta_1^2}\sin k(\eta_1-\eta_2)\Big]\phi'(\eta_1,\bm k)\nonumber\\
=&~ \Big[\frac{\sin k\eta_1}{k\eta_1} \phi(\eta_1,\bm k) + \frac{k\eta_1\cos k\eta_1-\sin k\eta_1}{k^3\eta_1^2}\phi'(\eta_1,\bm k)\Big] \times \Big[1+\mathcal O(\eta_2^2)\Big]\nonumber\\
=&~\Big[\phi(\eta_1,\bm k) - \frac13\eta_1\phi'(\eta_1,\bm k)\Big] \times \Big[1+\mathcal O(\eta_1^2)\Big]\times\Big[1+\mathcal O(\eta_2^2)\Big] \ .
\label{time_evol}
\end{align}

Since $\phi(\eta,\bm k)$ approaches a constant at late times, the derivative term is subdominant, and we can replace $\phi_r(\eta_2)$ by $\phi_r(\eta_1)$ in the leading IR contribution. This is a standard approximation in open systems and is frequently used to justify Markovian behaviour (see, e.g.,~\cite{Colas:2022hlq}).

Plugging this back into the interaction and integrating over $\eta_2$, we find:
\begin{align}
    S[\phi_a,\phi_r] \supset&~ \frac{\alpha^2}{2}\int_{-\infty}^{\eta_0}\frac{d\eta_1}{(-H\eta_1)^4}\int_{-\infty}^{\eta_0}\frac{d\eta_2}{(-H\eta_2)^4} \int_{\bm{k}_1,\cdots,\bm{k}_4,\bm{s}} \phi_a(\eta_1,\bm k_1)\phi_r(\eta_1,\bm k_2) \times G_R^\sigma(\bm s,\eta_1,\eta_2) \nonumber\\
    &\times \phi_r(\eta_2,\bm k_3)\phi_r(\eta_2,\bm k_4)+~t-\mathrm{and}\  u- \mathrm{channels}\nonumber\\
    =&~\frac{\alpha^2}{2H^2} \int_{-\infty}^{\eta_0} \frac{d\eta_1}{(-H\eta_1)^4}\int_{\bm{k}_1,\cdots,\bm{k}_4}\phi_a(\eta_1,\bm k_1)\phi_r(\eta_1,\bm k_2)\phi_r(\eta_1,\bm k_3)\phi_r(\eta_1,\bm k_4)\times \frac{4}{4\nu^2-9}+\cdots\ ,
    \label{eq_latetimeSretarded}
\end{align}
which reduces to a local $\lambda \phi^4/4$ interaction, with $\lambda = -\alpha^2/(2m^2)$. Higher-order terms involving $\dot{\phi}_r$ and spatial gradients are suppressed in the $-k\eta \ll 1$ limit.

We now consider the non-unitary term proportional to $\phi_r^2\phi_a^2$ that arises from the Keldysh propagator. At late times, this propagator behaves as
\begin{align}
    G_K(s,\eta_1,\eta_2)\sim \frac{iH^2}{4\pi} \left(\frac{s^2\eta_1\eta_2}{4}\right)^{-\nu}(\eta_1\eta_2)^{3/2} \Gamma(\nu)^2 \ .
\end{align}
implying that the interaction is non-local in space for all values of $\nu$. However, the time evolution approximation in Eq.~\eqref{time_evol} remains valid, so the interaction is still local in time. The resulting vertex takes the form:
\begin{align}
  S[\phi_a,\phi_r] \supset&~ \frac{i\alpha^2}{4\pi H^2}\frac{\Gamma^2(\nu)}{(3+2\nu)}\int_{-\infty}^{\eta_0}\frac{d\eta_1}{(-H\eta_1)^4}\int_{\bm{k}_1,\cdots,\bm{k}_4,\bm{s}}\left(\frac{s^2\eta_0\eta_1}{4}\right)^{-\nu}\left(\frac{\eta_1}{\eta_0}\right)^{3/2} \nonumber\\
  &\times \phi_r(\eta_1,\bm{k}_1)\phi_r(\eta_1,\bm{k}_2)\phi_a(\eta_1,\bm{k}_3)\phi_a(\eta_1,\bm{k}_4) + ~t-\mathrm{and}\  u- \mathrm{channels} \ ,
      \label{nonlocalvertex}
\end{align}
where note that the  time-dependent coefficient grows at late times. However, it contributes subdominantly to the four-point function compared to Eq.~\eqref{eq_latetimeSretarded} since $\Re(\nu) < 3/2$.

Assuming the Bunch–Davies vacuum, the scaling of this vertex at late times leads to:
\begin{align}
    \langle\phi_k^4\rangle\sim \alpha^2H^2\frac{(-s\eta_0)^{3-2\nu}}{k^6s^3}\ .
    \label{subd4pt:eq}
\end{align}
This contribution is subdominant relative to that from Eq.~\eqref{int:I3}, even though it arises from the same interaction. The reason is that the dominant contribution to the full four-point function comes from early times, where the integrand is finite, whereas Eq.~\eqref{subd4pt:eq} corresponds to the late-time (superhorizon) limit, where it vanishes.
With all of these elements in mind then we propose the EFT to be,
\begin{align}
      S_{\mathrm{EFT}}&=-\int d^3x\int dt\sqrt{-g} \left[\partial_\mu\phi_a\partial^\mu\phi_r+\lambda_{\mathrm{eff}}\phi_a\phi_r^3+\bar\lambda_{\mathrm{eff}}\phi_a^3\phi_r\right]\nonumber\\&\qquad+i\int d^3x\int d^3y \int dt a^3(t)  \ \phi_a^2(\bm{x},t)\lambda_\mathrm{NL}(\vert\bm{x}-\bm{y}\vert)\phi_r^2(\bm{y},t) \ .
      \label{eq:EFTsuperhorizonb}
\end{align}
In this case we do not expect all the coefficients to have the same behaviour on superhorizon scales, since the  $\phi_a$ and $\phi_r$ fields have a different scaling on superhorizon scales. As discussed in \eqref{scaling:fields} on superhorizon scales, we have:
\begin{align}
\frac{\phi_r}{\phi_a} \sim \frac{\langle \phi_r^2 \rangle}{\langle \phi_a \phi_r \rangle}  \sim \frac{1}{(k \eta)^3} \ .
\end{align}
This implies that interactions with a larger number of $\phi_r$ fields dominate in the IR. In particular, the $\phi_r^2\phi_a^2$ vertex in Eq.~\eqref{nonlocalvertex} yields a subleading contribution to the correlator, since $\Re\nu < 3/2$ for a massive field, even though the vertex itself becomes large at late times. 

\paragraph{EFT matching}
We now verify that the quartic coefficients of the EFT obtained in Eq.~\eqref{eq_latetimeSretarded} can also be derived by direct matching.
To this end, we compute the four-point correlation function of the UV theory in the superhorizon limit.
At leading order, the dominant contribution arises from the quartic vertex\eqref{int:I1}, yielding
\begin{align}
\langle\varphi_{\bm{k}_1}\varphi_{\bm{k}_2}\varphi_{\bm{k}_3}\varphi_{\bm{k}_4}\rangle'=- \frac{\alpha^2}{8m^2}\frac{k_1^3+k_2^3+k_3^3+k_4^3}{k_1^3k_2^3k_3^3k_4^3}\log(-k_T \eta_0)+\mathcal{O}(\alpha^2 H^2m^{-4}) \ .
\end{align}
In the regime where $k^2/a^2 \ll m^2/H^2 \ll 1$, the correlator no longer probes the short-distance vacuum structure of the UV theory.
Instead, its evolution is dominated by stochastic dynamics, as both the light and heavy fields have wavelengths exceeding the horizon size.
At horizon crossing, the leading contribution is controlled by the effective quartic coefficient inherited from integrating out the light field $\sigma$, and the correlation function behaves as in a local $\lambda \phi^4$ theory.

At later times—when $\log(-k_T\eta_0)\gg1$, corresponding to the regime of interest for stochastic inflation—the matching gives
\begin{align}
     \lambda_{\mathrm{eff}} = -\frac{\alpha^2H^2}{2m^4} \ .
\end{align}
For this expression to remain valid before perturbation theory breaks down, one must satisfy
\begin{align}
\log(-k_T\eta)\gg \frac{H^2}{m^2} \ .
\label{stocreg_cond}
\end{align}
In this case, the leading contribution to the four-point function originates from the local vertex discussed in the previous section.
This demonstrates that, in the stochastic regime, the EFT matching is performed at energies below the Hubble scale, provided that the condition~\eqref{stocreg_cond} holds.

Finally, regarding the validity of the EFT:
the perturbative expansion is controlled by the parameter
$\alpha\,\frac{H^2}{m^2}\,\log(-k_T\eta)$,
and once this quantity becomes of order unity, perturbation theory ceases to be reliable.
Even without additional self-interactions, quantum corrections then grow large and the EFT loses predictability.
While one might hope to resum these contributions using stochastic methods, we will show in Sec.~\ref{sec:bpt} that this is not possible: the UV completion is unbounded from below, and this instability is inherited by the probability distribution of $\phi$ after integrating out $\sigma$.

To match the mixed operator $\phi_a^2 \phi_r^2$, we note that the four-point function in the superhorizon limit (at finite time) develops non-analytic powers of $(-k\eta)$. The leading term in this expansion determines the coefficient of the corresponding operator, allowing for a direct matching of the effective coupling. 

\subsection{Decoherence}
The main difference between the various regimes of the effective action for $\lambda\phi^4$ that we have described lies in the behaviour of the off-diagonal components; in the stochastic regime, these off-diagonal terms are non-negligible. While we have shown that they contribute subdominantly to the four-point function in the superhorizon limit, they nevertheless play a crucial role. In particular, their presence indicates that the system is not in a pure state but is instead highly entangled with an environment composed of long-wavelength modes lying outside the horizon.

A natural way to quantify the impact of the off-diagonal components of the reduced density matrix is through the purity which varies between one and zero and is defined to be one for a pure state and decreases as the state becomes mixed.~\footnote{See~\cite{Colas:2022hlq,Colas:2022kfu,Colas:2024xjy,Bhattacharyya:2024duw,Colas:2024ysu,Burgess:2024heo,Lopez:2025arw,Cespedes:2025zqp,Burgess:2025dwm} for recent discussion about purity change in the context of inflation.} To compute this quantity, it is convenient to return to the $\pm$ basis of the Schwinger–Keldysh formalism. Our goal is to evaluate the reduced density matrix on shell. Upon integrating out the $\sigma$ field, we obtain
\begin{align}
    iS &=\frac{1}{2}\int_{\bm{k}}\left[\psi_2(\varphi^+_{\bm{k}})^2+\psi_2^*(\varphi^-_{\bm{k}})^2\right]\nonumber\\
&+\int_{\bm{k}_1,...\bm{k}_4}\left(\psi_{4}-\frac{(\psi_3^{\varphi\varphi\varsigma})^2}{2\mathrm{Re}\ \psi_2^\varsigma}\right)\varphi^+_{\bm{k}_1}\varphi^+_{\bm{k}_2}\varphi^+_{\bm{k}_3}\varphi^+_{\bm{k}_4}+\left(\psi_{4}^*-\frac{(\psi_3^{*\varphi\varphi\varsigma})^2}{2\mathrm{Re}\ \psi_2^\varsigma}\right)\varphi^-_{\bm{k}_1}\varphi^-_{\bm{k}_2}\varphi^-_{\bm{k}_3}\varphi^-_{\bm{k}_4}\nonumber\\
&+\int_{\bm{k}_1,...\bm{k}_4}\frac{\vert\psi_3^{\varphi\varphi\varsigma}\vert^2}{8\mathrm{Re}\ \psi_2^\varsigma}\varphi^+_{\bm{k}_1}\varphi^+_{\bm{k}_2}\varphi^-_{\bm{k}_3}\varphi^-_{\bm{k}_4}+\mathrm{perms} \ .
\label{effaction_intermediate}
\end{align}
where the wavefunction coefficients are given in App. \ref{app:wvfn}. The purity is defined as \begin{align}
     \gamma=\mathrm{Tr}\rho^2=\int d\phi_1\int d\phi_2\ e^{iS[\phi_1,\phi_2]+iS[\phi_2,\phi_1]}
 \end{align}
Using Eq.~\eqref{effaction_intermediate}, we find at leading order in the coupling $\alpha$~\cite{Pueyo:2024twm},
 \begin{align}
     \gamma=1-2\int_{\bm{q}}\frac{\vert\psi_3^{\varphi\varphi\varsigma}(\bm{s},\bm{q},-\bm{q}-\bm{s})\vert^2}{\mathrm{Re}\ \psi_2(s)\times\mathrm{Re}\ \psi_2(q)\times\mathrm{Re}\ \psi_2^\varsigma(\bm{s}+\bm{q})}+\mathcal{O}(\alpha^3) \ ,
 \end{align}
where we have also expanded the normalization of the density matrix up to first order in $\alpha^2$.
As expected, deviations from $\gamma=1$ originate from the off-diagonal components of the effective action in Eq.~\eqref{effaction_intermediate}. This provides a direct diagnostic of the non-unitarity of the effective theory. In the late-time limit, the purity becomes\footnote{
To see this, notice that the three-point wavefunction coefficient $\psi_3^{\vp\vp\varsigma}$ is dominated by its imaginary part at late time, $\psi_3^{\vp\vp\varsigma} \sim i\alpha/(3H^4\eta_0^3)$ (see Eq.~\eqref{eq_3ptIm}), while $\Re\psi_2^\varsigma(k) = -1/(2|\sigma(k,\eta_0)^2|)$ and $\Re\psi_2^\varphi(k) \sim -k^3/H^2$ (see Eqs.~\eqref{eq_powerspec} and \eqref{eq_masslesspowerspec}). Therefore, the purity in the late time becomes:
\begin{align}
\gamma -1\sim &~-2\times \frac{\alpha^2}{(3H^4\eta_0^3)^2} \times 2|\sigma(s,\eta_0)|^2 \times \int \frac{d^3\bm q}{(2\pi)^3}\frac{H^4}{q^3|\bm s+\bm q|^3}\ .
\end{align}
}
 \begin{align}
    1- \gamma=\frac{\alpha^2}{H^2}\frac{4^\nu\Gamma^2(\nu)}{9\pi^3}(-s \eta_0)^{-3-2\nu}\log(s\Lambda_{\mathrm{IR}}) \ ,
     \label{eq:purity}
     \end{align}
 where the result depends explicitly on the infrared cut-off $\Lambda_{\mathrm{IR}}$, since the loop integral involves two external massless legs.
This expression shows that when the mass of $\sigma$ satisfies $m^2/H^2\ll1$, the reduced state of the light field $\phi$ becomes mixed on superhorizon scales. The loss of purity directly traces back to the non-vanishing off-diagonal term $\phi_a^2\phi_r^2$ in the effective action. Conversely, when the heavy field limit applies, Eq.~\eqref{eq:purity} shows that the deviation from $\gamma=1$ is exponentially suppressed due to Boltzmann suppression of particle production.

Let us summarise the results of this section. We have shown that it is possible to integrate out a second field $\sigma$ even when its mass is lighter than the Hubble scale, provided that the condition in Eq.~\eqref{stocreg_cond} holds. The resulting effective theory contains a leading local operator of the form $\phi_a\phi_r^3$, together with two subleading terms: a local operator $\phi_r\phi_a^3$ and a non-local contribution $\phi_a^2\phi_r^2$. The latter encodes the entanglement between the superhorizon fluctuations of $\sigma$ and $\phi$, an effect that is also reflected in the decrease of the purity of the reduced density matrix as modes exit the horizon.

 A useful way to interpret this change of regime is by examining the momentum scaling of the fields at fixed time, through the rescaling of propagators $k \to \rho k$. Under this transformation,
$G_{R/A,K}(\rho k,\eta,\eta’) = \rho^{-\alpha_{r/a}}\,G_{R/A,K}(k,\eta,\eta’)\, $.
Thus on subhorizon scales, both components of the field, $\phi_r$ and $\phi_a$, have scaling dimensions $\alpha_{\phi_{r,a}} = 1$, corresponding to the standard Minkowski behaviour. In contrast, on superhorizon scales the scaling dimensions change to $\alpha_r = 3$ and $\alpha_a = 0$. This transition can be understood as the effect of lowering the cut-off: when the mass (or the cut-off) is large, both fields behave as in flat space and the effective theory is unitary; as the cut-off decreases toward the Hubble scale, the presence of a stochastic environment of superhorizon modes modifies the scaling behaviour. The resulting dynamics becomes effectively semiclassical, reflecting the loss of quantum coherence due to the interaction with long-wavelength modes.

This picture is fully consistent with the classical stochastic description of inflation~\cite{Baumgart:2019clc,Cespedes:2023aal}, where only the composite operator $\lambda\,\phi_a\phi_r^3$ contributes to the leading late-time behaviour of correlation functions. In that framework, each vertex insertion introduces a retarded propagator, and the corresponding diagrams reproduce the logarithmic enhancement of correlators that dominates at late times.
Let us briefly comment on the appearance of dissipative effects. As written, the EFT in \eqref{eq:EFTsuperhorizon} includes only a diffusion term, proportional to $\phi_a^2$. To generate genuine dissipation, one would need to break boost invariance, which could give rise to operators of the form $\dot{\phi}_a \phi_r^3$. However, given the scaling dimensions of the fields, such terms remain subleading compared to diffusion effects. It is nonetheless an interesting open question to determine how dissipative contributions could be disentangled from purely diffusive ones in the cosmological collider signal.. 

%=======================================
\section{Loop-Level Matching: $g\phi^2\sigma^2\rightarrow \lambda \phi^4$}
\label{sec:loop}
%=======================================
Let us now consider higher-order corrections to the effective field theory arising from more general couplings of the form $\phi^2\mathcal{O}$. The key difference compared to linear couplings is that when the interaction is at least quadratic in the second field, the effective action receives contributions both from integrating over the classical field profiles and from the quantum effective action itself. These contributions can be naturally interpreted as classical loops in the former case, and as quantum loops in the latter~\cite{Cespedes:2023aal}.

\subsection{1PI effective action of the density matrix}

To make this more concrete, let us examine the interaction $f(\sigma)=\frac{1}{2}g\sigma^2$ and derive the corresponding effective action. We first replace the heavy field $\sigma$ by its classical solution and then integrate over its field profile. Although the resulting integral in $\sigma$ is Gaussian, it produces a functional determinant that generates powers of $\phi$. Expanding this determinant to second order in the coupling yields and action up to quartic order in $\phi$.
\begin{small}
    \begin{align}
    \rho[\varphi_+,\varphi_-]\supset
    &\int^{\varphi_+} \mathcal{D}\phi_+\int^{\varphi_-}\mathcal{D}\phi_- \exp\left(iS_0[\phi_+]-iS_0[\phi_-]\right.\nonumber\\
    &\left.+\frac{g}{2}\int_{\bm{q,}\bm{k}}\int dt \sqrt{-g}\ \varphi^+_{\bm{k}}\frac{K_\sigma(q,t)K_\sigma(q,t)}{2\mathrm{Re{\psi}_2^\varsigma(q)}}\varphi^+_{\bm{k}}-\frac{g}{2}\int_{\bm{q,}\bm{k}}\int dt \sqrt{-g}\varphi^-_{\bm{k}}\frac{K^*_\sigma(q,t)K^*_\sigma(q,t)}{2\mathrm{Re}\psi_2^\varsigma(q)}\varphi^-_{\bm{k}}\right.\nonumber\\
   & +\frac{g^2}{4}\int_{\bm{k}_1,\cdots,\bm{k}_4,\bm{q}}\int dt\sqrt{-g}\int dt'\sqrt{-g}\nonumber\\
   &\times\left.\varphi_{\bm{k}_1}^+(t)\varphi_{\bm{k}_2}^+(t)K_\sigma(s,t)\left(G_\sigma(s+q,t,t')+\frac{K_\sigma(s,t)K_\sigma(s+q,t')}{2\mathrm{Re}\psi_2^\sigma}\right)K_\sigma(s,t')\varphi_{\bm{k}_3}^+(t')\varphi_{\bm{k}_4}^+(t')+\mathrm{c.c.}\right.\nonumber\\
   &\left.+\frac{g^2}{4}\int_{\bm{k}_1,\cdots,\bm{k}_4,\bm{q}}\int dt\sqrt{-g}\int dt'\sqrt{-g}\varphi_{\bm{k}_1}^+(t)\varphi_{\bm{k}_2}^+(t)\frac{K_\sigma(s,t)^2K_{\sigma}^*(s+q,t')^{2}}{4\mathrm{\Re}\psi_2}\varphi_{\bm{k}_3}^-(t)\varphi_{\bm{k}_4}^-(t) \right.\nonumber\\
   &\left.+~t-\mathrm{and}\  u- \mathrm{channels}+\mathrm{c.c.}\right) \ .
   \label{red_quadratic}
\end{align}
\end{small}Notice that all the resulting terms are constructed from products of tree-level propagators, arising solely from the nonlinearities of the equations of motion. For example, the quartic interaction in $\phi$ receives contributions both from the product of two disconnected fourth-order wavefunction coefficients and from the connected sixth-order coefficient. This structure shows that the corresponding vertices are classical rather than genuinely quantum, even though they take the form of loops. Indeed, all these terms follow from substituting the classical equation of motion for $\sigma$, and thus originate from a tree-level computation. Nevertheless, they appear as loop corrections once the power counting in $\hbar$ of \eqref{red_quadratic} is taken into account. Furthermore, there arises a mixed term—analogous to the one discussed earlier—that encodes the decoherence of the state, and that makes the density matrix non-unitary.

This is, of course, not yet the full loop computation, since we must still account for the genuine quantum corrections. Unlike the terms obtained from disconnected wavefunction coefficients, these contributions arise from the computation of the quantum wavefunction coefficients themselves. A systematic way to incorporate them is to correct the equation of motion for $\sigma$ to first order in $\hbar$, in close analogy with the construction of the 1PI effective action.
Let us begin by considering the partition function for a single-field theory of $\sigma$,
\begin{align}
    Z[J_+,J_-]\equiv e^{iW[J_+,J_-]}=\int d\varsigma\int^\varsigma \mathcal{D}\sigma_+\int^\varsigma \mathcal{D}\sigma_- e^{iS[\sigma_+]-iS[\sigma_-]+i\int d^4x (\sigma_+J_+\sigma_-J_-)} \ .
\end{align}
From this, the effective action is defined via the Legendre transform,
\begin{align}
    \Gamma[\sigma_+,\sigma_+]=W[J_+,J_-]-\int d^4x(\sigma_+ J_+-\sigma_-J_-) \ .
\end{align}
To first order in $\hbar$, this relation can be inverted, yielding the effective action
\begin{align}
\Gamma[\sigma_+,\sigma_+]=S[\sigma_+,\sigma_+]+\frac{i}{2}\log\mathrm{det} i\mathcal{D}^{-1}_{ij},\qquad \mathrm{where}\qquad iD^{-1}_{ij}\equiv\frac{\partial^2S}{\partial\sigma_i\partial\sigma_j} \ .
\label{eq:qeff}
\end{align}
The extension to the coupled system we are considering is straightforward, since the interaction is quadratic in both fields. Computing the propagator for $\sigma$ in the presence of the background $\phi$, we obtain
\begin{align}
    i\mathcal{D}^\sigma=\begin{pmatrix}
        \square_+-m^2-\frac{g^2}{2}\phi_+^2 & 0 \\
        0 & \square_--m^2-\frac{g^2}{2}\phi_-^2
    \end{pmatrix} \ ,
\end{align}
where the subscripts on the Box operators indicate the contour time ordering.
Note that the propagators must satisfy the path-integral boundary conditions, so that, for instance, $\square_+ - m^2 = (G_\sigma^{+})^{-1}$, and similarly for the lower branch of the contour. Expanding the determinant up to second order in  powers of the coupling $g$,
\begin{small}
\begin{align}
 \Gamma[\phi_+,\phi_+,\sigma_+,\sigma_-]&=   S[\phi_+,\phi_+,\sigma_+,\sigma_-]+\frac{ig}{2}\int_{\bm{k},\bm{q}}\int dt\sqrt{-g}\ \phi^+_kG_\sigma(q)\phi^+_k-\frac{ig}{2}\int_{\bm{k},\bm{q}}\int dt\sqrt{-g}\phi^-_kG^*_\sigma(q)\phi^-_k\nonumber\\
& +\frac{ig^2}{2}\int_{\bm{k}_1,\cdots,\bm{k}_4,\bm{q}}\int dt\sqrt{-g}\int dt'\sqrt{-g}\phi_{\bm{k}_1}^+(t)\phi_{\bm{k}_2}^+(t) G_\sigma(s,t)G_\sigma(\vert s+q\vert)\phi_{\bm{k}_3}^+(t')\phi_{\bm{k}_4}^+(t') \nonumber\\
&-\frac{ig^2}{4}\int_{\bm{k}_1,\cdots,\bm{k}_4,\bm{q}}\int dt\sqrt{-g}\int dt'\sqrt{-g}\phi_{\bm{k}_1}^-(t)\phi_{\bm{k}_2}^-(t)G_\sigma(s,t)G_\sigma(\vert s+q\vert)\phi_{\bm{k}_3}^-(t')\phi_{\bm{k}_4}^-(t')\nonumber\\
&+\mathcal{O}(g^3)  \ ,
\label{effective_action}
\end{align}
\end{small}The first term corresponds to the tree-level action. Since the interaction is quadratic in $\sigma$, the resulting effective action contains only interactions involving $\phi$. All of these new interactions are 1PI, in agreement with the standard effective field theory computations in quantum field theory. In fact, it is straightforward to verify that, at higher orders, the functional determinant generates exclusively one-loop diagrams with an arbitrary number of external $\phi$ legs.

A key feature of this construction is that the effective action contains only unitary interactions, as no cross terms appear between the forward and backward branches of the Schwinger–Keldysh contour. This property follows directly from the form of the $\sigma$ propagator, which does not mix the two branches. Consequently, all terms in the quantum effective action are manifestly unitary, provided this condition continues to hold.

Let us now consider tracing over the $\sigma$ field in the density matrix, this time using the effective action \eqref{effective_action}. In practice, this amounts to adding the terms in \eqref{red_quadratic} with the interactions from the quantum effective action \eqref{effective_action}. Making use of the propagator identities introduced in \eqref{props_indent}, we obtain
\begin{small}
\begin{align}
    \rho[\varphi_+,\varphi_-]&=\int^{\varphi_+} \mathcal{D}\phi_+\int^{\varphi_-}\mathcal{D}\phi_- \exp\left(-iS_0[\phi_+]+iS_0[\phi_-]\right.\nonumber\\
    &+\frac{g}{2}\int_{\bm{k},\bm{q}}\int dt\sqrt{-g}\phi^+_{k}(t)G_{++}(q,t,t')\phi^+_{k}(t')-\frac{g}{2}\int_{\bm{k},\bm{q}}\int dt\sqrt{-g}\phi^-_{k}(t)G_{--}(q,t,t')\phi^-_{k}(t')\nonumber\\
      & +\frac{g^2}{4}\int_{\bm{k}_1,\cdots,\bm{k}_4,\bm{q}}\int dt\sqrt{-g}\int dt'\sqrt{-g}\left.\sum_{\pm,\pm}\phi_{\bm{k}_1}^\pm(t)\phi_{\bm{k}_2}^\pm(t)G^\sigma_{\pm,\pm}(s,t,t')G^\sigma_{\pm,\pm}(s+q,t,t')\phi_{\bm{k}_3}^\pm(t')\phi_{\bm{k}_4}^\pm(t')\right.\nonumber\\
   &\left.+t-\mathrm{and}\  u- \mathrm{channels}\right) \ .
   \label{eq:rhoeff}
\end{align}    
\end{small}This agrees with the result one would obtain using the standard in-in correlator approach. Notice that the distinction between classical and quantum loops is generally absent, except for the terms that mix the two branches, which arise only from the semiclassical part of the action. This is a consequence of the specific form of the interaction. In more general situations, the effective action may also contain new $\sigma$ self-interactions. In such cases, one must first compute the quantum effective action and then construct the reduced density matrix \eqref{red_quadratic} by evaluating it at the new saddle point. This procedure can, in principle, generate off-diagonal contributions from the quantum effective action, although these are typically subleading compared to the terms that originate from the tree-level part of the action. 

Some additional comments about the density matrix are in order. First, observe that all three quartic vertices share the same propagator structure as in \eqref{effective_linear}, but with the internal propagators squared. This is a nontrivial feature, since in principle one could have obtained two different propagators inside the loop. Furthermore, the combinations of fields that appear seem to respect a $Z_2$ symmetry for each branch and there are no terms of the form $\phi_+\phi_-^3$ or $\phi_+^3\phi_-$.  This in agreement with the results of~\cite{Frangi:2025xss}, where a similar scenario was studied for a thermal theory.

\paragraph{Effective action}
As we discussed previously it is more instructive to do the EFT matching with the action written in the $a/r$ basis. In addition there are some other features which are more clearly seen when written in this way as we will discuss along the way. 

In this basis the two field action is given by,
\begin{small}\begin{align}
    S=-\int_{\bm{k}} \int dt a(t)^3\left[\partial_\mu\phi_a\partial^\mu\phi_r+\partial_\mu\sigma_a\partial^\mu\sigma_r+m^2\sigma_a\sigma_r+\frac{g}{8}\sigma_a^2\phi_a\phi_r+\frac{g}{2}\sigma_r^2\phi_a\phi_r+\frac{g}{8}\sigma_a\sigma_r\phi_a^2+\frac{g}{2}\sigma_a\sigma_r\phi_r^2\right] \ ,
    \label{2field_quadratic}
\end{align}
\end{small}Because the Keldysh propagator $G_K$ is explicitly non-time-ordered, the computation is more direct than in the wavefunction approach. In fact, it is enough to construct the effective action from \eqref{2field_quadratic}, with the caveat that the functional determinant must be evaluated without imposing the boundary conditions of the path integral. Importantly, since the propagator matrix already encodes diffusion through $G_K$, the resulting effective action is generically non-unitary.
\begin{figure}[h]
\begin{center}
\includegraphics[trim={3cm 2cm 2cm 0cm},scale=.6]{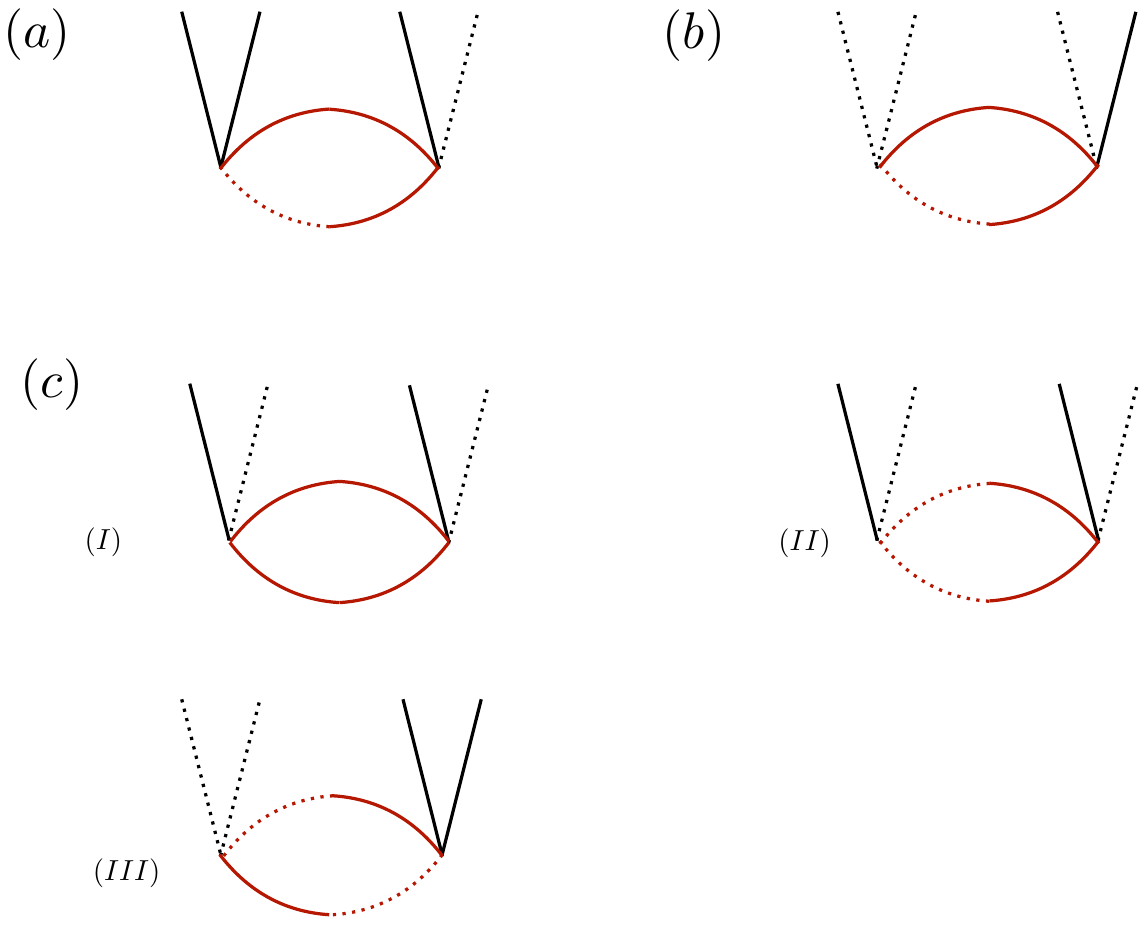}
\caption{\small Quartic vertices appearing at 1-loop. Lines in red are $\sigma$ fields and in black $\phi$ fields. (a) and (b) are the diagrams from the unitary theory while (c) is the non-unitary vertex that leads to decoherence.} \label{Fig:loopar}
\end{center}
\end{figure}

This structure closely parallels the $\phi^2\sigma$ coupling case, where the effective action was found to contain both unitary interactions (arising from the retarded/advanced propagators $G_{R/A}$) and non-unitary terms (originating from $G_K$), the latter encoding dissipation and noise in the reduced dynamics of $\phi$. By analogy, we therefore expect the effective action here to take the following general form,
\begin{align} S_{\mathrm{EFT}}=
      &-\int d^3x\int dt a^3(t) \bigg[\partial_\mu\phi_a\partial^\mu\phi_r+m_{\mathrm{eff}}^2 \phi_a\phi_r+\lambda_{\mathrm{eff}}\phi_a\phi_r^3+\bar{\lambda}_{\mathrm{eff}}\phi_a^3\phi_r\bigg]\nonumber\\&\qquad+i\int d^3x\int d^3y\int dt a^3(t)   \ \phi_a^2\lambda_\mathrm{NL}(\vert\vec{x}-\vec{y}\vert)\phi_r^2 \ ,
      \label{EFTloop}
\end{align}
where by dimensional analysis we expect that $m_\mathrm{eff}^2\sim\frac{gH^4} {m^2}$ and $\lambda_{\mathrm{eff}}\sim\bar\lambda_{\mathrm{eff}}\sim \frac{g^2H^4}{m^4}$. The non-local parameter $\lambda_{\mathrm{NL}}$ should also scale as $\frac{g^2H^4}{m^4}$ times some overall scale dependence that depends on the mass of the second field as in \eqref{nonlocalvertex}.

 At quadratic order the only new term is the effective mass $m_\mathrm{eff}^2$, generated by closing a single $\phi_a\phi_r^{3}$ vertex with a $G_K$ propagator; which we expect to be of order  $m_\mathrm{eff}^2\sim\frac{gH^2}{m^2}$. No other quadratic terms is allowed since their contribution vanishes because $G_{R/A}(k,t,t)=0$.

At quartic order, three types of diagrams contribute to Eq.~\eqref{EFTloop}, as shown in Fig.~\eqref{Fig:loopar}.
The first two types correspond to the vertices $\phi_a^3\phi_r$ and $\phi_a\phi_r^3$. These diagrams have identical internal structures: a loop built from the product of one retarded propagator $G_R$ and one Keldysh propagator $G_K$. Their coefficients scale as
$\lambda_{\mathrm{eff}} \sim \bar\lambda_{\mathrm{eff}} \sim \frac{g^2 H^4}{m^4}\,$,
Notice that this structure differs from the usual form in the $\pm$ basis, where loop diagrams are simply the square of the tree-level effective action propagators generated by the cubic vertex. The third type of diagram contributes to the non-unitary term $\phi_a^2\phi_r^2$.
Two of these diagrams are proportional to the square of the propagators $G_K$ and $G_R$, with their overall coefficient scaling as $\lambda_{\mathrm{NL}}\sim\frac{g^2 H^4}{m^4}$, multiplied by a scale-dependent factor determined by the mass of the heavy field, as discussed in Eq.~\eqref{nonlocalvertex}.

\begin{figure}[t]
\centering
\includegraphics[trim={3cm 14cm 3cm 0cm},scale=.6]{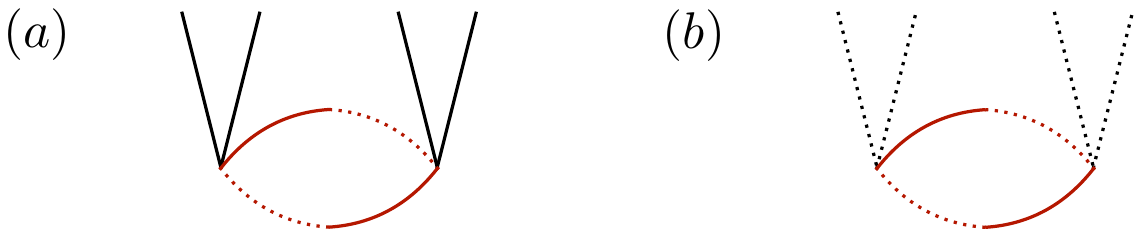}
\caption{\small Loop diagrams contributing at 1-loop to the four-point function. Lines in red are $\sigma$ fields and in black $\phi$ fields (a) contribution to the $\phi_r^4$ vanishes since it does not have support for any time, (b) contribution to the $\phi_a^4$ has the same propagator structure inside the loop and thus vanishes.} \label{Fig:looparvanish}
\end{figure}

There is another diagram that contributes to this vertex, as seen in Fig.\eqref{Fig:loopar} (labeled III), which, however, has a qualitatively different structure, involving a product of retarded and advanced propagators, $G_R$ and $G_A$. A similar structure appears in the diagrams shown in Fig.~\ref{Fig:looparvanish}, which we have not included in the effective action. The reason is that all these diagrams vanish identically, a consequence of the combined requirements of unitarity and causality.
Any diagram with four external $\phi_r$ legs necessarily contains an internal loop formed by one $G_R$ and one $G_A$, as illustrated in Fig.~\ref{Fig:looparvanish}.
Since $G_R(t,t’)$ has support only when $t>t’$, while $G_A(t,t’)$ is nonzero only when $t<t’$, their product vanishes once the external times are ordered.
This cancellation is formalized by the largest-time equation (LTE)~\cite{Veltman:1963th,Gao:2018bxz,Donath:2024utn}: whenever the latest time among all vertices sits on a retarded leg, the diagram cancels exactly against its advanced counterpart, ensuring conservation of probability.

Because the same propagator structure is obtained when exchanging $r\leftrightarrow a$, the would-be $\phi_a^4$ diagram cancels as well.
Thus, no purely $\phi_r^4$ or $\phi_a^4$ vertices appear in the effective action.
As a result, at one loop the effective action retains only the mixed $a/r$ interaction terms already present in Eq.~\eqref{action:areff}, with causality (via the LTE) and unitarity working hand-in-hand to enforce this restricted vertex structure.

\paragraph{Higher order operators}
Higher order operators with one closed loop will contribute with suppressed powers of $g^3/m^6$. At cubic order in the coupling, these contributions generate the vertices of a unitary sixth-order theory, namely $\phi_a^5\phi_r$, $\phi_a^3\phi_r^3$, and $\phi_a\phi_r^5$, together with two additional non-unitary vertices, $\phi_a^4\phi_r^2$ and $\phi_a^2\phi_r^4$. By the same causality argument as before, there is no $\phi_r^6$ vertex, since the propagators inside the loop have no overlapping time support. Likewise, the $\phi_a^6$ operator involves the same propagator structure and therefore also vanishes. Thus the additional terms in the effective action would have the following structure
\begin{small}\begin{align}
     S_{\mathrm{EFT}}&\supset\frac{1}{H^2}\int d^3x\int dt a^3(t) \left[\lambda^{(2)}\phi_a\phi_r^5+\bar\lambda_{\mathrm{eff}}^{(2)}\phi_a^5\phi_r\right]-\frac{i}{H^2}\int d^3 x\int d^3 y \int dt a^3(t)\phi_a^2(t,\bm{x})\lambda_{\mathrm{NL}}^{(2)}\phi_r^4(t,\bm{y}) \ ,
\end{align}
\end{small}where we have only included some of the operators and each of their couplings would scale as $\lambda^{(2)}\sim g^4 {H^6}/{m^6}$ and $\lambda_{\mathrm{NL}}^{(2)}$ is a non-local kernel. Notice that the non-local operators is still diffusive and it would be expected to modify the change in the purity at subleading order in the coupling.

\subsection{Operator matching}
In order to perform the matching to the EFT in Eq.~\eqref{EFTloop}, we now compute the four-point function arising from the exchange of the massive field $\sigma$. Analogously to the cubic interaction case, we expect the four-point function to contain an IR-divergent contribution that can be matched to a unitary quartic vertex, where in this case the coupling $\lambda_{\mathrm
{eff}}$ is positive, reflecting the fact that the potential is stable at large field values.

Unlike the cubic interaction, this diagram also contains a UV divergence, which must be regularised — here, we will use dimensional regularization (dim-reg) for that purpose.
The interaction in $(d+1)$-dimensional de Sitter space takes the following form,
\begin{align}
\mathcal{L}_{\mathrm{int}} = -\frac{g \mu_R^{3-d}}{4} a^{d+1}\phi^2\sigma^2 \ ,\label{eq_Lbubble}
\end{align}
where $d = 3 - \epsilon$ is the spatial dimension used in dimensional regularisation, and $\mu_R$ is the renormalisation scale. Without loss of generality we will first consider the case that $\sigma$ lies within the principal series and define the (positive real) mass parameter of $\sigma$:
\begin{align}
\mu \equiv \sqrt{\frac{m^2}{H^2} - \frac{d^2}{4}}.
\end{align}
The case where $\sigma$ is in the complementary series can be easily derived by simple analytic continuation.

The four-point correlator mediated by the bubble can be conveniently computed using spectral decomposition,\footnote{See Appendix \ref{app:bubble} for further details.} in which the correlator is expressed as an integral over an infinite tower of tree-level exchanges of fields with mass parameter $\mu’$. Concretely, we represent the correlator as a sum over these exchanges and then integrate over $\mu’$ with a measure given by $\mu’ \rho^d_\mu(\mu’)$, where $\rho^d_\mu(\mu’)$ is the spectral density. Notice that the UV divergence comes precisely from the integration over $\mu'$, and thus this one has to be regularised.

If we focus on the leading IR behaviour we obtain that the four-point function is given by
\begin{align}
\la\varphi_{\bm k_1}\varphi_{\bm k_2}\varphi_{\bm k_3}\varphi_{\bm k_4}\ra'_\text{IR}
=&~\frac{g^2H^4\sum k_i^3}{16\prod k_i^3}
\left[\wh\rho_\mu\left(\frac{-3i}2\right)
-\frac{1}{16\pi^2}\log \left(\frac{\mu_R^2}{H^2}\right)\right]
\log(-k_T\eta_0) \ ,\label{eq_loopIR}
\end{align}
where we have defined the renormalised spectral function:
\begin{align}
\wh \rho_{\mu}(\mu') \equiv \lim_{d\to3} \Big[ \rho^d_{\mu}(\mu')+\frac{1}{16\pi^2}\Big(\frac{2}{3-d}-\gamma_E+\log4\pi\Big)\bigg] \  ,
\end{align}
 which is finite in the limit when $d\to 3$ and where we have added a quartic counterterm to remove the UV divergence,
 \bge
\mathcal{L}_c = -\frac{\lam_c\mu_R^{3-d}}{4}a^{d+1}\phi^4 \ ,
\label{counterterm}
\ede
where $\lambda_c$ is fixed by the renormalisation condition to cancel the divergent part of the four-point function and $\mu_R$ is the renormalisation scale that also appears in Eq.~\eqref{eq_loopIR}.

\begin{figure}
    \centering
    \includegraphics[width=0.7\linewidth]{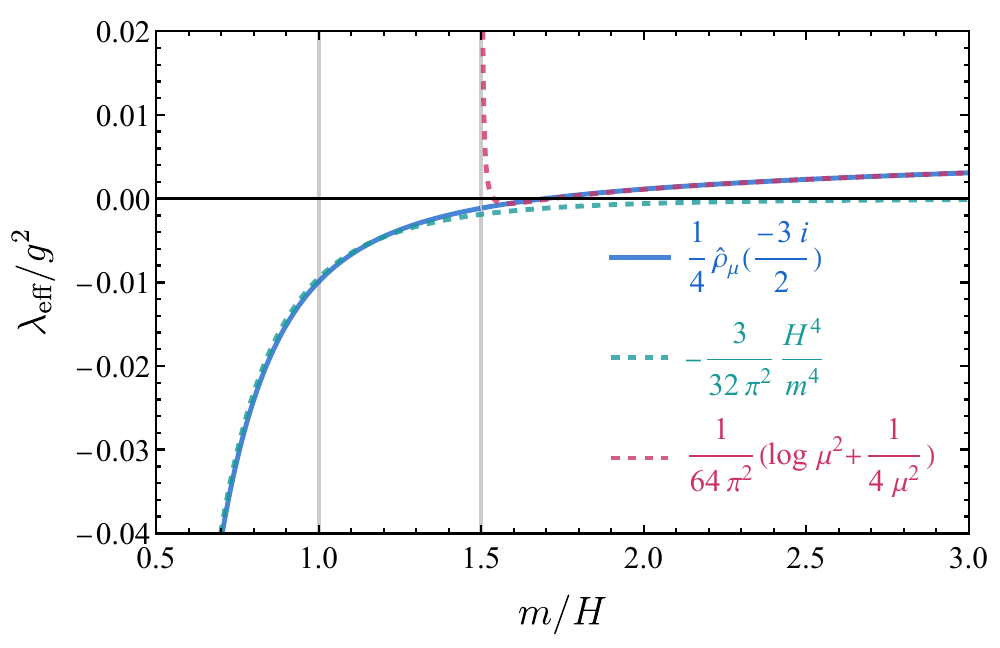}
    \caption{The effective quartic coupling $\lambda_{\mathrm{eff}}$ obtained by matching the four-point function in Eq.~\eqref{eq_loopIR}, with $\mu_R$ fixed to be $H$. The blue solid line represents the exact result, the green dashed line corresponds to the coefficient derived from the stochastic description in Eq.~\eqref{eq:stochasticquarticcoupling}, and the pink dashed line illustrates the interpolation valid in the large-mass limit.}
    \label{fig:quartic}
\end{figure}

Notice in general that the four-point function \eqref{eq_loopIR} contains both IR and UV divergences, where the UV divergence has been renormalised using $\overline{\text{MS}}$.
It is interesting to compute the quartic coefficient directly from the correlator. As we have discussed, the EFT depends on the mass of the integrated field, which in this case means that the coefficient depends directly on the renormalised spectral function $\widehat\rho_\mu(\mu')$ evaluated at $\mu'=-3i/2$. We plot in Fig.~\ref{fig:quartic} the quartic coupling as the mass changes. Notice that the coupling changes its sign depending on the mass of $\sigma$, signaling an important difference between the two EFT regimes. We will analyse both regions separately in order to better characterise these regimes.

\paragraph{Large mass limit}
In the large mass limit we require that the EFT reproduces a unitary quartic vertex, which imposes the relation $\lambda^{(2)}_{\mathrm{eff}}=\lambda^{(1)}_{\mathrm{eff}}/4$. 
and where the mixing term $\phi_a^2\phi_r^2$ is exponentially suppressed.

At large mass the effective quartic coupling could be obtained by using the WKB approximation and the heat kernel method~\cite{Bunch:1979uk}. We obtain that the IR part \eqref{eq_loopIR} in the large mass limit becomes:
\begin{align}
\lim_{\mu\gg1}\la\varphi_{\bm k_1}\varphi_{\bm k_2}\varphi_{\bm k_3}\varphi_{\bm k_4}\ra'_\text{IR} = &~
\frac{g^2H^4\sum k_i^3}{16\prod k_i^3}\times \frac{1}{16\pi^2}\left[ \log\left(\frac{H^2\mu^2}{\mu_R^2}\right) + \frac{1}{4\mu^2}\right] \times \log(-k_T\eta_0) \ .\label{IR_heavymasslimit}
\end{align}
where the second term comes from the curvature, and its coefficient is determined by numerical fit.\footnote{The coefficient for the $1/\mu^2$ term is slightly different from \cite{Xianyu:2022jwk}. We have checked that our result is a better fit.} Notice that in this case the coupling is positive and thus the theory is stable.

In order to do the matching to the EFT coefficient from \eqref{EFTloop} we just compare the result from the IR from the correlator in \eqref{IR_heavymasslimit} to the parameter obtained from computing the correlator in the unitary $\lambda\phi^4$ theory. We then get that 
\begin{align}
    \lambda_{\mathrm{EFT}}=\frac{g^2}{64\pi^2}\left(\log\frac{H^2\mu^2}{\mu_R^2}+\frac{1}{4\mu^2}\right) \ .
    \label{couplingmatchingloop}
\end{align}
The result depends on the choice of renormalisation scale $\mu_R$, which in standard flat-space quantum field theory is typically set equal to the heavy mass $\mu$ in order to cancel the logarithmic terms. If instead $\mu_R$ is chosen far from $m\sim H\mu$, the resulting logarithms can become large, making the correction terms dominant and potentially invalidating the perturbative expansion—for instance, by changing the sign of the effective coupling. For this reason, it is standard practice to choose $\mu_R$ close to the physical threshold $m\sim H\mu$, and then evolve parameters to lower energy scales using the renormalisation group flow.

A crucial difference between flat-space QFT and cosmology lies in the freedom to choose this renormalisation scale. In QFT, one can, in principle, probe the theory at any energy by adjusting the experimental setup, allowing $\mu_R$ to be chosen arbitrarily. In contrast, cosmological correlators are measured at a fixed physical scale set by the Hubble parameter $H$, corresponding to the moment of horizon crossing. This fixes the observational scale and limits the flexibility in choosing $\mu_R$. Although logarithmic terms are somewhat suppressed in this case, they can still be sizable if there exists a large hierarchy between $m$ and $H$.

In fact, when $m^2 \gg H^2$, the logarithmic term dominates over the finite terms in the loop correction, and the coupling becomes effectively dependent only on the microscopic coupling $g$, matching the flat-space result. The subleading term proportional to $H^2/m^2$ can be interpreted as a curvature correction due to the de Sitter background—it arises because the propagators are defined on de Sitter space—but in this regime, it remains subdominant.

An additional and important distinction from flat-space QFT is that, in cosmology, the RG flow does not necessarily preserve unitarity. Even if the effective theory is matched at a scale near $\mu$ where it is manifestly unitary, evolving it down to the observational scale $H$ generally generates non-unitary terms. This results in a non-zero coefficient for the non-unitary operator $\lambda_{\mathrm{NL}}$, as shown in~\cite{Frangi:2025xss}. However, in the regime where $m^2 \gg H^2$, these effects are exponentially suppressed due to the damping of off-diagonal density matrix components by a factor $\exp(-\pi \mu)$.

Because of this, it is often convenient to simply match the EFT directly at the Hubble scale, $\mu_R = H$ This approach works well in scenarios where correlators freeze after horizon exit. However, the situation changes in the theories of interest here, which exhibit IR divergences that cause correlations to grow with time until they reach equilibrium. In such settings, matching at horizon exit may not be sufficient, and one may instead need to match the EFT at equilibrium—a procedure that would require a detailed understanding of the RG flow precisely at that late-time scale.

\paragraph{Small mass limit}

When the mass scale $m$ falls below the Hubble scale $H$, the structure of the effective action changes qualitatively. In this regime, stochastic dynamics dominate, and a fully local and unitary description ceases to be valid — except in the limit where the infrared (IR) divergent terms provide the leading contribution. In this case, the scaling of fields changes, and the dynamics approaches a semiclassical limit.

In this small-mass limit, the IR contribution that we have computed dominates only when
$\log(-k\eta)\gg {m^2}/{H^2}$.
This corresponds precisely to the stochastic regime, where the coarse-graining cut-off lies below the horizon scale. The effective theory is then described by the action in Eq.~\eqref{EFTloop}. Although an analytic expression cannot be obtained directly from the spectral decomposition, Fig.~\ref{fig:quartic} shows that the quartic coupling becomes negative in this limit. This behaviour reflects the fact that the spectral function $\rho$ is non-positive for fields belonging to the complementary series. In the next section, we will verify that this small-mass scaling coincides with the result of an effective coupling from the stochastic description
\begin{align}
\lambda_{\mathrm{eff}} = -\frac{3}{32\pi^2}\frac{g^2 H^4}{m^2} \ .
\end{align}

The emergence of a negative coupling as the mass is lowered below the horizon scale is, at first sight, surprising, since it seems to indicate an instability of the theory on superhorizon scales. However, as we will show in the next section, this interpretation is incorrect: higher-order operators restore stability at large field amplitudes.

The change in sign of the coupling instead signals a different phase in the physical regime of the theory—one in which the dynamics are dominated by diffusion, and the semiclassical limit governs the behaviour of the system. In this phase, the boundary contributions of the quartic interaction (see Eq.~\eqref{red_quadratic}) dominate over the quantum effective action. Nevertheless, a full understanding of this sign reversal requires a more detailed analysis, which we leave for future work.

Finally, let us comment on the matching procedure in this regime. The matching scale $\mu_R = H$ is no longer appropriate, as it lies above the cut-off of the effective theory. To minimize large logarithms, the matching should instead be performed at a scale closer to the mass of the integrated-out field $\sigma$. As discussed previously, there remains a small ambiguity since the precise matching scale depends on where physical observables are evaluated. However, this ambiguity is subleading and does not affect the qualitative result—the sign change of the quartic coupling remains robust.
\paragraph{Validity of the EFT description}
Let us comment on the validity of the effective field theory. Since we integrate out the field $\sigma$, the parameters of the EFT are organised in powers of $gH^2/m^2$. For a fixed coupling $g < 1$, the theory becomes weakly coupled when the mass is large, $m^2 \gg H^2$. In this regime, the dynamics reduce to a quartic theory, and perturbation theory remains valid as long as $\phi \ll H/\sqrt{\lambda}$.

When perturbation theory eventually breaks down, certain classes of diagrams can still be resumed, by using the stochastic formalism,  allowing one to obtain results that extend beyond the naive perturbative expansion. However, this breakdown does not imply ultraviolet sensitivity: higher-order corrections—particularly those associated with the heavy field—lie beyond the domain that the stochastic description captures.

Since the second field $\sigma$ decays on subhorizon scales, it is conceivable that its residual effects could be associated with a different semiclassical saddle, such as those analyzed in~\cite{Celoria:2021vjw}. A detailed exploration of this possibility is left for future work.

As the mass decreases, however, the expansion parameter $gH^2/m^2$ grows, and consistency of the EFT requires
$\frac{gH^2}{m^2} \ll 1$.
Perturbation theory further imposes a stricter condition
\begin{align}
\frac{g^2 H^2}{m^2}\log(-k\eta) \ll 1 \ ,
\end{align}
since otherwise higher-loop corrections become significant and the perturbative description breaks down. In this low-mass regime, the dynamics can instead be captured using the stochastic formalism, as we will discuss in the next section.

Finally, it is worth emphasizing that the limit $m \to 0$ is ill-defined within the EFT: the parameter $gH^2/m^2$ eventually exceeds unity, signaling the onset of strong coupling and the breakdown of the effective description.

Let us now discuss the inclusion of self-interactions in the potential $V(\phi)$, which we have so far neglected. In the absence of a shift symmetry, a quartic self-coupling $\lambda\phi^4$ is naturally expected, and in most situations it contributes positively to the effective quartic coupling. Its presence would also modify the matching procedure, since one would need to include the corresponding one-loop correction from the massless interaction. This computation can be carried out using the same techniques developed here, but a more careful analysis would be required to determine whether new singularities appear when the spectral decomposition is evaluated at $\nu = 3/2$.
%=======================================
\section{Beyond Perturbation Theory}
\label{sec:bpt}
%=======================================

Even within the stochastic regime, there exists a point where perturbation theory breaks down. For instance, in the case of a quartic coupling, this occurs when the correction to the two point function from the quartic vertex becomes large,
$\frac{g H^2}{m^2} \log(k \eta) \gtrsim 1$,
at which stage higher-order terms become as important as the leading-order contribution, rendering the perturbative expansion invalid.

An alternative way of understanding this is by considering a coarse-grained theory for the long modes where the short modes are treated in perturbation theory. As the coarse-graining scale $\Lambda \to 0$, the number of long-wavelength modes grows without bound, leading to a loss of perturbative control. In this limit, the correct description is no longer a standard perturbative expansion but rather a stochastic framework, which resums IR divergences non-perturbatively and provides a well-defined late-time behaviour~\cite{Starobinsky:1986fx,Starobinsky:1994bd}.

\subsection{Stochastic computation}

To illustrate these ideas concretely, let us perform the stochastic computation for the original theory \eqref{action} of two scalar fields in de Sitter space.
In the limit where both fields are light $(m \ll H)$, one can derive the two-field Fokker–Planck equation for the joint probability distribution $P[\phi, \sigma; t]$ of the coarse-grained field. In this section, $\phi$ and $\sigma$ represent the long-wavelength fluctuations in the stochastic regime.
The resulting equation reads as follows
\begin{small}
\begin{align}
    \frac{\partial P[\phi,\sigma]}{\partial t}&=\frac{1}{3H}\frac{\partial}{\partial\phi}\Big(\left(\phi f(\sigma)+\partial_\phi V(\phi,\sigma)\right) P[\phi,\sigma]\Big)+\frac{1}{3H}\frac{\partial}{\partial\sigma}\left(\left(m^2\sigma+\frac12\phi^2\partial_\sigma f(\sigma)+\partial_\sigma V(\phi,\sigma)\right)P[\phi,\sigma]\right)\nonumber\\
    &\qquad+\frac{H^3}{8\pi^2}\frac{\partial^2P[\phi,\sigma]}{\partial\phi^2}+\frac{H^3}{8\pi^2}\frac{\partial^2P[\phi,\sigma]}{\partial\sigma^2}  \ .
    \label{eq:FP}
\end{align}
\end{small}As pointed out in~\cite{Cespedes:2023aal}, Eq.~\eqref{eq:FP} only incorporates the tree-level contribution to the wavefunction. While quantum corrections can be systematically included order by order, the comparison between our perturbative computations and the solutions of Eq.~\eqref{eq:FP} should be understood strictly within this tree-level approximation. This also indicates that perturbation theory must eventually be reorganised, as we will explain in detail later.

Rather than attempting to solve the full Fokker–Planck equation for all fields, we note that it admits an equilibrium solution of the form
\begin{align}
    P[\phi,\sigma]\sim\exp\left(-\frac{8\pi^2}{3H^4}\left(\frac{1}{2}m^2\sigma^2+\frac{1}{2}\phi^2f(\sigma)+V(\phi,\sigma)\right)\right) \ .
\end{align}
To directly compare this result with our previous perturbative calculations, we will integrate over $\sigma$ in the special cases where $g(\sigma)$ is either linear or quadratic, and where $V(\phi,\sigma)=0$. Generalizations to more complicated forms of $g(\sigma)$ and non-zero $V(\phi,\sigma)$ will be discussed later.

\paragraph{Case a. $f(\sigma)=\alpha\sigma$ } Here the equilibrium distribution becomes Gaussian in $\sigma$, and we can perform the path integral over $\sigma$ explicitly. The resulting marginal distribution for $\phi$ is:
\begin{align}
P[\phi]\sim\exp\left(\frac{4\pi^2}{3H^4}\frac{\alpha^2}{m^2}\phi^4\right) \ .
\end{align}
The resulting probability distribution corresponds to that of an effective quartic theory with a negative coupling,
$\lambda_{\mathrm{eff}} \sim -{\alpha^2}/{(2m^2)}$,
analogous to the effective theory analyzed in Section~\ref{sec:EFTdS}. However, because the original potential is unbounded from below, the probability distribution becomes non-normalizable, signaling an instability. As a consequence, the theory cannot be consistently defined beyond the perturbative regime, and any physical interpretation must be restricted to the domain where the perturbative expansion remains valid.

\paragraph{{Case b. $f(\sigma)={(g/2)}\sigma^2$} }
A more interesting and physically consistent case is a quadratic potential. In this scenario, the integral over $\sigma$ can still be performed analytically, yielding:
\begin{align}
P[\phi] \propto \left( \frac{H^2}{2m^2 + g \phi^2} \right)^{1/2}
\propto \exp\left[ -\frac{1}{2} \log\left( \frac{2m^2 + g \phi^2}{H^2} \right) \right] \ ,
\label{PDF:quartic}
\end{align}
which is shown in Fig. \ref{fig:pdfs}a).
One might worry about the normalisation of this probability distribution, as the integral of Eq.~\eqref{PDF:quartic} over $\phi$ would diverge at infinity but this divergence will be cured so long as we add a small mass for $\phi$ field.
The appearance of the logarithm closely resembles the structure of the functional determinant in Eqs.~\eqref{red_quadratic} and \eqref{eq:qeff}. Since the solution of the Fokker–Planck equation is non-perturbative in the fields, integrating out $\sigma$ in this way effectively resums all one-particle-irreducible (1PI) diagrams with internal $\sigma$ lines only. Thus, the resulting expression in Eq.~\eqref{PDF:quartic} keeps track of the full set of such diagrams at once.

However, this result is not equivalent to the effective action obtained in Eq.~\eqref{eq:rhoeff}. The reason is that the effective action contains both classical and quantum contributions, whereas Eq.~\eqref{PDF:quartic} contains only classical terms—specifically, those that originate from Eq.~\eqref{red_quadratic}.
To see this explicitly, we can write assume that the PDF has the canonical form for the equilibrium with a  potential given by \begin{align}
    U(\phi)=\frac{3H^4}{16\pi^2}\log\left(\frac{2m^2}{H^2}+\frac{g}{H^2}\phi^2\right) \ .
    \label{eq:effectivepot}
\end{align}
from this we can compute the coefficients after expanding in $\phi$. This yields the following expansion
\begin{align}
    U(\phi) = \frac{3H^4}{16\pi^2}\log\left(\frac{2m^2}{H^2}\right) + \frac{3gH^4}{32m^2}\phi^2 - \frac{3g^2H^4}{128m^4}\phi^4 + \cdots
\end{align}
which defines a controlled effective expansion in $\phi$, valid as long as $g\phi^2 \ll m^2$. From this expression, the effective quartic coupling can be read off as
\begin{align}
    \lambda_\mathrm{eff}\sim-\frac {3}{32\pi^2}\frac{g^2H^4}{m^4} \ ,
    \label{eq:stochasticquarticcoupling}
\end{align}
which has the opposite sign compared to the one obtained via loop matching in \eqref{couplingmatchingloop}.

Notice that this coupling exactly matches the result obtained by the loop integral Eq.~\eqref{eq_loopIR} in the regime when $m^2/H^2\ll 1$. 
The sign difference between the large and low masses can be understood as follows. As shown in~\cite{Cespedes:2023aal}, the stochastic approach resums the classical non-linearities that dominate on superhorizon scales. These correspond precisely to the terms appearing in Eq.~\eqref{red_quadratic}, which arise directly from the classical equations of motion. In contrast, the quantum effective action in Eq.~\eqref{effective_action} contains contributions that are protected from IR divergences: the internal propagators appearing in loops vanish as their momenta go to zero. This suppression prevents quantum terms from growing strongly at late times, while the classical contributions continue to build up.
Consequently, at very large scales, the classical terms resumed by the Fokker–Planck equation dominate, while quantum corrections remain subleading and must be added perturbatively, order by order.
Furthermore, notice that even though the effective quartic coupling is negative, the full probability distribution, once considering higher order non-linearities, as in Eq.~\eqref{PDF:quartic} remains well-defined and bounded, representing a renormalizable probability distribution. This contrasts with the case of a truly unstable potential, where the distribution would be non-normalizable.

\subsection{Tail of the distribution}
Another important feature of the probability distribution in Eq.~\eqref{PDF:quartic} is that its tail decays much more slowly than that of a standard quartic theory. This happens because the PDF is valid for a larger range of fields values than the quartic theory. As a result, for larger values of $\phi$ the equilibrium distribution no longer resembles that of a quartic theory; instead, it exhibits qualitatively different behaviour. In this regime, the tail becomes sensitive to the UV completion of the theory, as emphasized in~\cite{Cohen:2021jbo,Cohen:2022clv}.

We can estimate the point at which the effect of the tail becomes important with the following argument. The validity of the effective description of a $\lambda \phi^4$ theory can be estimated by comparing the quartic interaction to the kinetic term. Using the effective coupling in Eq.~\eqref{couplingmatchingloop}, this criterion gives
\begin{align}
\frac{\phi}{H} \ll  \lambda_{\mathrm{eff}}^{-1/4} \sim g^{-1/2} \frac{m}{H} \ .
\label{eq:validityquartic}
\end{align}
In the probability distribution given by Eq.~\eqref{PDF:quartic}, the regime where the mass term dominates corresponds to field values small enough that an expansion around the Gaussian saddle point is valid. In this regime, the perturbative description holds, and the effective theory remains weakly coupled. However, for larger field values, the perturbative expansion gradually breaks down, and the condition in Eq. \eqref{eq:validityquartic} eventually fails. In terms of the distribution, this occurs when the second term inside the logarithm begins to dominate. At that point, it is no longer possible to expand in the small coupling: all diagrams contribute at the same order, signaling a breakdown of perturbation theory. In this regime, one expects the correlation functions of $\phi$ to become  enhanced, similar to the behaviour observed in~\cite{Panagopoulos:2019ail,Panagopoulos:2020sxp,Achucarro:2021pdh}. Computing the rest of the coefficients from the potential in \eqref{eq:effectivepot}, leads to the following expansion,
\begin{align}
    c^{(n)}\equiv\frac{1}{n!}\left.\frac{d^{n}U}{d\phi^n}\right\vert_{\phi=0}=-\frac{3 \times 2^{-3-n/2}}{n\pi^2}\frac{H^4g^{n/2}}{m^n}\cos\left(\frac{n\pi}{2}\right) \ ,\qquad n=1,2,3,
    \cdots
\end{align}
thus clearly for $\phi$ larger than $H\lambda_{\mathrm{eff}}^{1/4}$ all terms become of a similar size, and need to be considered. 

As the mass of $\sigma$ decreases, the range of validity of the perturbative expansion shrinks further, and it becomes necessary to systematically include higher-order diagrams to maintain a consistent description. In particular, when $g H^2/m^2 \gtrsim 1$, the perturbative EFT breaks down entirely, as the theory enters a strongly coupled regime. In the stochastic description, this corresponds to taking the limit $m \to 0$, which, while still formally allowed, renders the Gaussian saddle point expansion ill-defined. In this case, the probability distribution becomes singular near $\phi = 0$.

Of course, the example considered here is simplified: in realistic models, one would expect higher-order self-interactions in both $\sigma$ and $\phi$. These interactions could dynamically generate a mass, regularizing the distribution around $\phi = 0$, as occurs in standard $\lambda \phi^4$ theories. Nevertheless, it is remarkable that the stochastic framework remains applicable even in this heavily non-perturbative regime. The presence of a heavy tail in the distribution encodes precisely this non-perturbative structure.

\begin{figure}
    \centering
    \includegraphics[width=0.9\linewidth]{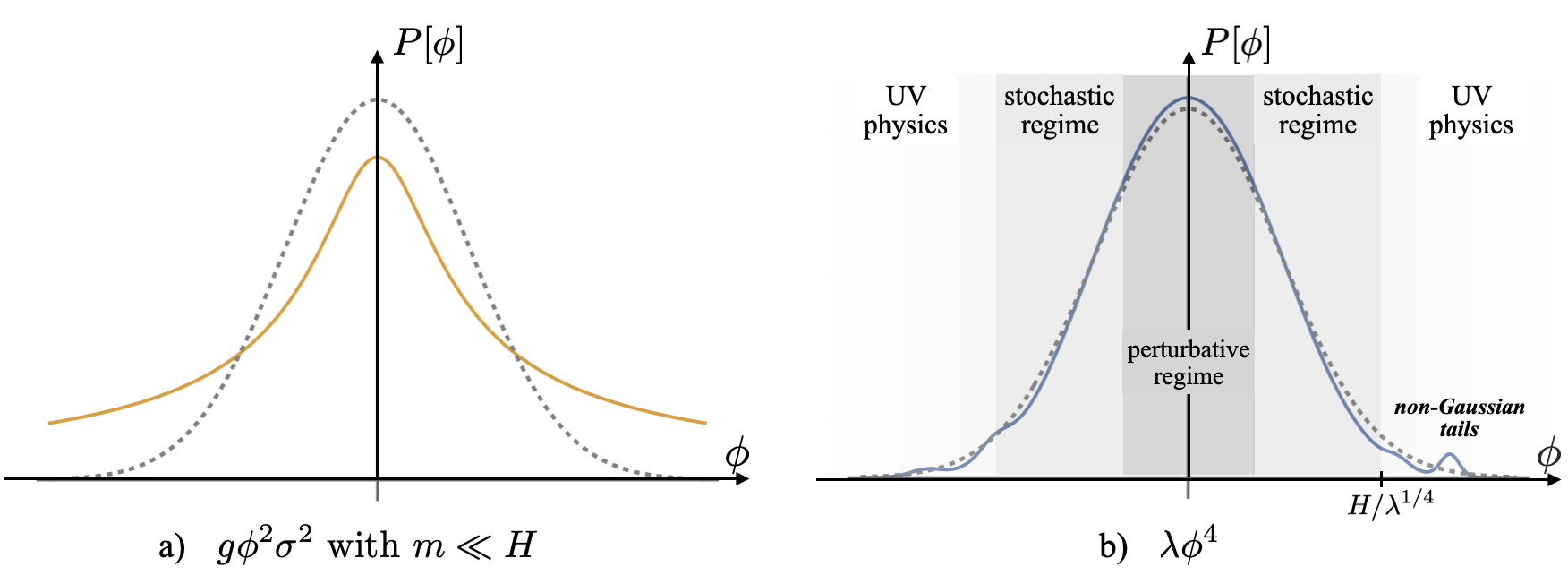}
    \caption{The probability distributions of $\phi$ by tracing out $\sigma$: a) heavy tail \eqref{PDF:quartic} in the small mass limit; b) illustration of the non-Gaussian tail in the large mass limit where the effective $\lambda\phi^4$ breaks down and UV physics becomes relevant. The dashed gray lines are the Gaussian distributions. }
    \label{fig:pdfs}
\end{figure}

\paragraph{Validity of the Fokker-Planck equation}
Let us now clarify in which regimes the stochastic description gives the correct account for the physics of the problem. Let us begin by considering the following equations of motion
\begin{align}
\ddot \phi + 3H \dot\phi + f(\sigma)\phi + \frac{k^2}{a^2}\phi &= 0 \ , \\
\ddot \sigma + 3H \dot\sigma + m^2 \sigma + \frac{1}{2} \partial_\sigma f(\sigma) \phi^2 + \frac{k^2}{a^2}\sigma &= 0  \ .
\end{align}
We assume that $\sigma$ behaves as a stochastic field. In this case, its spatial gradients are negligible compared to all other scales:
\begin{align}
\frac{k^2}{a^2} \ll m^2 \ll H^2 \ ,
\end{align}
where we have momentarily ignored the interaction $f(\sigma)$.

Let us now examine the first equation. Since $\phi$ is massless, for it to be treated as a stochastic variable its physical momentum must be smaller than the effective mass induced by its coupling to $\sigma$:
\begin{align}
\frac{k^2}{a^2} \ll f(\sigma) \ll H^2  \ .
\end{align}
To make this more explicit, consider the case $f(\sigma) = g\sigma^2/2$. If $\sigma$ has reached equilibrium, then its variance satisfies $\langle \sigma^2 \rangle \sim H^4/m^2$, and the effective mass for $\phi$ becomes
\begin{align}
\frac{k^2}{a^2} \lesssim \frac{g H^4}{m^2} \ .
\label{eq:momentumregime}
\end{align}

The term on the right-hand side is precisely the mass generated by quantum corrections in the effective action, cf. Eq.~\eqref{EFTloop}. It also controls the EFT expansion parameter, and thus should remain smaller than $H^2$ as long as the modes lie within the regime of validity of the EFT. Notice that even when the field $\sigma$ is not at equilibrium the regime still applies as the inequality in Eq.~\eqref{eq:momentumregime} is an upper bound for the physical momenta.

To reach an equilibrium distribution for $\phi$, perturbation theory must already have broken down—otherwise the stochastic regime never becomes relevant. Perturbation theory fails when loop corrections to the two-point function of $\phi$ become order one, that is,
\begin{align}
    \frac{gH^2}{m^2}\log(-k\eta)\gtrsim 1 \ .
\end{align}
For both conditions (EFT validity and stochasticity) to be simultaneously satisfied, there must exist a range of momenta such that
\begin{align}
    e^{-\frac{m^2}{H^2}g^{-1/2}}\lesssim \frac{k}{aH}\lesssim \mathrm{min}\left( \frac{H^2g^2}{m^2},\frac{m^2}{H^2}\right)\ll 1,\qquad\mathrm{where} -k\eta\leq 1 \ ,
\end{align}
This is precisely the regime in which what we have referred to as the stochastic regime is valid: the field $\sigma$ behaves as a stochastic variable, while $\phi$ can still be treated perturbatively. As time evolves, however, the physical momentum $k/a$ continues to redshift. Once it drops below the exponential lower bound above, perturbation theory also breaks down for $\phi$, and a fully stochastic treatment for both fields becomes necessary.

Let us briefly comment on the inclusion of a potential $V(\phi,\sigma)$. The main effect of self-interactions for $\sigma$ would be to modify the expansion parameter $g^2 H^2/m^2$, which will now depend on the different coefficients of the expansion. However, this cannot be larger than one; otherwise, the expansion fails, and thus the results we have proposed will not change qualitatively. Introducing a self-interaction for $\phi$ primarily affects the equilibrium distribution, particularly the behaviour of its tails. At large field values, the self-coupling dominates over the logarithmic corrections discussed here. It should be noted, however, that the validity of this new interaction is limited to the range defined by the EFT; beyond this regime, the predictions cannot be trusted. In order to obtain a robust and controlled late-time tail, all self-interactions of $\phi$ must remain suppressed—such as in the presence of a shift symmetry~\cite{Panagopoulos:2019ail,Achucarro:2021pdh}.

\paragraph{Large mass limit} In the end, let us also take a look at the behaviour of the tail in the large mass limit analysed in Section \ref{sec:loop}. This case successfully reproduces $\lambda\phi^4$ as an effective theory, within and beyond perturbation theory.
An illustration of the probability distribution is shown in Fig. \ref{fig:pdfs}b).
In the non-perturbative regime, as $m\gg H$, the $\sigma$ field is not stochastic. Thus, the Fokker-Planck equation \eqref{eq:FP} reduces to the one for the single-field $\lambda\phi^4$. However, as we approach the tail of the distribution, the theory becomes strongly coupled and the single field $\lambda\phi^4$ breaks down for $\phi\gtrsim H/\lambda^{1/4}$. 
This suggests that the effects of UV physics become significant when fluctuations get large, and we need new tools beyond the stochastic formalism to describe the possibly nontrivial behaviour of the non-Gaussian tail.

%=======================================
\section{Discussions and Outlook}
\label{sec:conclussions}
%=======================================
In this work we have looked into the effective field theories (EFTs) of a light scalar field in de Sitter space under different choices of cut-off, ranging from subhorizon to superhorizon scales. Our investigation describes the low-energy EFT of a light field coupled to a heavy sector, whose mass sets the physical cut-off of the theory. We identified two qualitatively distinct regimes, summarised in Table~\ref{tab:summary}.

When the cut-off is much larger than the Hubble scale $H$, the EFT remains approximately unitary and can be described by a conventional Wilsonian expansion. In contrast, when the cut-off lies below $H$, the resulting EFT describes a mixed state that encodes the loss of quantum coherence due to the integration of superhorizon modes. Although the coarse-graining scale is still larger than the horizon $1/H$, the dominant interaction remains a local operator, while subleading vertices become non-local and account for decoherence effects. This non-unitary phase exhibits a modified scaling behaviour, signaling the emergence of a semiclassical regime as the cut-off is lowered.

One important feature of the different EFT regimes is that the effective couplings may change sign as the cut-off scale is lowered. We explicitly observe this in a model where a heavy field, coupled to the light sector, is integrated out at one loop. The resulting effective quartic coupling is positive for heavy masses above the horizon scale but becomes negative as the mass is reduced below $H$. Remarkably, this low-mass limit coincides with the quartic vertex obtained from the effective potential derived via the stochastic approach, after marginalising over the heavy field in the equilibrium distribution. This agreement provides a concrete link between the Wilsonian and stochastic descriptions of infrared dynamics in de Sitter space.  Moreover it allows us to distinguish between different UV realisations of $\lambda\phi^4$.

Our findings point to the existence of distinct dynamical phases in de Sitter space and to the appearance of a diffusion-dominated regime at low cut-offs. This regime provides a natural bridge to the stochastic description of inflation, which becomes valid precisely in this limit.

Several directions for future work emerge from our analysis:
\begin{itemize}
\item We derived the EFT by integrating out a heavy scalar field. It would be interesting to investigate the structure of the EFT when considering fields with higher spin or composite operators. Another complementary direction would be to repeat the analysis by explicitly coarse-graining short-wavelength modes, which would make the connection with open-system approaches more direct. An explicit construction of an EFT along these lines has been presented in~\cite{Li:2025azq}, and it would be interesting to understand its relation to our framework. Moreover, it would be important to clarify how dissipation emerges from integrating out degrees of freedom and what impact it has on the EFT—particularly through the appearance of non-local operators. For related analyses, in which dissipation and diffusion arise effectively as local operators, see for example~\cite{Creminelli:2023aly,Salcedo:2024smn}.
\item In Section~\ref{sec:loop} we observed that the quartic coupling changes sign as the heavy-field mass decreases. 
One interesting question is about the notion of positivity in cosmology. Differing from QFT in flat spacetime, there would be more subtleties about the sign of couplings in the cosmological EFTs.
It would be worthwhile to examine whether other observables exhibit similar non-trivial behaviour. Recent studies have suggested that the anomalous dimensions of spinning fields may violate unitarity in certain limits, and our methods could be extended to probe such phenomena~\cite{Green:2023ids}.
\item Another future direction is to go beyond perturbation theory and establish a new description for non-Gaussian tails. One recent attempt along this line is the semiclassical method \cite{Celoria:2021vjw,Creminelli:2024cge}. As shown in concrete examples here, the tail of the probability distribution is expected to be sensitive to the microscopic details of the low-energy EFT and the stochastic formalism may not be able to capture the behaviour of fluctuations with very large amplitudes.
It would be very interesting to investigate how different UV realisations of $\lambda\phi^4$ manifest their effects in the highly non-perturbative regime of the probability distribution.
\item Finally, our analysis focused on models without a shift symmetry. In realistic inflationary settings the inflaton often possesses a non-linearly realised shift symmetry, which forbids the leading IR-divergent contributions. In that case our organising principle would need to be modified, yet a semiclassical regime is still expected to emerge on superhorizon scales. Characterising this limit within our approach would provide a direct link between stochastic and EFT descriptions.
\end{itemize}

\vskip12pt
\paragraph{Acknowledgements} 
%=======================================

We would like to thank Santiago Agüi Salcedo, Qi Chen, Andrew Cohen, Thomas Colas, Sadra Jazayeri, Greg Kaplanek, Arttu Rajantie, Guanhao Sun, Xi Tong, Zhong-Zhi Xianyu for insightful discussions.
SC would like to thank Gonzalo Palma and the physics department at FCFM, Universidad de Chile for their hospitality during part of this work. ZQ would like to thank Thomas Colas and Xi Tong for collaborations on relevant projects. DGW wishes to thank Yanjiao Ma, Xiangwei Wang, Yi Wang and Wenqi Yu for discussions on the tree-level computation in Section \ref{sec:EFTdS} from a different perspective. Part of this project took place during the programme CoBALt held at the Institut Pascal at Université Paris-Saclay with the support of the program “Investissements d’avenir” ANR-11-IDEX-0003-01.

SC is supported in part by the STFC Consolidated Grants ST/T000791/1 and ST/X000575/1 and by a
Simons Investigator award 690508.
ZQ is supported by NSFC under Grant No.\ 12275146, the National Key R\&D Program of China (2021YFC2203100), and the Dushi Program of Tsinghua University.
DGW is partially supported by a Rubicon Postdoctoral Fellowship from the Netherlands Organisation for Scientific Research (NWO), the EPSRC New Horizon grant EP/V048422/1, and the Stephen Hawking Centre for Theoretical Cosmology. 

\clearpage
\phantomsection
\clearpage
\phantomsection
\appendix
\section{Various Propagators}
\label{app:propagators}
%%%%%%%%%%%%%%%%%%%%%%%%%%%%%%%%%%%%%%%%%%%%%%%%%%%%%%%%%%%%%%%%%%%%%%%%%%%%%%%%%%%%%%%%%%%%%%%%%%%%%%%%t
In this work, we make use of various versions of propagators, so we clarify and summarise our notation below. These propagators can be built up from the mode function of the field, so without loss of generality, let us consider a self-interacting theory of a scalar field $\sigma$ with mass $m$ in de Sitter space, and we work with the conformal time for convenience.

As in standard QFT textbooks, the field operator $\sigma_{\bm k}(\eta)$ can be expanded by creation/annihilation operators with the coefficients called the mode function $\sigma(k,\eta)$:
\begin{align}
\si_{\bm k}(\eta) = \si(k,\eta) a_{\bm k} + \si^*(k,\eta)a_{-\bm k}^\dagger\ .
\end{align}
In particular, the mode function satisfies the Klein-Gordon equation in de Sitter background,
\begin{align}
    \left(\eta^2 \pd_\eta^2 - 2\eta\pd_\eta + k^2\eta^2 + \frac{m^2}{H^2} \right)  \si(k,\eta) = 0 \ ,
\end{align}
and is normalised by the Wronskian condition,
\begin{align}
    \si(k,\eta)\si'^*(k,\eta)-\si^*(k,\eta)\si'(k,\eta)=iH^2\eta^2\ .
    \label{eq_Wronski}
\end{align}
Once we specify the Bunch-Davies initial condition, we can solve the mode function up to an irrelevant phase:
\begin{align}
    \si(k,\eta) = \frac{H\sqrt\pi}{2}e^{-\pi\mu}(-\eta)^{3/2}\mathrm H^{(1)}_{i\mu}(-k\eta),\qquad \mu\equiv\sqrt{\frac{m^2}{H^2}-\frac94}\ .
\end{align}
Notice that we have introduced the mass parameter $\mu$ that is positive real for $\si$ lying in the principal series with mass $m>3H/2$.  For $\si$ in the complementary series with $m<3H/2$, $\mu \to i \sqrt{9/4-m^2/H^2}$ becomes pure imaginary but the expression still applies. In particular, for the massless scalar field $\phi$, we have $m=0$ and $\mu=3i/2$, and the mode function reads:
\begin{align}
    \phi(k,\eta) =\frac{H}{\sqrt{2k^3}}(1+ik\eta)e^{-ik\eta}\ .
\end{align}
\subsection{Propagators for wavefunction coefficients}
The wavefunction for a given state $|\Psi\rangle$ is defined as its inner product with the field eigenstates $|\varsigma(\eta_0)\rangle$ at a given time slice,
\begin{align}
    \Psi[\varsigma] \equiv \la \varsigma(\eta_0)| \Psi \ra \ ,
\end{align}
and it is well-known that the wavefunction can be computed by a path integral with boundary conditions:
\begin{align}
    \Psi[\varsigma] = \int_{\sigma(\eta=-\infty)=0}^{\sigma(\eta=\eta_0)=\varsigma} \mathcal D\sigma \,e^{iS[\sigma]} \ ,
    \label{eq_defWF}
\end{align}
where $S$ is the action of the theory, and in the past infinity we have assumed the vacuum condition. 
For a free theory, the wavefunction is Gaussian.
Therefore, it is convenient to expand the wavefunction around a Gaussian distribution so long as perturbativity is assumed,
\begin{align}
    \Psi[\varsigma] = \exp\left[
    \frac12 \int_{\bm k_1,\bm k_2} \psi_2(\bm k_1,\bm k_2)\, \varsigma_{\bm k_1}\varsigma_{\bm k_2} + \sum_{n=3}^\infty \frac{1}{n!}\int_{\bm k_1,\cdots,\bm k_n}\psi_n(\bm k_1,\cdots,\bm k_n)\,\varsigma_{\bm k_1}\cdots\varsigma_{\bm k_n}
\label{def_wvcoe}
    \right] \ ,
\end{align}
where $\psi_n$ is the $n$-point wavefunction coefficient, and the integral measurement is defined as:
\begin{align}
    \int_{\bm k_1,\cdots,\bm k_n} \equiv \int \prod_{i=1}^n \frac{d^3\bm k_i}{(2\pi)^3}\,\times (2\pi)^3\delta(\bm k_1+\cdots+\bm k_n)\ .
\end{align}

For tree diagrams, the path integral can be calculated via the saddle point approximation, that is, the action for the classical field configuration $\sigma_\text{cl}$,
\begin{align}
    \Psi[\varsigma] = e^{iS[\sigma_\text{cl}]}\ ,\qquad \left.\frac{\delta S[\sigma]}{\delta\sigma}\right|_{\sigma=\sigma_\text{cl}} = \int d^4x  \left[ \sqrt{-g} ( \Box - m^2)\sigma_\text{cl}+  \left.\frac{\delta \mathcal L_\text{int}}{\delta \sigma}\right|_{\sigma=\sigma_\text{cl}} \right] = 0\ . \label{eq_treeWF}
\end{align}
One can then easily solve $\sigma_\text{cl}$ from the EoM and the boundary conditions using the Green's function method,
\begin{align}
\sigma_\text{cl}(\eta,\bm k) = \varsigma(\eta_0,\bm k)  K(k,\eta) + i\int d\eta'  \left.\frac{\delta \mathcal L_\text{int}}{\delta \sigma}\right|_{\sigma=\sigma_\text{cl}(\eta')} G(k,\eta,\eta')\ ,
\label{eq_clasolution}
\end{align}
with the bulk-to-boundary propagator $K$ satisfying the Klein-Gordon equation:
\begin{align}
(\Box_k-m^2) K(k,\eta) = 0\ ,\qquad \lim_{\eta\to-\infty}K(k,\eta) = 0\ ,\qquad \lim_{\eta\to0}K(k,\eta)=1\ ,
\end{align}
and the bulk-to-bulk propagator $G$ being the Green's function:
\begin{align}
(\Box_k-m^2) G(k,\eta_1,\eta_2) = \frac{i}{\sqrt{-g}}\delta(\eta_1-\eta_2),\qquad \lim_{\eta_{1,2}\to-\infty}G(k,\eta_1,\eta_2)= \lim_{\eta_{1,2}\to0} G(k,\eta_1,\eta_2) = 0.
\end{align}
In particular, the bulk-to-boundary and the bulk-to-bulk propagators can be expressed in terms of mode functions,
\begin{align}
    &K(k,\eta)= \frac{u^*(k,\eta)}{u^*(k,\eta_0)}\ ,\\
    &G(k,\eta_1,\eta_2) = u(k,\eta_1)u^*(k,\eta_2)\theta(\eta_1-\eta_2) + (\eta_1\leftrightarrow\eta_2) - u^*(k,\eta_1)u^*(k,\eta_2)\frac{u(k,\eta_0)}{u^*(k,\eta_0)}\ .
\end{align}
By inserting the classical solution \eqref{eq_clasolution} back into Eq.~\eqref{eq_treeWF} and comparing it with Eq.~\eqref{def_wvcoe}, one can obtain the $n$-point wavefunction coefficients, see e.g. \cite{Cespedes:2023aal}. And finally, the equal-time $n$-point correlation function can be derived from the wavefunction,
\begin{align}
\langle \si_{\bm k_1}\cdots \si_{\bm k_n} \rangle_{\eta_0} = \int d\varsigma \, \varsigma_{\bm k_1}\cdots \varsigma_{\bm k_n} \left| \Psi[\varsigma]\right|^2\ .
\end{align}

\subsection{Propagators for in-in correlator}
Alternatively, the $n$-point correlators can also be computed via the Schwinger-Keldysh or the in-in formalism mentioned in Sec.\ \ref{sec:DensityMatrix}, where we double the field operator $\sigma$ to be $\sigma_\pm$ and the $n$-point correlator is expressed as the following path integral,
\begin{align}
\langle \si_{\bm x_1}\cdots \si_{\bm x_n} \rangle_{\eta_0} = \int_\Omega^\sigma \mathcal D\sigma_+ \int_\Omega^\sigma \mathcal D \sigma_-\, \sigma_\pm(\bm x_1,\eta_0)\cdots\sigma_\pm(\bm x_n,\eta_0)e^{iS[\sigma_+] - i S[\sigma_-]}\ .
\label{eq_+-PI}
\end{align}
where the bulk fields appearing in the path integral could be either $\sigma_+$ or $\sigma_-$ since they take the same value $\sigma$ at the boundary.
As flat-space amplitudes, the in-in correlator can be computed following a well-established Feynman rule with four types of in-in propagators, differing by the time-orderings,
\begin{align}
&G_{-+}(k,\eta_1,\eta_2) =  \langle  \sigma_-(\eta_1,\bm k) \sigma_+(\eta_2,-\bm k) \rangle' = \sigma(k,\eta_1)\sigma^*(k,\eta_2)\ ,\\
&G_{+-}(k,\eta_1,\eta_2) =  \langle \sigma_+(\eta_1,\bm k) \sigma_-(\eta_2,-\bm k) \rangle' = \sigma^*(k,\eta_1)\sigma(k,\eta_2)\ ,\\
&G_{++}(k,\eta_1,\eta_2) =  \langle \sigma_+(\eta_1,\bm k) \sigma_+(\eta_2,-\bm k) \rangle' = G_{-+}(k,\eta_1,\eta_2) \theta(\eta_1-\eta_2) + G_{+-}(k,\eta_1,\eta_2) \theta(\eta_2-\eta_1)\ ,\\
&G_{--}(k,\eta_1,\eta_2) =  \langle \sigma_-(\eta_1,\bm k) \sigma_-(\eta_2,-\bm k) \rangle' = G_{+-}(k,\eta_1,\eta_2) \theta(\eta_1-\eta_2) + G_{-+}(k,\eta_1,\eta_2) \theta(\eta_2-\eta_1)\ ,
\end{align}
and the correlators receive contributions from all possible diagrams where each vertex could be in either forward or backward branches of the Schwinger-Keldysh contour, composed of $\sigma_+$ or $\sigma_-$ respectively. We further introduce boundary propagators just for convenience,
\begin{align}
K_+(k,\eta) \equiv G_{+\pm}(k,\eta,\eta_0)\ ,\qquad  K_-(k,\eta) \equiv G_{-\pm}(k,\eta,\eta_0)\ ,
\end{align}
where the $\pm$ subscript for the boundary vertex does not matter. When external fields are massless, we can usually (but not always) safely take the limit $\eta_0\to 0$ for the boundary propagators $K_\pm$.
We recommend readers refer to \cite{Chen:2017ryl} for further details and a simple diagrammatic rule.

\subsection{Propagators for advanced/retarded basis}
The path integral \eqref{eq_+-PI} can also be written in the advanced/retarded basis of fields. In particular, we can change the integral variables from $\sigma_\pm$ to $\sigma_{a/r}$ similar to Eq.~\eqref{eq_+-2ar}, and we obtain:
\begin{align}
\langle \si_{\bm x_1}\cdots \si_{\bm x_n} \rangle_{\eta_0} = \int_\text{BD}^0 \mathcal D\sigma_a \int_\text{BD}^\sigma \mathcal D \sigma_r\, \sigma_r(\bm x_1,\eta_0)\cdots\sigma_r(\bm x_n,\eta_0)e^{iS[\sigma_a,\sigma_r]}\ ,
\label{eq_arPI}
\end{align}
where the action for the density matrix can be derived from the original action,
\begin{align}
S[\sigma_a,\sigma_r] = S\left[\sigma_r + \frac{\sigma_a}2\right] - S\left[\sigma_r - \frac{\sigma_a}2 \right]\ .
\end{align}
In particular, the free theory action is,
\begin{align}
S_0[\sigma_a,\sigma_r] = - \frac12 \int d^4x\sqrt{-g}
\begin{pmatrix}
\phi_r & \phi_a
\end{pmatrix}
\begin{pmatrix}
0 & -\Box_{\epsilon^+}+m^2 \\
-\Box_{\epsilon^-}+m^2 & i f(k) \epsilon
\end{pmatrix}
\begin{pmatrix}
\phi_r \\ \phi_a
\end{pmatrix}\ ,
\end{align}
where $\epsilon \to 0^+$ is the prescription, $f(\partial_i^2)$ is a positive function encoding the occupation of state~\cite{Kamenev:2009jj}, and by $\Box_{\epsilon^\pm}$ we mean $\partial_\eta$ in $\Box = g^{\mu\nu}\partial_\mu\partial_\nu$ should be replaced by $\partial_\eta\pm \epsilon $.  With this $i\epsilon$-prescription taken into account, the differential operator in the free action is invertible, whose inverse gives the advanced/retarded and Keldysh propagators. Turning to momentum space, we have
\begin{align}
\begin{pmatrix}
G_K & G_R \\
G_A & 0
\end{pmatrix} = \frac{1}{\sqrt{-g}} \begin{pmatrix}
0 & -\Box_{\epsilon^+}+m^2 \\
-\Box_{\epsilon^-}+m^2 & i f(k) \epsilon
\end{pmatrix} ^{-1}\ ,
\end{align}
that gives,
\begin{align}
&G_R(k,\eta_1,\eta_2) = i \left[ \si(k,\eta_1)\si^*(k,\eta_2) - \si^*(k,\eta_1)\si(k,\eta_2)\right]\theta(\eta_1-\eta_2)\ ,\label{eq_GR}\\
&G_A(k,\eta_1,\eta_2) = i \left[ \si^*(k,\eta_1)\si(k,\eta_2) - \si(k,\eta_1)\si^*(k,\eta_2)\right]\theta(\eta_2-\eta_1)\ ,\label{eq_GA}\\
&G_K(k,\eta_1,\eta_2) = \frac i2 \left[ \si(k,\eta_1)\si^*(k,\eta_2) + \si^*(k,\eta_1)\si(k,\eta_2)\right]\ ,\label{eq_GK}
\end{align}
where we have taken the limit $\epsilon\to 0^+$. 
One can then easily compute correlation functions by the corresponding Feynman rule, where external fields are always $\sigma_r$, and all propagators are equipped with an additional factor of $-i$ by construction, see e.g.~\cite{Salcedo:2024smn}.

\section{Four-Point Function from Tree-Level Exchange}
\label{app:UVcomputation}
Let we consider the following action with the cubic interaction:
\begin{align}
    &S[\phi,\sigma] = S_0[\phi] + S_0[\sigma] + S_\text{int}[\phi,\sigma]\ ,\label{eq_S}\\
    &S_0[\phi] = \int d^4x\sqrt{-g}\left[-\frac12 (\partial_\mu\phi)^2\right]\ ,\qquad 
    S_0[\sigma] =\int d^4x\sqrt{-g}\left[
    -\frac12\sigma(-\Box+m^2)\sigma\right]\ ,\\
    &S_\text{int}[\phi,\sigma] = -\frac{\alpha}2\int d^4x\sqrt{-g} \phi^2\sigma\ .
\end{align}
Since the action is quadratic in $\sigma$, we can integrate it out in the partition function,
\begin{align}
    Z = \int \mathcal D\phi \mathcal D\sigma e^{iS[\phi,\sigma]}\ .
\end{align}
To do so, we define a shifted field $\bar\sigma$,
\begin{align}
    \bar\sigma(x) \equiv \sigma(x) +\frac{\alpha}2 \int d^4y\frac{\delta^4(x-y)}{-\Box+m^2}\phi^2(y)\ .
\end{align}
Thus, in terms of $\bar\sigma$, the action becomes:
\begin{align}
    S[\phi,\sigma] =&~S_0[\phi] + S_0[\bar\sigma]
    +\frac{\alpha^2}8\int d^4x d^4y\sqrt{-g}  \phi^2(x) \frac{\delta^4(x-y)}{-\Box+m^2} \phi^2(y)\nonumber\\
    \equiv&~ S_0[\phi,\bar\sigma] + \bar S_\text{int}[\phi]\ ,
\end{align}
and the partition function of the system becomes:
\begin{align}
    Z =&~ \int\mathcal D\phi \mathcal D\bar\sigma e^{i S[\phi,\sigma]}\nonumber\\
    =&~\int\mathcal D\bar\sigma e^{iS_0[\bar\sigma]} \times \int\mathcal D\phi e^{i S_0[\phi] + i \bar S_\text{int}[\phi]}\ .
\end{align}
Notice that the shift $\sigma\to\bar\sigma$ does not affect the integral measure.
As one can see, since the path integral of $\bar\sigma$ is Gaussian, the system can be described by a single field theory with the effective action $S_0[\phi] + \bar S_\text{int}[\phi]$, where the interaction term $\bar S_\text{int}[\phi]$ is nonlocal due to $\Box$ in the denominator. Nevertheless, when $m^2 \gg |\Box^2|$, we can perform the large mass expansion to obtain a local EFT.

On the other hand, when the mass is not large, we can apply another trick to rewrite the interaction. Using the identity:
\begin{equation}
    \frac{\delta^4(x-y)}{-\Box+m^2} = \frac{\delta^4(x-y)}{m^2} + \frac{\delta^4(x-y)}{m^2}\frac{\Box}{-\Box+m^2}\ ,
\end{equation}
we can rewrite the interaction as:
\begin{align}
    \bar S_\text{int}[\phi] =& \frac{\alpha^2}8\int d^4xd^4y\sqrt{-g}\phi^2(x)\frac{\delta^4(x-y)}{-\Box+m^2}\phi^2(y)\nonumber\\
    =&~\frac{\alpha^2}{8m^2}\left[
    \int d^4x\sqrt{-g}\phi^4(x) + \int d^4xd^4y\sqrt{-g}\phi^2(x)\frac{\delta^4(x-y)}{-\Box+m^2} \Box\phi^2(y)\right]\ .
\end{align}
To make the result more symmetric in $x\leftrightarrow y$, we can play the trick again but now acting $\Box$ on position $x$, and we obtain:
\begin{align}
    \bar S_\text{int}[\phi]
    =&~\frac{\alpha^2}{8m^2}\left[\int d^4x\sqrt{-g}\left(\phi^4+\frac{1}{m^2}\phi^2\Box\phi^2\right) + \frac{1}{m^2}
    \int d^4xd^4y\sqrt{-g}\Box\phi^2(x)\frac{\delta^4(x-y)}{-\Box+m^2} \Box\phi^2(y)\right]\ .
\end{align}
Finally, one can perform another shift:
\begin{align}
    \tilde \sigma(x) \equiv \bar\sigma(x) - \frac{\alpha^2}{2m^2}\int d^4y \frac{\delta^4(x-y)}{-\Box+m^2}\Box\phi^2(y)\ ,
\end{align}
and after changing the integral variable $\bar\sigma\to\tilde\sigma$ and renaming $\tilde\sigma\to\sigma$, we can rewrite the partition function back into the form of a two-field system:
\begin{align}
    Z = \int\mathcal D\phi\mathcal D\sigma e^{i\tilde S[\phi,\sigma]}\ ,
\end{align}
with the action:
\begin{align}
    \tilde S[\phi,\sigma] =
    S_0[\phi]+S_0[\sigma] + \frac{\alpha^2}{8m^2}\int d^4x\sqrt{-g}\left(\phi^4+\frac{1}{m^2}\phi^2\Box\phi^2\right)
    -\frac{\alpha}{2m^2}\int d^4x\sqrt{-g}\Box\phi^2\sigma\ .
\end{align}

The new action is equivalent to the original, where the only term without derivatives is the $\phi^4$ operator, which is responsible for the IR divergence of the four-point correlator of $\phi$. Furthermore, at the tree level, $\phi$ is the on-shell external field and thus satisfies the equation of motion:
\begin{equation}
    \Box\phi^2 = 2(\partial_\mu\phi)^2 + 2\phi\Box\phi = 2(\partial_\mu\phi)^2\ .
\end{equation}
Therefore, for the tree-level four-point correlator, the original theory is equivalent to the following: 
\begin{equation}
    S_\text{equiv}[\phi,\sigma] = S_0[\phi]+S_0[\sigma] + \int d^4x\sqrt{-g}\left[\frac{\alpha^2}{8m^2}\phi^4+\frac{\alpha^2}{4m^4}\phi^2(\partial_\mu\phi)^2\right]
    -\frac{\alpha}{m^2}\int d^4x\sqrt{-g}(\partial_\mu\phi)^2\sigma\ .\label{eq_Seq}
\end{equation}

Notice that we have not used the explicit form of the metric during the derivation. That is to say, the two actions \eqref{eq_S} and \eqref{eq_Seq} are equivalent in any spacetime. As a consistent check, let us compute the tree-level scattering amplitude $\mathcal M $ of the process $\phi\phi\to\phi\phi$ in flat space, with $\phi$ massless and $\sigma$ of mass $m$. From the original action \eqref{eq_S}, we have:
\begin{align}
    \mathcal M = (-i\alpha)^2\left(\frac{-i}{k_s^2+m^2}+\frac{-i}{k_t^2+m^2}+\frac{-i}{k_u^2+m^2}\right)\ ,
\end{align}
where $k_s^2 = (k_1+k_2)^2$ is the exchanged momentum in $s$-channel, and similar for $k_t$ and $k_u$.\footnote{In this flat space example we have abused notation with $k_{i,s,t,u}$ denoting four-momenta.}
From the equivalent action \eqref{eq_Seq} we have:
\begin{align}
    \mathcal M =&~ 4!\times \frac{i\alpha^2}{8m^2} + 4\times \frac{i\alpha^2}{4m^4}\Big[(-k_1\cdot k_2) + 5~\mathrm{perms}\Big]\nonumber\\
    &+ 4\times\left(\frac{-i\alpha}{m^2}\right)^2\left[\frac{-i(-k_1\cdot k_2)(-k_3\cdot k_4)}{k_s^2+m^2}+t-\mathrm{and}\  u- \mathrm{channels}\right]\ .\label{eq_amplitude}
\end{align}
Using the on-shell condition $k_i^2=0$, one can easily check that the two expressions are equal.

Now let us go back to the de Sitter case. With this equivalent action, it is easy to see that there are three contributions to the tree-level four-point correlators:
\begin{align}
    \langle \varphi_{\bm k_1}\varphi_{\bm k_2}\varphi_{\bm k_3}\varphi_{\bm k_4}\rangle' = \mathcal I_1 + \mathcal I_2 + \mathcal I_3\ .
\end{align}
Here $\mathcal I_1$ comes from the contact $\phi^4$ interaction:
\begin{align}
    \mathcal I_1 = &~4!\times \sum_{\pm}\frac{\pm i\alpha^2}{8m^2}\int_{-\infty}^{\eta_0} \frac{d\eta}{(-H\eta)^4}K_\pm(k_1,\eta)K_\pm(k_2,\eta)K_\pm(k_3,\eta)K_\pm(k_4,\eta)\nonumber\\
    =&-\frac{\alpha^2H^4}{8m^2\prod k_i^3}
    \sum k_i^3\log(-k_T\eta_0)\nonumber\\
    & + \frac{\alpha^2H^4}{8m^2\prod k_i^3k_T}\left\{ (1-\gamma)\sum k_i^4+\sum_{i\neq j}\left[(2-\gamma) k_i^3k_j + k_i^2k_j^2\right] + \frac12\sum_{i\neq j\neq \ell}k_i^2k_jk_\ell- \prod k_i\right\}\ .
\end{align}
Notice that in this calculation we cannot directly set $\eta_0\to 0$ in the boundary propagators $K_\pm$ as it will contribute to a finite term in the final result.

Similarly, $\mathcal I_2$ comes from the contact $\phi^2(\partial_\mu\phi)^2$ interaction. Since there is no IR divergence now, we can safely set $\eta_0\to 0$ for the boundary propagators, with corrections vanishing in the late-time limit,
\begin{align}
    \mathcal I_2 = &~4\times \sum_{\pm}\frac{\pm i\alpha^2}{4m^4}\int_{-\infty}^{\eta_0} \frac{d\eta}{(-H\eta)^2}K_\pm(k_1,\eta)K_\pm(k_2,\eta)\Big[ -K_\pm'(k_3,\eta)K_\pm'(k_4,\eta)\nonumber\\
    & -\bm k_3\cdot \bm k_4 K_\pm(k_3,\eta)K_\pm(k_4,\eta)\Big] + 5~\mathrm{perms}\nonumber\\
    =&~\frac{\alpha^2H^6\sum k_i^3}{16m^4\prod k_i^3}\ .
\end{align}
Notice that there is no associated $k_T$ pole, indicating the corresponding scattering amplitude in flat space, i.e., the second term on the RHS of \eqref{eq_amplitude}, is vanishing.

Finally, $\mathcal I_3$ comes from the exchange diagram with the $(\partial_\mu\phi)^2\sigma$ interaction:
\begin{align}
    \mathcal I_3 =&~ 4\times \sum_{\pm,\widehat\pm}\frac{(\pm i)(\widehat\pm i)\alpha^2}{m^4}\int_{-\infty}^{\eta_0}\frac{d\eta_1}{(-H\eta_1)^2}\frac{d\eta_2}{(-H\eta_2)^2}
    \left[-K_\pm'(k_1,\eta_1)K_\pm'(k_2,\eta_1)-\bm k_1\cdot\bm k_2 K_\pm(k_1,\eta_1)K_\pm(k_2,\eta_1)\right] \nonumber\\
    &\times G_{\pm\widehat\pm}(s,\eta_1,\eta_2)\times\left[-K'_{\widehat\pm}(k_3,\eta_2)K'_{\widehat\pm}(k_4,\eta_2)-\bm k_3\cdot\bm k_4 K_{\widehat\pm}(k_3,\eta_2)K_{\widehat\pm}(k_4,\eta_2)\right] + t-\mathrm{and}\  u- \mathrm{channels}\nonumber\\
    =&~\frac{4\alpha^2H^4}{m^4}\sum_{\pm,\widehat\pm}(\pm i)(\widehat\pm i)\int_{-\infty}^{\eta_0}\frac{d\eta_1}{\eta_1^2}\frac{d\eta_2}{\eta_2^2}  \frac{\mathcal O_{12} e^{\pm i k_{12}\eta_1}}{4k_1^3k_2^3}\times  G_{\pm\widehat\pm}(s,\eta_1,\eta_2)\times \frac{\mathcal O_{34} e^{\widehat\pm i k_{34}\eta_2}}{4k_1^3k_2^3} +t-\mathrm{and}\  u- \mathrm{channels}\nonumber\\
    =&~\frac{\alpha^2H^6}{4m^4\prod k_i^3 s}\mathcal O_{12}\mathcal O_{34} \mathcal I\left(\frac{s}{k_{12}},\frac{s}{k_{34}}\right)+t-\mathrm{and}\  u- \mathrm{channels}\ .
\end{align}
Here we have defined the weight-shifting operators:
\begin{equation}
    \mathcal O_{ij} \equiv k_i^2k_j^2\partial_{k_{ij}}^2 - \bm k_i\cdot\bm k_j\left(1-k_{ij}\partial_{k_{ij}} + k_ik_j\partial_{k_{ij}}^2\right)\ ,
\end{equation}
and the dimensionless seed integral $\mathcal I$, whose analytical result can be found in e.g. \cite{Arkani-Hamed:2018kmz,Qin:2022fbv}:
\begin{align}
    \mathcal I(u,v) \equiv&~ \frac{s}{H^2}\sum_{\pm,\widehat\pm}(\pm i)(\widehat\pm i)\int_{-\infty}^{\eta_0}\frac{d\eta_1}{\eta_1^2}\frac{d\eta_2}{\eta_2^2}   G_{\pm\widehat\pm}(s,\eta_1,\eta_2) e^{\pm i k_{12}\eta_1\widehat\pm i k_{34}\eta_2}\nonumber\\
    =&~\frac{1+i\sinh\pi\mu}{2\pi}\left[
    \mathbf F_+(u)\mathbf F_+(v) + \mathbf F_+(u)\mathbf F_-(v)\right]+ (\mu\to-\mu)\nonumber\\
    &+\sum_{m,n=0}^\infty\frac{(-1)^n(n+1)_{2m}}{2^{2m+1}(\frac n2+\frac14+\frac{i\mu}2)_{m+1}(\frac n2+\frac14-\frac{i\mu}2)_{m+1}}u^{2m+1}\left(\frac{u}{v}\right)^{n}\ ,
\end{align}
where we have defined $\mu\equiv \sqrt{m^2/H^2-9/4}$ and have assumed $k_{12}>k_{34}$ for the convergence of the series representation and defined the homogeneous solution to the bootstrap equation:
\begin{equation}
    \mathbf F_\pm(r) \equiv r^{1/2}\left(\frac{r}{2}\right)^{\pm i\mu}\Gamma\left(\frac12\pm i\mu\right)\Gamma\left(\mp i\mu\right) {}_2\mathrm F_1\left[
    \begin{matrix}
        \frac14\pm\frac{i\mu}2,\frac34\pm\frac{i\mu}2\\
        1\pm i\mu
    \end{matrix}\middle|r^2
    \right]\ .
\end{equation}

\section{Four-Point Function from Bubble Loop}
\label{app:bubble}
Working in $(d+1)$-dimensional dS spacetime, the four-point correlator generated from the interaction \eqref{eq_Lbubble} reads:
\begin{align}
\la\varphi_{\bm k_1}\varphi_{\bm k_2}\varphi_{\bm k_3}\varphi_{\bm k_4}\ra'_d
=&~
g^2\mu_R^{2\epsilon}\sum_{\pm,\widehat\pm}(\pm i)(\widehat\pm i)\int_{-\infty}^{\tf} \frac{d\eta_1}{(-H\eta_1)^{d+1}}\frac{d\eta_2}{(-H\eta_2)^{d+1}}K_{\pm}(k_1,\eta_1)K_{\pm}(k_2,\eta_1)K_{\widehat\pm}(k_3,\eta_2)K_{\widehat\pm}(k_4,\eta_2) \nonumber\\
& \times \frac12\int \frac{d^d\bm q}{(2\pi)^d}G_{\pm\widehat\pm}^{(\mu)}(q;\eta_1,\eta_2)G_{\pm\widehat\pm}^{(\mu)}(|\bm s -\bm q|;\eta_1,\eta_2) +t-\mathrm{and}\  u- \mathrm{channels}\ .
\label{eq_BubbleFeyn}
\end{align}
Notice that the symmetry factor $1/2$ is introduced for a bubble composed of identical particles, and we add a superscript $(\mu)$ on the $(d+1)$-dimensional bulk-to-bulk propagator to denote its mass via the equality $\mu\equiv \sqrt{m^2/H^2-9/4}$. Without loss of generality, we assume $m>3H/2$ and thus $\mu$ is real, but the final result of the four-point function for $0<m<3H/2$ can be directly obtained by analytical continuation.

Most of the computational difficulty comes from the loop integral. Fortunately, it can be circumvented by making use of the spectral decomposition. We encourage readers to refer to~\cite{Marolf:2010zp,Xianyu:2022jwk} for the derivation and technical details, and here we only list the results.

Using the spectral decomposition, we can rewrite the loop integral as a spectral integral of a single bulk-to-bulk propagator of mass $\mu'$, weighted by a spectral function:
\begin{align}
\frac12 \int \frac{d^d\bm s}{(2\pi)^d}G_{\pm\widehat\pm}^{(\mu)}(s,\eta_1,\eta_2)G_{\pm\widehat\pm}^{(\mu)}(|\bm s -\bm q|;\eta_1,\eta_2) = H^{2d-4}\int_{-\infty}^{+\infty} \frac{\mu' d\mu'}{2\pi i}\rho^d_\mu(\mu')G^{(\mu')}_{\pm\widehat\pm}(s,\eta_1,\eta_2)\ ,\label{eq_SpecDec}
\end{align}
where the spectral function $\rho_{\mu}^d(\mu')$ is given by:\footnote{There is an equivalent spectral function (K\"all\'en-Lehmann spectral density) with a simpler form without involving generalized hypergeometric functions~\cite{Loparco:2023rug}. However, we still choose to use the expression in Eq.~\eqref{eq_SpecDec} since here the UV divergence is manifest, as will be explained below.}
\begin{align}
\rho_{\mu}^d(\mu') =&~\frac1{(4\pi)^{(d+1)/2}}\frac{\cos[\pi(\frac d2-\ii\mu)]}{\sin(-\pi \ii \mu)}     
\frac{\Gamma(\frac{3-d}2)\Gamma(\frac d2-i\mu)}{\Gamma(\frac{2-d}2-i\mu)} \\
    &~\times {}_7\mathcal{F}_6\left[\begin{matrix}
        \frac{2-d}2+\ii \mu'-\ii \mu, \frac{3-d/2+\ii \mu'-\ii \mu}2,  \frac{2-d}2, \frac{2-d}2-\ii \mu, \frac{2-d}2+\ii \mu', \frac{\ii \mu'-2\ii \mu+d/2}2, \frac{\ii \mu'+d/2}2 \\
        \frac{1-d/2+\ii \mu'-\ii \mu}2, 1+\ii \mu'-\ii \mu, 1+\ii \mu', 1-\ii \mu, \frac{4+\ii \mu'-3d/2}2, \frac{4+\ii \mu'-2\ii \mu-3d/2}2
    \end{matrix}\middle|1\right]\n\\
    &~+(\mu\to-\mu)\ ,
\end{align}
with the dressed generalised hypergeometric function defined as:
\begin{align}
{}_p\mathcal F_q\left[ \begin{matrix}  a_1,\cdots,a_p \\ b_1,\cdots,b_q
\end{matrix}\middle| z\right] \equiv \frac{\prod \Gamma(a_i)}{\prod \Gamma(b_j)} \times {}_p\mathrm F_q\left[ \begin{matrix}  a_1,\cdots,a_p \\ b_1,\cdots,b_q
\end{matrix}\middle| z\right]\ .
\end{align}
Inserting the spectral decomposition \eqref{eq_SpecDec} back into Eq.~\eqref{eq_BubbleFeyn}, we obtain:
\begin{align}
\la\varphi_{\bm k_1}\varphi_{\bm k_2}\varphi_{\bm k_3}\varphi_{\bm k_4}\ra'_d
=&~g^2H^2\left(\frac{\mu_R}{H}\right)^{2\epsilon}
\int_{-\infty}^{+\infty} \frac{\mu' d\mu'}{2\pi i}\rho^d_\mu(\mu')
\bigg[ \sum_{\pm,\widehat\pm}(\pm i)(\widehat\pm i) \int_{-\infty}^{\tf} \frac{d\eta_1}{(-H\eta_1)^{d+1}}\frac{d\eta_2}{(-H\eta_2)^{d+1}}\nonumber\\
&~ K_{\pm}(k_1,\eta_1)K_{\pm}(k_2,\eta_1)\times G^{(\mu')}_{\pm\widehat\pm}(s,\eta_1,\eta_2)  \times K_{\widehat\pm}(k_3,\eta_2)K_{\widehat\pm}(k_4,\eta_2) \bigg]\nonumber\\
&+t-\mathrm{and}\  u- \mathrm{channels}\ .
\end{align}

This expression spells out the UV and IR divergences.
On the one hand, the UV divergence comes from the loop integral, as one would expect, and is manifest as the pole at $d=3$ of the factor $\Gamma((3-d)/2)$ in the spectral function. In particular, we have~\cite{Xianyu:2022jwk}:
\begin{align}
\lim_{d\to 3} \rho_\mu^d(\mu') = - \frac{1}{8\pi^2\epsilon}\ .
\end{align}
To deal with this UV divergence, we adopt the $\overline{\mathrm{MS}}$ renormalisation scheme and introduce the renormalised spectral function:
\begin{align}
\wh \rho_{\mu}(\mu') \equiv \lim_{d\to3} \Big[ \rho^d_{\mu}(\mu')+\frac{1}{16\pi^2}\Big(\frac{2}{3-d}-\gamma_E+\log4\pi\Big)\bigg]\ .\label{eq_def_rhohat}
\end{align}
Therefore, replacing the bare spectral function by the renormalised one while keeping trace of the renormalisation scale $\mu_R$, we obtain the renormalised four-point correlator:\footnote{Notice that the counterterm in $(d+1)$ dimension should be proportional to $\mu_R^{3-d}a^{d+1}\phi^4 \sim (1+\epsilon\log \mu_R) a^{d+1}\phi^4$, which would contribute an additional factor of $-\log(\mu_R^2/H^2)/(16\pi^2)$ to the renormalised correlator \eqref{eq_4ptbubble_intermediate} after cancelling the divergence in $\rho_\mu^d(\mu')$ manifest in Eq.~\eqref{eq_def_rhohat}.}
\begin{align}
\la\varphi_{\bm k_1}\varphi_{\bm k_2}\varphi_{\bm k_3}\varphi_{\bm k_4}\ra'_{\mathrm{re}}
=&~g^2H^2
\int_{-\infty}^{+\infty} \frac{\mu' d\mu'}{2\pi i}
\left[\wh\rho_\mu(\mu')
-\frac{1}{16\pi^2}\log \left(\frac{\mu_R^2}{H^2}\right)\right]
\bigg[ \sum_{\pm,\widehat\pm}(\pm i)(\widehat\pm i) \int_{-\infty}^{\tf} \frac{d\eta_1}{(-H\eta_1)^{4}}\frac{d\eta_2}{(-H\eta_2)^{4}}\nonumber\\
&~ K_{\pm}(k_1,\eta_1)K_{\pm}(k_2,\eta_1)\times D^{(\mu')}_{\pm\widehat\pm}(s,\eta_1,\eta_2)  \times K_{\widehat\pm}(k_3,\eta_2)K_{\widehat\pm}(k_4,\eta_2) \bigg]\nonumber\\
&+t-\mathrm{and}\  u- \mathrm{channels}\ .
\label{eq_4ptbubble_intermediate}
\end{align}

On the other hand, notice that the term in the second square brackets is nothing but a four-point correlator from an exchange of a massive particle with mass $m'=H\sqrt{\mu'^2+9/4}$ via the direct interaction $\phi^2\sigma/2$, which has been calculated in App.\ \ref{app:UVcomputation}. In particular, if we only focus on the leading IR behavior, we have:
\begin{align}
\la\varphi_{\bm k_1}\varphi_{\bm k_2}\varphi_{\bm k_3}\varphi_{\bm k_4}\ra'_{\mathrm{IR}}
= &~g^2
\int_{-\infty}^{+\infty} \frac{\mu' d\mu'}{2\pi i}
\left[\wh\rho_\mu(\mu')
-\frac{1}{16\pi^2}\log\left(\frac{ \mu_R^2}{H^2}\right)\right]
\bigg[ -\frac{H^4\sum k_i^3}{8(\mu'^2+9/4)\prod k_i^3}\log(-k_T\eta_0)\bigg]\ .
\end{align}
To evaluate this integral, we close the contour in the lower half-plane and apply the residue theorem. Notice that there is no pole on the lower half-plane of $\mu'$ for the spectral function $\rho_\mu^d(\mu')$ or $\widehat\rho_\mu(\mu')$, so the only pole for the integrand is a single simple pole at $\mu’=-3i/2$ from the factor $1/(\mu'^2+9/4)$, with the residue:
\bge
\text{Res}\Big|_{\mu'=- 3i/2} \Big(\frac{1}{\mu'^2+9/4}\Big) = \frac i3\ .
\ede
Plugging this into the contour integral, we find,
\begin{align}
\la\varphi_{\bm k_1}\varphi_{\bm k_2}\varphi_{\bm k_3}\varphi_{\bm k_4}\ra'_\text{IR}
=&~\frac{g^2H^4\sum k_i^3}{16\prod k_i^3}
\left[\wh\rho_\mu\left(\frac{-3i}2\right)
-\frac{1}{16\pi^2}\log\left(\frac{\mu_R^2}{H^2}\right)\right]
\log(-k_T\eta_0)\ .
\end{align}
%%%%%%%%%%%%%%%%%%%%%%%%%%%%%%%%%%%%%%%%%%%%%%%%%%%%%%%%%%%%%%%%%%%%%%%%%%%%%%%%%%%%%%%%%%%%%%%%%%%%%%%%

\section{Wavefunction Approach}
\label{app:wvfn}

In this appendix, we will compute some useful tree-level $n$-point wavefunction coefficients, including the two-point coefficient $\psi_2^\varsigma$ for a free scalar $\sigma$, the three-point coefficient $\psi_3^{\varphi\varphi\varsigma}$ with interaction $\phi^2\sigma$, and the four-point coefficient for $\phi$ from the exchange of $\sigma$ via the same interaction. We will mainly focus on the real parts of these wavefunction coefficients since it is their real parts that finally contribute to the correlator.
%%%%%%%%%%%%%%%%%%%%%%%%%%%%%%%%%%%%%%%%%%%%%%%%%%%%%%%%%%%%%%%%%%%%%%%%%%%%%%%%%%%%%%%%%%%%%%%%%%%%%%%%
\subsection{Two-point wavefunction coefficents}
Let us start from the two-point wavefunction coefficient for a free scalar $\sigma$. Inserting the classical solution \eqref{eq_clasolution} into the saddle-point approximation \eqref{eq_treeWF}, and comparing it with Eq.~\eqref{def_wvcoe}, we obtain the two-point function for $\sigma$ after integrating by parts and making use of EoM,
\begin{align}
\psi_2^\varsigma =&~\left. i a^2(\eta) K(k,\eta)\partial_\eta K(k,\eta) \right|_{\eta_0} = \left. i a^2(\eta) \frac{\sigma'^*(k,\eta)}{\sigma^*(k,\eta)} \right|_{\eta_0}\ ,
\end{align}
whose real part reads:
\begin{align}
\Re \psi_2^\varsigma = \Re\left[
 \left. i a^2(\eta) \frac{\sigma'^*(k,\eta)}{\sigma^*(k,\eta)} \right|_{\eta_0}
\right]= -\frac{a^2(\eta_0)}{|\sigma(k,\eta_0)|^2}\times  \Im\left[\si(k,\eta_0)\si'^*(k,\eta_0)\right]\ .
\end{align}
Now notice the Wronskian condition \eqref{eq_Wronski},
\bge
\si(k,\eta)\si'^*(k,\eta) - \si^*(k,\eta)\si'(k,\eta) =  \frac{i}{a^2(\eta)}
\ede
which implies that:
\bge
\Im\Big[\si(k,\eta_0)\si'^*(k,\eta_0)\Big] = \frac{1}{2a^2(\eta_0)}\ ,
\ede
and we immediately obtain the real part of the power spectrum:
\bge
\Re \psi_2^\varsigma = -\frac{1}{2|\si(k,\eta_0)|^2}\ .
\label{eq_powerspec}
\ede
In particular, for massless scalar $\phi$, we have:
\bge
\Re\psi^\varphi_2 = -\frac{k^3}{H^2}\ ,
\label{eq_masslesspowerspec}
\ede
whose inverse gives the standard power spectrum up to a factor of 2. One can also verify that we have the following identities:
\begin{align}
    G^{\sigma}(k,t,t')+\frac{K^\sigma(k,t)\sigma(k,t)}{2\mathrm{\Re}\psi^\sigma_2}&=G_{++}^\sigma(k,t,t')\ ,\\
     (G^{\sigma})^*(k,t,t')+\frac{K_\sigma^*(k,t)K^*_\sigma(k,t)}{2\mathrm{\Re}\psi^\sigma_2}&=G_{--}^\sigma(k,t,t')\ ,\\
     \frac{K_\sigma(k,t)K^*_\sigma(k,t)}{2\mathrm{\Re}\psi^\sigma_2}&=G_{+-}^\sigma(k,t,t')\ ,
     \label{props_indent}
\end{align}

\subsection{Three-point wavefunction coefficients}
Let us now proceed to consider the three-point wavefunction coefficient $\psi_3^{\varphi\varphi\varsigma}$ with the interaction $-\alpha \phi^2\sigma/2$. It is easy to see that $\psi_3^{\varphi\varphi\varsigma}$ is dominant by its imaginary part:
\begin{align}
\Im\psi_3^{\varphi\varphi\varsigma} =&~ \Im\bigg[ -i\alpha \int_{-\infty}^{\eta_0} \frac{d\eta}{(-H\eta)^4}\,K_\phi(k_1,\eta)K_\phi(k_2,\eta)K_\si(k_3,\eta)\bigg] = \frac{\alpha}{3H^4\eta_0^3} + \cdots\ .
\label{eq_3ptIm}
\end{align}
where we do the late-time expansion to the integrand in the last step and only keep the leading order. On the other hand,
the real part of $\psi_3^{\varphi\varphi\varsigma}$ is given by:
\begin{align}
\Re\psi_3^{\varphi\varphi\varsigma} =&~ \Re\bigg[ -i \alpha \int_{-\infty}^{\eta_0} \frac{d\eta}{(-H\eta)^4}\,K_\phi(k_1,\eta)K_\phi(k_2,\eta)K_\si(k_3,\eta)\bigg]\nonumber\\
=&~\Re\bigg[-i \alpha \int_{-\infty}^{\eta_0} \frac{d\eta}{H^4\eta^4}\,K_\phi(k_1,\eta)K_\phi(k_2,\eta)\frac{\si^*(k_3,\eta)}{\si^*(k_3,\eta_0)}\bigg]\nonumber\\
=&~\frac{-\alpha}{H^4|\si(k_3,\eta_0)|^2}\Re\bigg[  \si(k_3,\eta_0) \times i \int_{-\infty}^{\eta_0} \frac{d\eta}{\eta^4}\,K_\phi(k_1,\eta)K_\phi(k_2,\eta)\si^*(k_3,\eta)\bigg]\ .
\label{eq_3pt_intermediate}
\end{align}

To extract the dominant term in the IR, we can again do a late-time expansion of the time integrand, keeping the most divergent terms.
Let us consider the case that $\sigma$ has a mass $m<3H/2$ and define $\nu\equiv \sqrt{9/4-m^2/H^2}$. We will find:
\bge
\frac{1}{\eta^4}K_\phi(k_1,\eta)K_\phi(k_2,\eta) = \frac{1}{\eta^4}+\frac{k_1^2+k_2^2}{2\eta^2} +i\frac{k_1^3+k_2^3}{3\eta} + \cdots\ ,
\ede
and
\bge
\Re\Big[ i \sigma(k_3,\eta_0)\sigma^*(k_3,\eta)\Big] = \frac{H^2\csc\pi\nu}{4\nu} \Im\Big[e^{i\pi\nu}(-\eta_0)^{3/2-\nu}(-\eta)^{3/2+\nu} + e^{-i\pi\nu}(-\eta_0)^{3/2+\nu}(-\eta)^{3/2-\nu} \Big] + \cdots\ ,
\ede
and thus the leading term of the real part in the bracket of Eq.~\eqref{eq_3pt_intermediate} is:
\begin{align}
&~ \frac{H^2\csc\pi\nu}{4\nu}\Im \Big[ e^{i\pi\nu}(-\eta_0)^{3/2-\nu}\int^{\eta_0} d\eta\,(-\eta)^{-5/2+\nu} + e^{-i\pi\nu}(-\eta_0)^{3/2+\nu}\int^{\eta_0} d\eta\,(-\eta)^{-5/2-\nu} \Big]\nonumber\\
=&~ \frac{H^2\csc\pi\nu}{4\nu}\Im \Big[ \frac{e^{i\pi\nu}}{3/2-\nu} + \frac{e^{-i\pi\nu}}{3/2+\nu} \Big]\nonumber\\
=&~ \frac{2H^2}{9-4\nu^2}\nonumber\\
=&~\frac{H^4}{2m^2}\ .
 \end{align}
 Therefore, we finally have:
 \bge
 \Re \psi_3^{\varphi\varphi\varsigma} = \frac{-\alpha}{H^4|\si(k_3,\eta_0)|^2} \times \frac{H^4}{2m^2} \times \Big[1+\mathcal O(\eta_0^2)\Big] = \frac{\alpha}{m^2}\Re\psi_2^\varsigma \times \Big[1+\mathcal O(\eta_0^2)\Big]\ .
 \ede
In particular, the leading contribution of the three-point correlator is:
 \bge
\la \varphi_{\bm k_1} \varphi_{\bm k_2} \varsigma_{\bm k_3}\ra' = - \frac{\Re \psi_3^{\varphi\varphi\varsigma}}{4\Re \psi_2^{\varphi}\times \Re \psi_2^{\varphi} \times \Re \psi_2^{\varsigma}} =-  \frac{\alpha}{m^2} \times \frac{H^4}{4k_1^3k_2^3}\ ,
 \ede
which is free from IR divergence, so long as $m\neq 0$.

\subsection{Four-point wavefunction coefficients}
Finally, we go into the final target, namely the $s$-channel four-point wavefunction coefficient still with the interaction $-\alpha\phi^2\sigma/2$:
\bge
\la \varphi_{\bm k_1} \varphi_{\bm k_2} \varphi_{\bm k_3} \varphi_{\bm k_4} \ra'_s = \frac{1}{8[\Re\psi_2^{\varphi}]^4}\Big[ \Re\psi_4 - \frac{\Re\psi_3^{\varphi\varphi\varsigma}\times \Re\psi_3^{\varsigma\varphi\varphi}}{\Re\psi_2^\varsigma}\Big]\ ,
\ede
where we already know that:
\bge
\frac{1}{8[\Re\psi_2^{\varphi}]^4} = \frac{1}{8k_1^3k_2^3k_3^3k_4^3}\ ,\qquad \frac{\Re\psi_3^{\varphi\varphi\varsigma}\times \Re\psi_3^{\varsigma\varphi\varphi}}{\Re\psi_2^\varsigma} = \frac{\alpha^2\psi_2^\si}{m^4} = -\frac{\alpha^2}{m^4} \times \frac{1}{2|\si(s,\eta_0)|^2}\ .
\ede

The only remaining part is the connected $\psi_4$:
\begin{align}
\Re\psi_4 =&~ \Re \bigg[ -\alpha^2\int_{-\infty}^{\eta_0} \frac{d\eta_1}{H^4\eta_1^4}\frac{d\eta_2}{H^4\eta_2^4}K_1K_2K_3K_4 \times G_\si(s,\eta_1,\eta_2)
\bigg]\nonumber\\
=&~\Re \bigg\{ - \frac{\alpha^2}{H^8}\int_{-\infty}^{\eta_0} \frac{d\eta_1}{\eta_1^4}\frac{d\eta_2}{\eta_2^4}K_1K_2K_3K_4\nonumber\\
& \times  \Big[ \si(s,\eta_1)\si^*(s,\eta_2)  \theta(\eta_1-\eta_2) + (\eta_1\leftrightarrow \eta_2) - \si^*(s,\eta_1)\si^*(s,\eta_2)\frac{\si(s,\eta_0)}{\si(s,\eta_0)^*}\Big]
\bigg\}\nonumber\\
=& ~\frac{\alpha^2}{H^8}\int_{-\infty}^{\eta_0} \frac{d\eta_1}{\eta_1^4}\frac{d\eta_2}{\eta_2^4}\, \Im\Big[ K_1K_2K_3K_4\Big] \Im\Big[\si(s,\eta_1)\si^*(s,\eta_2)\Big] \text{sn}(\eta_1-\eta_2)\nonumber\\
&-\frac{\alpha^2}{H^8}\int_{-\infty}^{\eta_0} \frac{d\eta_1}{\eta_1^4}\frac{d\eta_2}{\eta_2^4}\, \Re\Big[ K_1K_2K_3K_4 \Big]\Re\Big[\si(s,\eta_1)\si^*(s,\eta_2)\Big]\nonumber\\
&-\frac{\alpha^2}{H^8}\Re \Big[i \int_{-\infty}^{\eta_0} \frac{d\eta_1}{\eta_1^4}\,K_1K_2K_\si \times i \int_{-\infty}^{\eta_0}\frac{d\eta_2}{\eta_2^4}\,K_3K_4K_\si\Big] \times |\si(s,\eta_0)|^2\ ,
\end{align}
where we have used shorthand:
\bge
K_{1,2}\equiv K_\phi(k_{1,2},\eta_1),\qquad K_{3,4}\equiv K_\phi(k_{3,4},\eta_2).
\ede
Notice that the last term above, combining with the disconnected part, gives:
\begin{align}
&-\Re \Big[\ii \int_{-\infty}^{\eta_0} \frac{\di\eta_1}{\eta_1^4}\,K_1K_2K_\si \times \ii \int_{-\infty}^{\eta_0}\frac{\di\eta_2}{\eta_2^4}\,K_3K_4K_\si\Big] \times |\si(s,\eta_0)|^2 - \frac{\Re\psi_3^{\phi\phi\si}\times \Re\psi_3^{\si\phi\phi}}{\Re\psi_2^\si}\nonumber\\
=&-\Re \Big[\ii \int_{-\infty}^{\eta_0} \frac{\di\eta_1}{\eta_1^4}\,K_1K_2K_\si \times \ii \int_{-\infty}^{\eta_0}\frac{\di\eta_2}{\eta_2^4}\,K_3K_4K_\si\Big] \times |\si(s,\eta_0)|^2\nonumber\\
& +2 |\si(s,\eta_0)|^2 \times \Re \Big[\ii \int_{-\infty}^{\eta_0} \frac{\di\eta_1}{\eta_1^4}\,K_1K_2K_\si \Big]\times \Re \Big[ \ii \int_{-\infty}^{\eta_0}\frac{\di\eta_2}{\eta_2^4}\,K_3K_4K_\si\Big]\nonumber\\
=&~|\si(s,\eta_0)|^2 \times \Re\Big[\int_{-\infty}^{\eta_0} \frac{\di\eta_1}{\eta_1^4}\,K_1K_2K_\si \times \int_{-\infty}^{\eta_0} \frac{\di\eta_2}{\eta_2^4}\,K_1^*K_2^*K_\si^*\Big]\nonumber\\
=&~\int_{-\infty}^{\eta_0}\frac{\di\eta_1}{\eta_1^4}\frac{\di\eta_2}{\eta_2^4}\,\Re\Big[K_1K_2K_3^*K_4^* \times  \si^*(s,\eta_1)\si(s,\eta_2)\Big]\nonumber\\
=&~\int_{-\infty}^{\eta_0}\frac{\di\eta_1}{\eta_1^4}\frac{\di\eta_2}{\eta_2^4}\,\Re\Big[K_1K_2K_3^*K_4^*\Big]\Re\Big[\si(s,\eta_1)\si^*(s,\eta_2)\Big]\nonumber\\
& + \int_{-\infty}^{\eta_0}\frac{\di\eta_1}{\eta_1^4}\frac{\di\eta_2}{\eta_2^4}\,\Im\Big[K_1K_2K_3^*K_4^*\Big]\Im\Big[\si(s,\eta_1)\si^*(s,\eta_2)\Big]\ .
\end{align}

So in total, we have:
\begin{align}
\label{eq_4ptfromWF}
&~\Re\psi_4 - \frac{\Re\psi_3^{\phi\phi\si}\times \Re\psi_3^{\si\phi\phi}}{\Re\psi_2^\si}\nonumber\\
=& ~\frac{\alpha^2}{H^8}\int_{-\infty}^{\eta_0} \frac{\di\eta_1}{\eta_1^4}\frac{\di\eta_2}{\eta_2^4}\, \Im\Big[ K_1K_2K_3K_4\Big] \Im\Big[\si(s,\eta_1)\si^*(s,\eta_2)\Big] \text{sn}(\eta_1-\eta_2)\nonumber\\
&-\frac{\alpha^2}{H^8}\int_{-\infty}^{\eta_0} \frac{\di\eta_1}{\eta_1^4}\frac{\di\eta_2}{\eta_2^4}\, \bigg( \Re\Big[ K_1K_2K_3K_4\Big] - \Re\Big[ K_1K_2K_3^*K_4^*\Big] \bigg)\Re\Big[\si(s,\eta_1)\si^*(s,\eta_2)\Big]\nonumber\\
& +\frac{\alpha^2}{H^8} \int_{-\infty}^{\eta_0} \frac{\di\eta_1}{\eta_1^4}\frac{\di\eta_2}{\eta_2^4}\,\Im\Big[K_1K_2K_3^*K_4^*\Big]\Im\Big[\si(s,\eta_1)\si^*(s,\eta_2)\Big]\ .
\end{align}

Now focusing on the IR-divergent piece, we do the late-time expansion. Still assuming $m<3H/2$, we have:
\begin{align}
\Re\Big[\si(s,\eta_1)\si^*(s,\eta_2) \Big]
=&~\frac{H^2}{4\pi}\Big(\frac{s}{2}\Big)^{-2\nu}(-\eta_1)^{3/2-\nu}(-\eta_2)^{3/2-\nu}\Gamma(\nu)^2 + \cdots\ ,\\
\Im\Big[\si(s,\eta_1)\si^*(s,\eta_2) \Big] =& - \frac{H^2}{4\nu} \Big[ \Big(\frac{\eta_1}{\eta_2}\Big)^{-\nu} - \Big(\frac{\eta_2}{\eta_1}\Big)^{-\nu}\Big] (-\eta_1)^{3/2}(-\eta_2)^{3/2} \Big[ 1+ \mathcal O(\eta_1^2,\eta_2^2)\Big]\ ,\\
\Im\Big[K_1K_2K_3K_4\Big] =&~\Big(\frac{k_1^3+k_2^3}3 \eta_1^3 + \frac{k_3^3+k_4^3}3 \eta_2^3\Big) \Big[ 1+ \mathcal O(\eta_1^2,\eta_2^2)\Big]\ ,\\
\Im\Big[K_1K_2K_3^*K_4^*\Big] =&~\Big(\frac{k_1^3+k_2^3}3 \eta_1^3 - \frac{k_3^3+k_4^3}3 \eta_2^3\Big) \Big[ 1+ \mathcal O(\eta_1^2,\eta_2^2)\Big]\ ,\\
\Re\Big[K_1K_2(K_3K_4-K_3^*K_4^*)\Big] =&~ \mathcal O(\eta_1^3\eta_2^3)\ ,
\end{align}
we find only the first term in \eqref{eq_4ptfromWF} will diverge in the late-time limit, giving rise to a $\log(-k_T\eta_0)$:
\begin{align}
& ~\int_{-\infty}^{\eta_0} \frac{\di\eta_1}{\eta_1^4}\frac{\di\eta_2}{\eta_2^4}\, \Im\Big[ K_1K_2K_3K_4\Big] \Im\Big[\si(s,\eta_1)\si^*(s,\eta_2)\Big] \text{sn}(\eta_1-\eta_2)\nonumber\\
=&~\int_{-\infty}^{\eta_0} \frac{\di\eta_1}{\eta_1^4}\frac{\di\eta_2}{\eta_2^4}\,\Big(\frac{k_1^3+k_2^3}3 \eta_1^3 + \frac{k_3^3+k_4^3}3 \eta_2^3\Big) \nonumber\\
&\times \frac{-1}{4\nu} \Big[ \Big(\frac{\eta_1}{\eta_2}\Big)^{-\nu} - \Big(\frac{\eta_2}{\eta_1}\Big)^{-\nu}\Big] (-\eta_1)^{3/2}(-\eta_2)^{3/2}\text{sn}(\eta_1-\eta_2)\nonumber\\
=&-\frac{4(k_1^3+k_2^3+k_3^3+k_4^3)}{3(9-4\nu^2)}\log(-k_T\eta_0)\nonumber\\
=&-\frac{H^2}{3m^2}(k_1^3+k_2^3+k_3^3+k_4^3)\log(-k_T\eta_0)\ .
\end{align}
Therefore, the IR divergent part for the correlator is:
\bge
\la \varphi_{\bm k_1} \varphi_{\bm k_2} \varphi_{\bm k_3} \varphi_{\bm k_4} \ra'_s = \frac{1}{8[\Re\psi_2^{\phi}]^4}\Big[ \Re\psi_4 - \frac{\Re\psi_3^{\phi\phi\si}\times \Re\psi_3^{\si\phi\phi}}{\Re\psi_2^\si}\Big] = - \frac{\alpha^2H^4(k_1^3+k_2^3+k_3^3+k_4^3)}{24k_1^3k_2^3k_3^3k_4^3m^2}\log(-k_T\eta_0)\ .
\ede
Together with contributions from the $t$- and $u$-channels, we find that the result perfectly matches what we have in the main text.
\addcontentsline{toc}{section}{References}
\bibliographystyle{utphys}
\bibliography{references.bib}

\end{document}